\documentclass[fleqn,usenatbib]{mnras}

\usepackage[T1]{fontenc}

\DeclareRobustCommand{\VAN}[3]{#2}
\let\VANthebibliography\thebibliography
\def\thebibliography{\DeclareRobustCommand{\VAN}[3]{##3}\VANthebibliography}

\usepackage{graphicx}	
\usepackage{amsmath}	
\usepackage{amssymb}	
\usepackage{xcolor}
\usepackage{cancel}
\usepackage{newtxtext,newtxmath}
\usepackage{arydshln}
\usepackage{longtable}
\usepackage{ulem}
\usepackage{soul}

\title[SN2021bnw and SN2021fpl Polarimetry]{Optical polarization and spectral properties of the H-poor superluminous supernovae SN 2021bnw and SN 2021fpl}

\author[F. Poidevin et al.]{F. Poidevin$^{1,2}$,\thanks{E-mail: fpoidevin@iac.es (IAC)}
 C. M. B. Omand$^{3}$,
R\'eka K\"onyves-T\'oth $^{4,5,6}$,
  I. P\'{e}rez-Fournon$^{1,2}$,
  R. Clavero$^{1,2}$,
S. Geier$^{1,2,7}$,
  \newauthor 
C. Jimenez Angel$^{1,2}$,
R. Marques-Chaves$^{8}$,
  R. Shirley$^{9,10}$
\\
$^{1}$ Instituto de Astrofis\'{i}ca de Canarias, 38200 La Laguna,Tenerife, Canary Islands, Spain\\
$^{2}$ Departamento de Astrof\'{\i}sica, Universidad de La Laguna (ULL), 38206 La Laguna, Tenerife, Spain\\
$^{3}$ The Oskar Klein Centre, Department of Astronomy, Stockholm University, AlbaNova, SE-10691 Stockholm, Sweden\\
$^{4}$ Konkoly Observatory,  CSFK, Konkoly-Thege M. \'ut 15-17, 
Budapest, 1121, Hungary\\
$^{5}$ Department of Experimental Physics, Institute of Physics, University of
Szeged, D\'om t\'er 9, Szeged, 6720 Hungary\\
${6}$ ELTE Eötvös Loránd University, Gothard Astrophysical Observatory, Szombathely, Hungary\\
$^{7}$ GRANTECAN: Cuesta de San Jos\'{e} s/n, 38712 Bre{\~n}a Baja, La Palma, Spain\\
$^{8}$ Geneva Observatory, University of Geneva, Chemin Pegasi 51, CH-1290 Versoix, Switzerland\\
$^{9}$ Astronomy Centre, Department of Physics \& Astronomy, University of Southampton, Southampton, SO17 1BJ, UK\\
$^{10}$ Institute of Astronomy, University of Cambridge, Madingley Road, Cambridge, CB3 0HA, UK\\
}

\date{Accepted XXX. Received YYY; in original form ZZZ}

\pubyear{2023}

\begin{document}
\label{firstpage}
\pagerange{\pageref{firstpage}--\pageref{lastpage}}
\maketitle

\begin{abstract}
New optical photometric, spectrocopic and imaging polarimetry data are combined with publicly available data to study some of the physical properties of the two H-poor superluminous supernovae (SLSN) SN 2021bnw and SN 2021fpl. For each SLSN, the best-fit parameters obtained from the magnetar model with \texttt{MOSFiT} do not depart from the range of parameter obtained on other SLSNe discussed in the literature. A spectral analysis with \texttt{SYN++} shows that SN 2021bnw is a W Type, Fast evolver, while SN 2021fpl is a 15bn Type, Slow evolver. The analysis of the polarimetry data obtained on SN 2021fpl at four epochs (+1.8, +20.6, +34.1 and +43.0 days, rest-frame) shows $> 3\sigma$ polarization detections in the range 0.8--1 $\%$.
A comparison of the spectroscopy data suggests that SN 2021fpl underwent a spectral transition a bit earlier than SN 2015bn, during which, similarly, it could have underwent a polarization transition. The analysis of the polarimetry data obtained on SN 2021bnw do not show any departure from symmetry of the photosphere at an empirical diffusion timescale of $\approx$ 2 (+81.1 days rest-frame). This result is consistent with those on the sample of W Type SLSN observed at empirical diffusion timescale $\le$ 1 with that technique, even though it is not clear the effect of limited spectral windows varying from one object to the other. Measurements at higher empirical diffusion timescale may be needed to see any departure from symmetry as it is discussed in the literature for SN 2017egm. 
\end{abstract}




\section{Introduction} \label{intro}

H-poor superluminous supernovae have been discovered about 15 years ago \citep[see reviews by][]{howell2017, moriya2018, gal-yam2019, chen2021}. 
While the spectral classification frequency of such transients is quite low ($\approx1\%$) with respect to other types of supernovae \citep[e.g.][]{nicholl2021}, spectroscopic and photometric follow-ups of Hydrogen-poor superluminous supernovae (SLSN-I) by several surveys like, among others, the Palomar Transient Factory \citep[PTF,][]{rau2009} and its intermediate version (iPTF), the Dark Energy Survey \citep[DES][]{angus2021} and more recently the Zwicky Transient Facility \citep[ZTF][]{bellm2019}, provide a large amount of data allowing more and more discoveries and time evolution studies of individual objects \citep[e.g.][]{kumar2020,konyvestoth2020}, as well as statistical analysis of relatively large samples \citep[e.g.][]{quimby2018,angus2021,konyvestoth21, chen2022_arxiv1,chen2022_arxiv2}. From the light curves and spectra analyses, several models have been proposed to explain the mechanisms behind the radiate total energies of $\approx$ 10$^{51}$ ergs from such objects. The current best models are those of a central engine consisting of fallback accretion on a compact object \citep[][]{dexter2013}, or the spin-down of a newly formed magnetar \citep[][]{kasen2010, woosley2010}. 
More recently, \citet{konyvestoth21} found that hydrogen-poor SLSNe can be divided in two distinct groups according to their pre-maximum spectra. The members of the first group show the W-shaped absorption feature well discussed in the literature, and identified as due to O II. This feature is not observed in the second group, whose spectral shape are more similar to that of SN 2015bn. Each of these groups can also be classified as Fast or Slow evolvers based on the photospheric velocities around maximum, even though this classification may be limited by the size of the sample and the lack of object having intermediate photospheric velocities.  

Polarimetry follow-ups of SLSN-I, on the other hand, are quite 
scarce. So far, no more than ten sources have been observed with linear polarimetry 
\citep[][]{leloudas2015, brown2016, inserra2016, leloudas2017, cikota2018, maund2019, maund2020, maund2021, lee2019, lee2020, saito2020, Poidevin2022, pursiainen2022}, among which three have also been observed with spectropolarimetry \citep[2015bn, 2017egm and 2018bsz by][respectively]{inserra2016, saito2020, pursiainen2022}, and two have been probed with circular 
polarimetry \citep[OGLE16dmu and LSQ14mo by][]{cikota2018}. The main difficulty encountered by such surveys comes from the Galactic contribution and possible contributions from the SLSNe hosts possibly combined with light dilution effects by the hosts \citep[see][]{leloudas2022}. Such problems can in principle be tackled with spectropolarimetry if specific spectral windows in the spectra can be used to isolate the host galaxy contribution to get intrinsic polarization estimates of the SLSN \citep[see][]{saito2020}. Alternatively, if the host galaxy is bright enough, one could measure its intrinsic level of polarization months or years after the SLSN has faded away. This would give an integrated estimate of the level of polarization of the Galaxy that could be removed afterward from polarimetry measurements of the SLSN. If such measurements are not possible or reach observation limit sensitivities, one could still rely on the low level of polarization measured on many systems before maximum light. Most of the current available measurements obtained so far with linear polarimetry show that once the Galactic interstellar polarization (ISP) contributions have been taken into account, one can expect low degrees of polarization of order fractions of a percent before maximum light. Such results suggest that the host contribution is very low and that the light of the SLSN pervading the host ISM is not significantly polarised by magnetically aligned dust grains \citep[see e.g. discussion in][]{Poidevin2022}. From that point, any increase in the degree of polarization measured at later phases could be attributed to a loss of the symmetry of the photosphere. Such measurements have been obtained on SN 2015bn with linear polarimetry \citep[][]{leloudas2017} and were found consistent with the results obtained with spectropolarimetry \citep[][]{inserra2016}. The increase of polarization is observed along a uniform polarization angle direction which is interpreted as the axis of symmetry of the inner photosphere. Alternatively, a change of polarization could come from the interaction of the photosphere with a possible disk-like, clumpy circumstellar medium (CSM) with emitting regions. This has been observed on SN 2018bsz and discussed by \cite{pursiainen2022}.

In this work we present the results obtained on the second and third target from a ten hour linear polarimetry survey conducted mainly during 2021, and designed to explore the frequency of objects like SN 2015bn and SN 2017egm. Along this work we use the \textit{Planck} 2018 Flat $\Lambda-$CDM cosmology model ($\Omega_{0}=0.31, H_{0}=67.7$ km$/$s) \citep[][]{planck2018i}

\section{H-poor SLSN sample and hosts associations}  \label{slsnsample}

Basic information about the two SLSN-I, SN 2021bnw and SN 2021fpl, and their possible hosts is given in Table~\ref{tab:slsn}.

\begin{table*}
	\centering
	\caption{Information about SN 2021bnw and SN 2021fpl. Photometry and coordinates of the host candidate of SN 2021bnw are from the Dark Energy Camera Legacy Survey (DECaLS). Upper magnitude limits at the position of SN 2021fpl are from Pan-STARRS1.}
	\label{tab:slsn}
	\begin{tabular}{lcccclccccc} 
          \hline
          Object & RA$_{\rm SN}$ & Dec$_{\rm SN}$ & redshift  & $\mu$ &Host & RA$_{\rm HOST}$ & Dec$_{\rm HOST}$ & g & r & z  \\
          Name & $[^{\circ}]$  & $[^{\circ}]$ & & [mag] & candidate &$[^{\circ}]$ & $[^{\circ}]$ &[mag] & [mag] & [mag]  \\
          \hline
          2021bnw &  163.467375 & +12.55805 & 0.098 & -38.24 & Brick: 1635p125, Objid: 1948 &  163.4678 & 12.5581 & 19.79 & 19.31 & 19.01  \\
          2021fpl &  303.577583 & -18.182381 & 0.115 & -38.59 & Orphan SLSN-I ? & RA$_{\rm SN}$ & Dec$_{\rm SN}$ & >  23.6 & > 23.0 & > 22.6 \\
          \hline
	\end{tabular}
\end{table*}

\subsection{SN 2021bnw}

SN 2021bnw (ZTF object ZTF21aagpymw, Pan-STARRS 1 object PS21ajy, ATLAS object ATLAS21dpf, Gaia object Gaia21caf) was discovered by
\citet{fremling2021} on February 03, 2021
from ZTF \footnote{Zwicky Transient Facility, {\tt
    https://www.ztf.caltech.edu}.} public alerts \citep[][]{bellm2019}. The discovery
magnitude obtained with the ZTF-cam mounted on the Palomar 1.2
meter Oschin was of 21.07 mag in the g-filter (AB system). 
The transient was classified the following day as a SLSN-I at a redshift z$=0.098$ by \citet{magee2021} \citep[see also][]{terwel2021} from the analysis of a spectrum obtained by
the extended Public ESO Spectroscopic Survey of Transient Objects
\citep[ePESSTO;][]{smartt2017} collaboration.
The spectrum is publicly available on the Transient Name Server
  \footnote{Transient Name Server, {\tt
    https://www.wis-tns.org/2021bnw}} (TNS). 
    
    SN 2021bnw is very likely associated
with the galaxy observed in the Dark Energy Camera Legacy Survey (DECaLS) \footnote{Legacy Survey, {\tt
    https://www.legacysurvey.org/}}, Data
Release 9 (DR9) at position $($RA, Dec$) = (163.4678^{\circ},
+12.5581^{\circ}$, J2000), of apparent magnitudes g=19.79, r=19.31, z=19.01 mag, i.e. of absolute magnitudes -18.45, -18.93 and -19.23 mag, respectively, assuming it is at the same redshift as the SLSN at z=0.098.
This galaxy has also been observed in the Sloan Digitized Sky Survey (SDSS) imaging 
Data Release 9 (DR9) as is identified as SDSS $J105352.17+123328.5$ but the photometry obtained on that source may be unreliable \footnote{{\tt http://skyserver.sdss.org/dr17/VisualTools/explore/summary
}}.


\subsection{SN 2021fpl}

SN 2021fpl (ZTF object ZTF21aaxwpyv, Pan-STARRS 1 object PS21evf, ATLAS object ATLAS21iao, Gaia object Gaia21ckf) was discovered by \citet{tonry2021} on March 16, 2021
from the Asteroid Terrestrial-impact Last Alert System (ATLAS) survey. 
The discovery magnitude 
was of 18.192 mag in the ATLAS Orange-filter (AB system). 
The transient was classified on 26 April, 2021 as a SLSN-I at a redshift z$=0.115$ by \citet{deckers2021a} \citep[see also][]{deckers2021b} from the analysis of a spectrum obtained with the
Liverpool Telescope (LT) SPectrograph for the Rapid Acquisition of Transients (SPRAT) \citep{2004SPIE.5489..679S,piacscik2014}.
This spectrum is publicly available on the Transient Name Server
  \footnote{Transient Name Server, {\tt
    https://www.wis-tns.org/2021fpl}} (TNS).

The field of view of SN 2021fpl was observed by the Panoramic Survey Telescope and Rapid Response System (PanSTARRS-1) \citep{chambers2016} making this SLSN looking as an orphan SN, or with no clear association. Magnitude upper limits from PanSTARSS-1 forced photometry are displayed in Table~\ref{tab:slsn} for indication.
Footprints from the DeCALS DR9 and SDSS DR17 surveys do not cover the field of view of SN 2021fpl.

\section{Data}

\subsection{Photometry} \label{dataphot}

The compilation of the photometry of SN 2021bnw and SN 2021fpl 
used in this work are listed in 
Tables~\ref{tab:photometry_2021bnw}, ~\ref{tab:photometry_2021bnw_2} 
and ~\ref{tab:photometry_2021bnw_3}, and in
Tables~\ref{tab:photometry_2021fpl}, 
~\ref{tab:photometry_2021fpl_2}, ~\ref{tab:photometry_2021fpl_3} and
~\ref{tab:photometry_2021fpl_4}, respectively.

\begin{figure}
\begin{center}
\vspace*{2mm}
\centering
\hspace*{0.cm}
\includegraphics[width=85mm,angle=0]{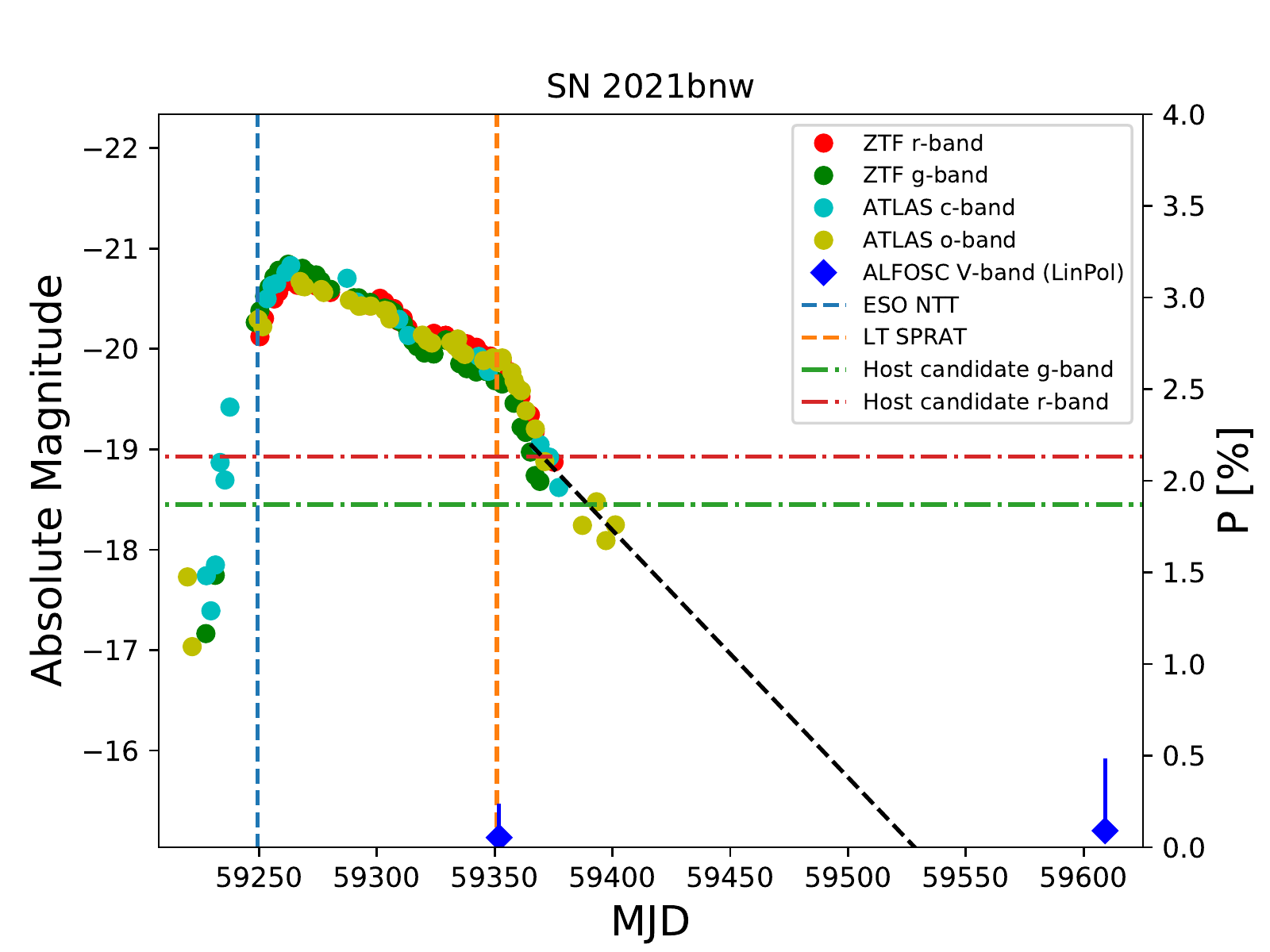}
\includegraphics[width=85mm,angle=0]{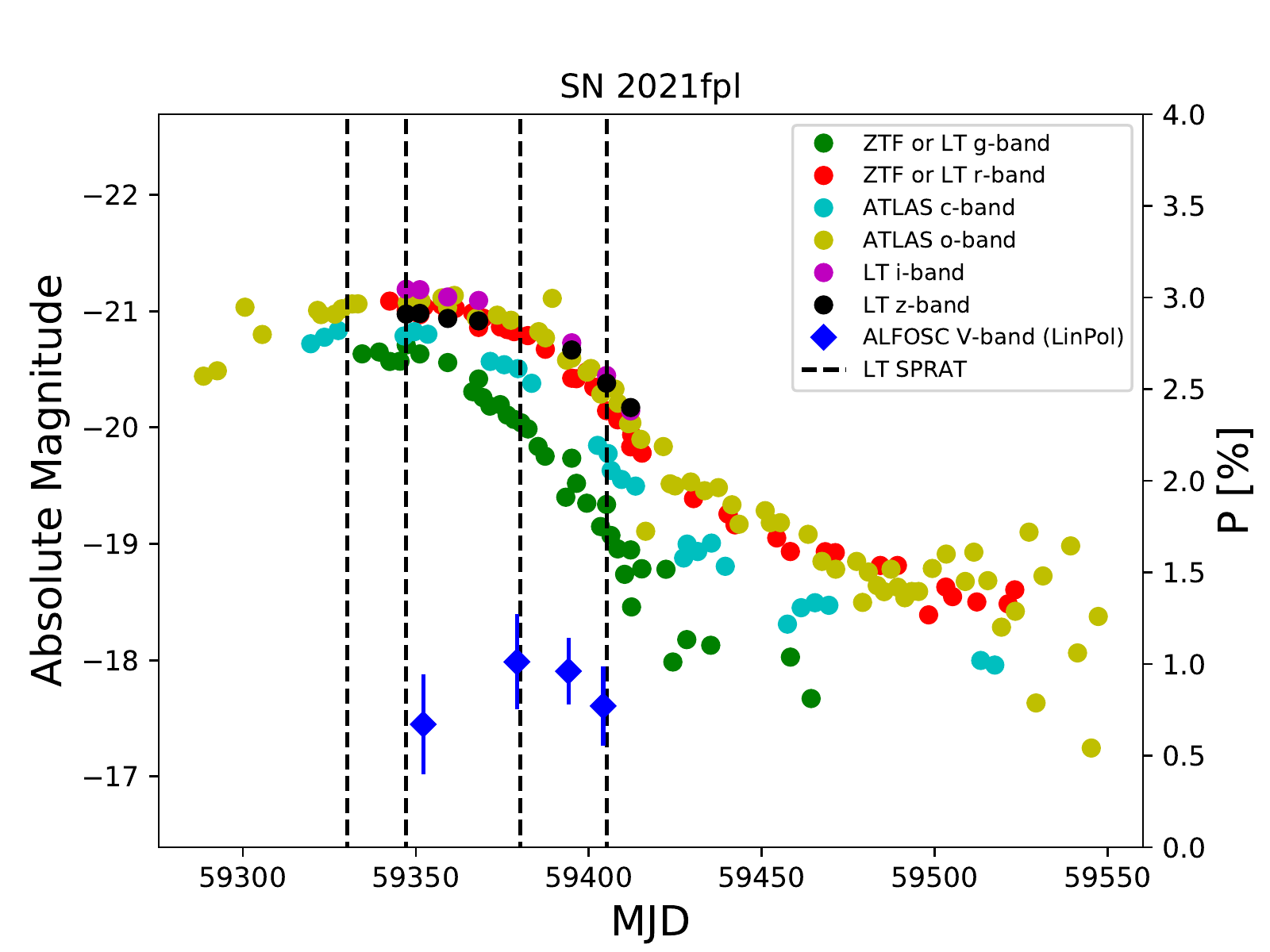}
\vspace*{0.cm}
\caption{Top: ZTF g- and r-band, and ATLAS c- and o-band absolute light curve of SN 2021bnw. The epochs when spectroscopy was obtained are shown with dashed-lines. The absolute magnitudes of the host candidate are shown with the horizontal dashed-lines. A linear fit to the last 8 points of the light curve is shown with the dark dashed-line. The linear polarization degree, $P$, is also shown by the diamond symbol. The scale is given on the right-axis. Bottom: same as Top for SN 2021fpl with additional photometry obtained from the Liverpool Telescope in g-, r-, i- and z-bands (see Section~\ref{ltioo} and Table~\ref{tab:ioolog}).}
\label{fig:LCandPOL}
\end{center}
\end{figure}

\subsubsection{ZTF} \label{ztf}

The ZTF g- and r-band public photometry data obtained on 
SN 2021bnw  (ZTF21aagpymw), and on SN 2021fpl (ZTF21aaxwpyv) 
were all retrieved from the Lasair broker 
\footnote{{\tt https://lasair.roe.ac.uk/object/}} \citep[][]{smith2019}. 

\subsubsection{LT IO$:$O} \label{ltioo}

Additional photometry observations were taken on SN 2021fpl 
in -u, -g, -r, -i and -z bands with the Liverpool 
Telescope IO:O instrument. The data were reduced using 
the \texttt{AutoPhOT} package
\footnote{{\tt https://github.com/Astro-Sean/autophot}}
\citep[AUTOmated PHotometry Of Transients;][]{brennan2022}. 
The main steps of the \texttt{AutoPhOT} pipeline are as follows. 
The pipeline builds a model of the Point Spread 
Function (PSF) in an image from bright isolated sources in the field, 
or uses aperture photometry if no suitable sources are present. 
The instrumental magnitude is then measured from a fit of the PSF 
to the transient. The instrumental magnitude is calibrated onto the 
standard system (AB magnitudes for Sloan-like filters) from 
an estimate of the zero point in each image which is obtained from
comparisons with catalogued standards in the field. For \it ugriz 
\rm filters, the zero point was calculated from magnitudes of sources
in the field observed by the SkyMapper Southern Observatory 
\citep[][]{skymapper_dr2}. A signal-to-noise ratio (SNR) of 5 was used to filter out bad photometry estimates. 
The log of the observations is displayed in Table~\ref{tab:ioolog}.

\subsubsection{ATLAS} \label{atlas}

The ATLAS forced photometry data were retrieved from the ATLAS 
public server \footnote{{\tt
    https://fallingstar-data.com/forcedphot/}} \citep[][]{tonry2018}.
The data were clipped and binned using the publicly available code
plot$\_$atlas$\_$fp.py \footnote{{\tt https://gist.github.com/thespacedoctor/
    86777fa5a9567b7939e8d84fd8cf6a76}.}
ATLAS data of SN 2021bnw were grouped in bins of one day. ATLAS data of 
SN 2021fpl were showing a bit more dispersion than those of 2021bnw 
and for that reason were binned in bins of two days.

\subsubsection{Light Curves} \label{LCs}

The absolute magnitude light curves of SN 2021bnw and of SN 2021fpl are displayed in Figure~\ref{fig:LCandPOL}, top and bottom, respectively. 
For each source, the distance modulus, $\mu = m - M$, displayed in 
Table~\ref{tab:slsn} was used to convert the apparent magnitude (see
Tables in Appendix D) to absolute magnitude. Both SLSNe peak at 
absolute magnitudes close to -21 mag. The public ZTF data and the 
public stacked and binned ATLAS data available for each of these sources are displayed in both plots. The additional photometry obtained on SN
2021fpl with the LT IO$:$O is displayed on the bottom plot in
Figure~\ref{fig:LCandPOL}.
Spectroscopy was obtained at several epochs on each source and will be
discussed in Section~\ref{dataspec}. The epochs of spectroscopy are marked with the vertical dashed lines in each plots.
Polarimetry, was also obtained at several epochs, as will be discussed 
in \ref{datapol}. The fractions of polarization are shown by blue 
diamonds symbols (the polarization scales are given in the right-axis 
of each plot). 

\subsection{Spectroscopy} \label{dataspec}

A summary of the spectroscopy obtained on SN 2021bnw and SN 2021fpl is given in Table~\ref{tab:speclog}.

\begin{figure}
\begin{center}
\vspace*{2mm}
\centering
\hspace*{0.cm}
\includegraphics[width=85mm,angle=0]{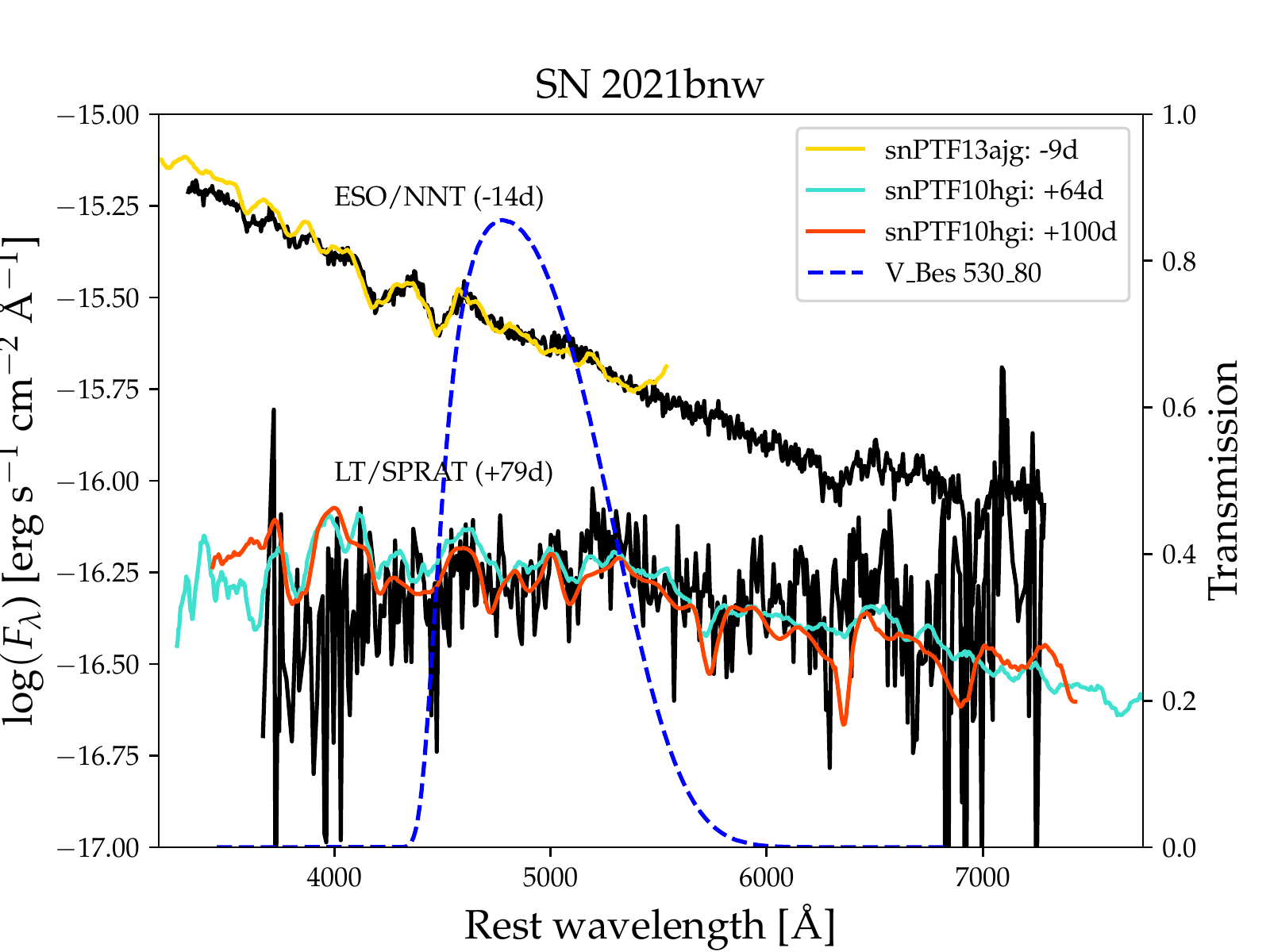}
\includegraphics[width=85mm,angle=0]{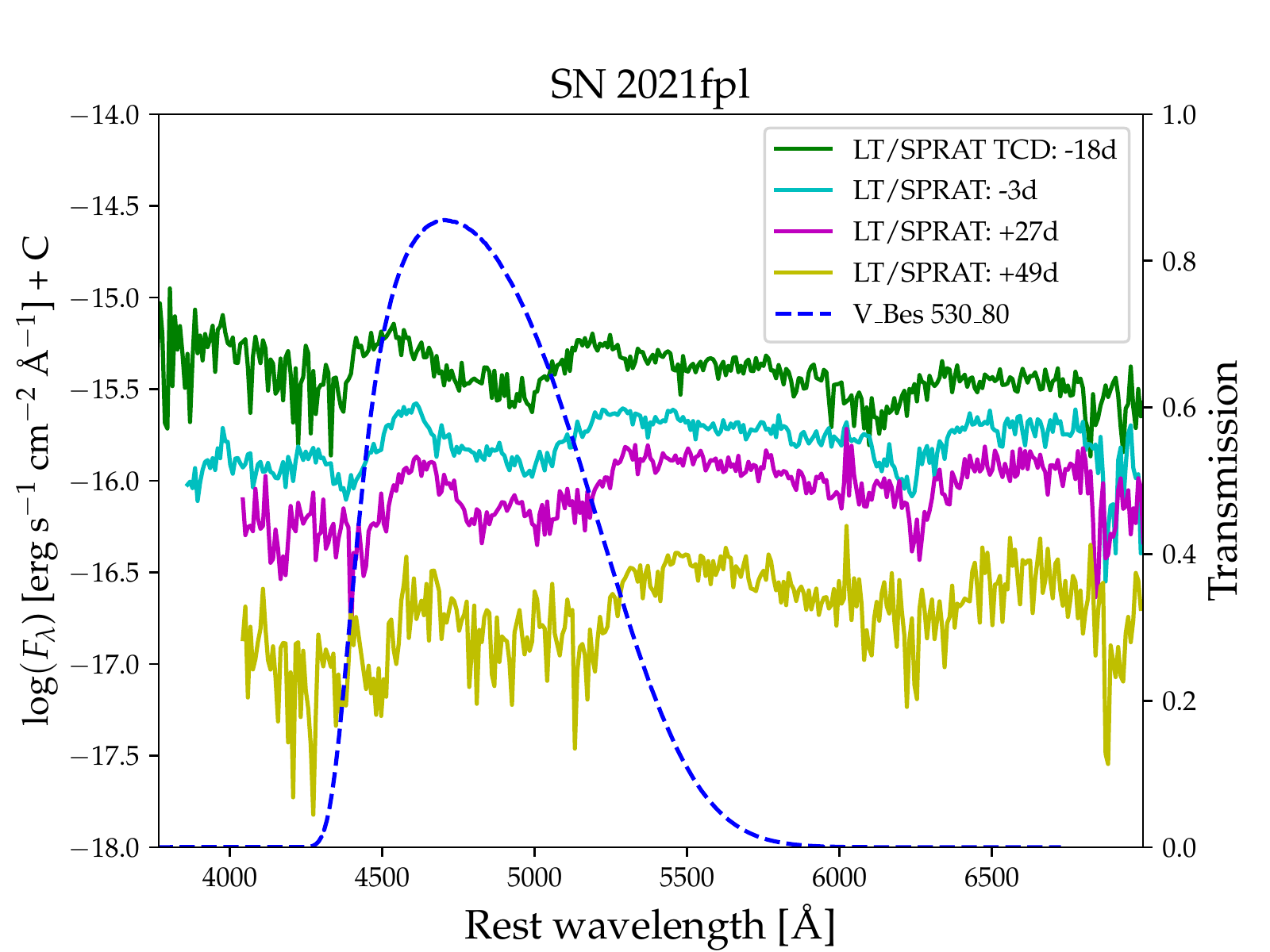}
\vspace*{0.cm}
\caption{Spectra obtained on SN2021bnw (top) and SN2021fpl (bottom). The transmission curves of the V-band filter
  mounted on the ALFOSC is shown with the blue solid curve. The blue dashed curve shows the wavelength range over which polarimetry is obtained in the rest frame of each supernova.}
\label{fig:SPEC_AND_FILT}
\end{center}
\end{figure}

\subsubsection{SN 2021bnw}

Two spectra are available on this source. We retrieved the publicly available spectrum obtained at early phase (Figure~\ref{fig:LCandPOL}, top) from TNS \footnote{Transient Name Server, {\tt https://www.wis-tns.org/2021bnw}}. A second spectrum was obtained with the LT SPRAT. 
This spectrum was obtained about one day before polarimetry was obtained on SN 2021bnw. 
The two spectra obtained on SN 2021bnw are shown in Figure~\ref{fig:SPEC_AND_FILT}, Top.
Also shown in the Figure are the transmission of the V-
filter mounted on the Alhambra Faint Object Spectrograph and
Camera (ALFOSC\footnote{ALFOSC, {\tt http://www.not.iac.es/instruments/alfosc/}}) used to get linear polarimetry data discussed in the next section. The solid line shows the filter in the observer frame while the dashed-line shows the imprint of the filter in the rest-frame of the supernova. We tested the classification and redshift of the two spectra of SN 2021bnw from a fitting template analysis with SNID, the SuperNova IDentification code,  
\footnote{{\tt
    https://people.lam.fr/blondin.stephane/software/snid/}} \citep[][]{blondin2007}. 
To do so we used the \cite{quimby2018} spectra database ingested in our custom SNID template library. The best match solutions displayed in Figure~\ref{fig:SPEC_AND_FILT}, Top, are consistent with the classification reported on TNS \footnote{Transient Name Server, {\tt https://www.wis-tns.org/2021bnw}}.

\subsubsection{SN 2021fpl}

Four spectra are available on that source. The first publicly available LT SPRAT spectrum used by \citet{deckers2021a} to classify SN 2021fpl was downloaded from TNS \footnote{Transient Name Server, {\tt https://www.wis-tns.org/2021fpl}}.  We obtained 3 more spectra on that source with LT SPRAT. All the spectra are displayed Figure~\ref{fig:SPEC_AND_FILT}, Bottom.

\subsection{Polarimetry} \label{datapol}

\begin{figure}
\begin{center}
\vspace*{2mm}
\centering
\hspace*{0.cm}
\includegraphics[width=80mm,angle=0]{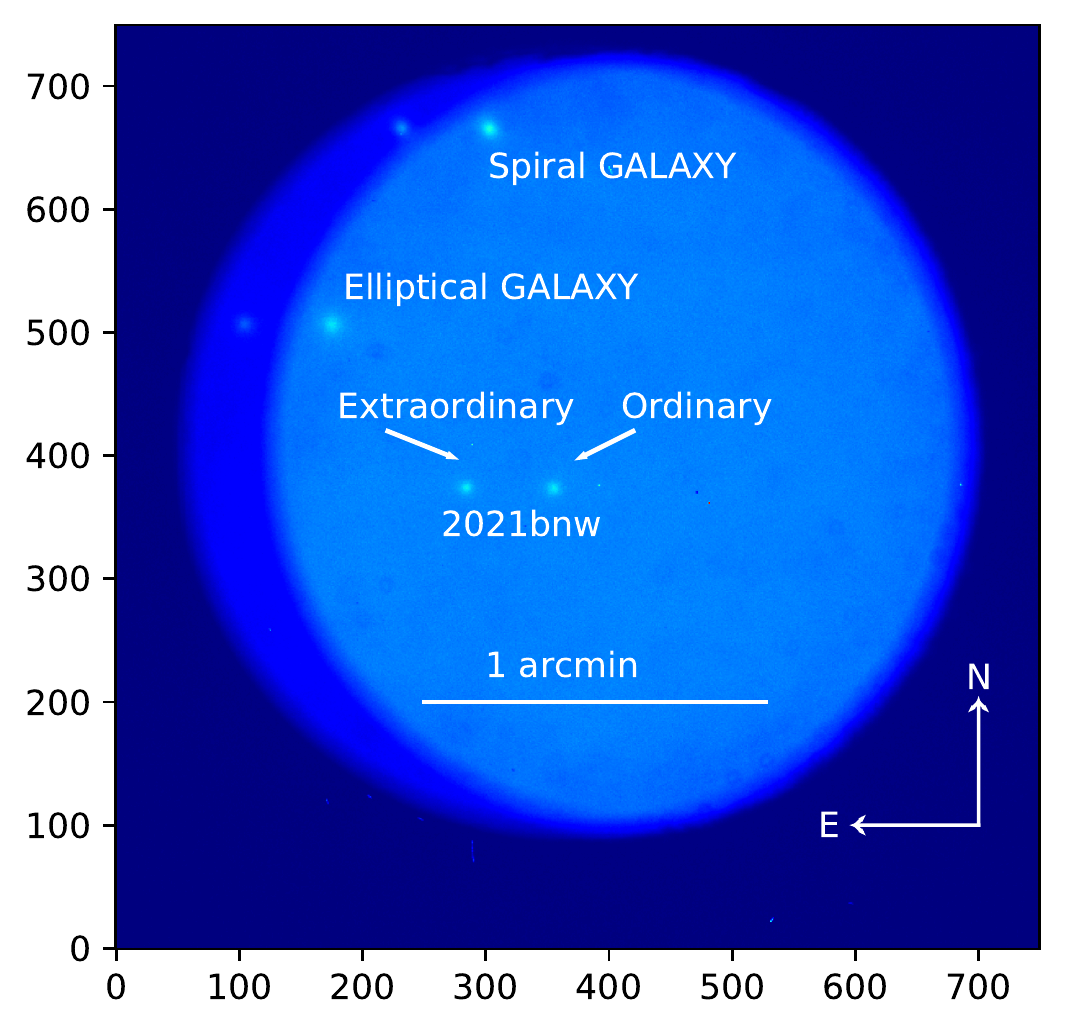}
\includegraphics[width=80mm,angle=0]{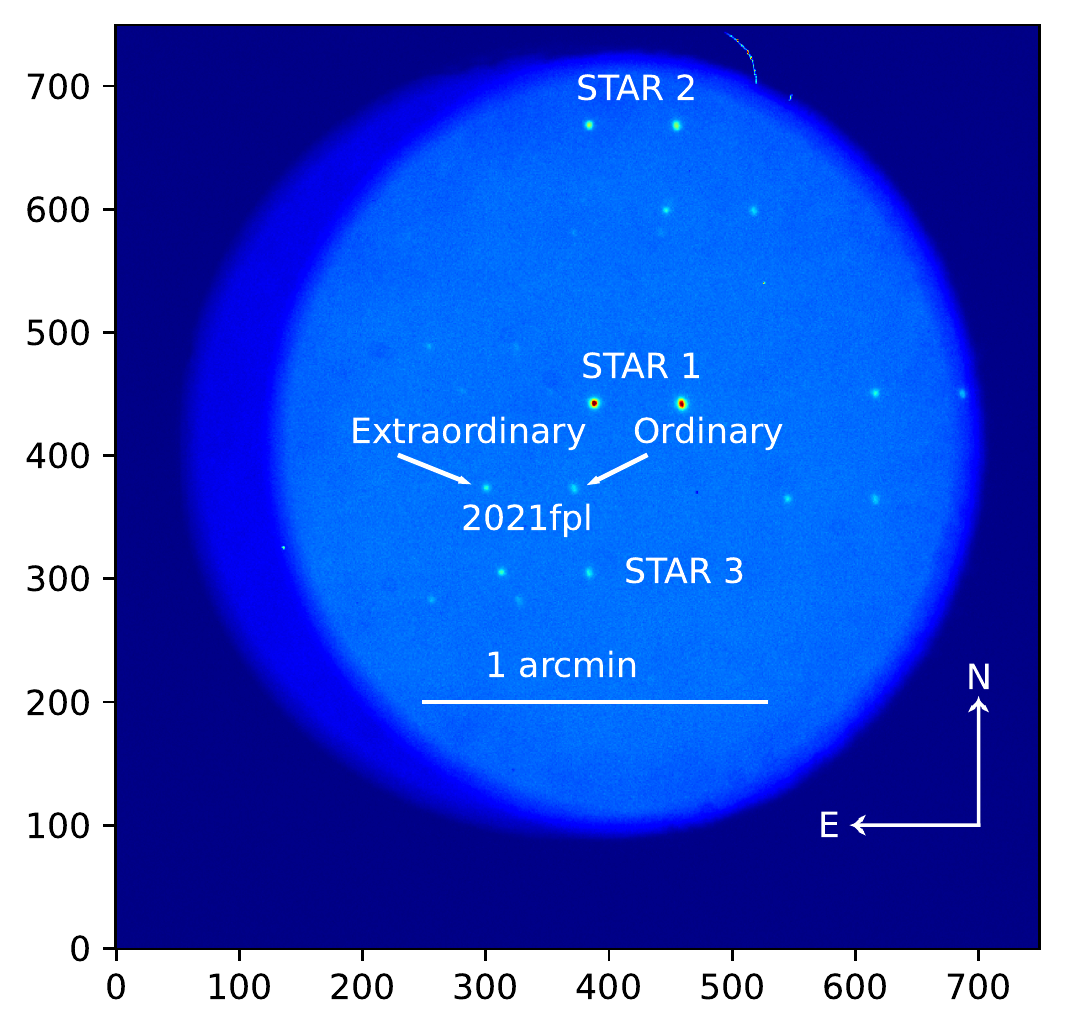}
\vspace*{0.cm}
\caption{Top: Linear polarimetry of SN 2021bnw with ALFOSC in one of the
  V-band flat-field and bias corrected data frame (file ALEe170044.fits) obtained with the half-wave plate at a position angle of $0.0^{\circ}$. Each pixel embeds the number of counts obtained after an exposure of 450 seconds. Imaging polarimetry was acquired through half-wave
plates positions angles at $0.0^{\circ}$, $22.5^{\circ}$,
$45.0^{\circ}$ and $67.5^{\circ}$.  The calcite plate splits the light
from the several objects into Ordinary images and
Extraordinary images separated by about $15^{\arcsec}$ from each other. Bottom: same as top for SN 2021fpl on one of the bias corrected 
data frame (file ALEf280162.fits) 
obtained after an exposure time of 90 seconds.}
\label{fig:images}
\end{center}
\end{figure}

Polarimetry on SN 2021bnw and SN 2021fpl was obtained with ALFOSC on the Nordic Optical Telescope (NOT). The observations log of the imaging polarimetry is displayed in Table~\ref{tab:pollog}. 
Linear polarimetry is made using a half wave plate 
in the FAPOL unit and a calcite plate
mounted in the aperture wheel. The calcite plate provides the
simultaneous measurement of the ordinary and the extraordinary
components of two orthogonal polarized beams (see
Figure~\ref{fig:images}, top and bottom). The half wave
plate can be rotated in steps of 22.5$^{\circ}$ from 0$^{\circ}$ to
$337.5^{\circ}$. As a standard, 4 angles are used ($0^{\circ},
22.5^{\circ}, 45^{\circ}$, and $67.5^{\circ}$), which we used during
our observations as referred to with the factor 4 used in the
exposure time calculations displayed in Table~\ref{tab:pollog}.
The data reduction of the polarimetry data was done using the same 
method and pipeline discussed in \citet{Poidevin2022}. The photometry of the ordinary and extra-ordinary beams was done using aperture photometry. When taking multiple sequences of 4 HWP angles (e.g. SN 2021flp was observed 6*4*90sec) the polarization was calcutated by summing-up the fluxes from the ordinary and extra-ordinary beams to minimize uncertainty propagation. Indeed, if the polarization was calculated individually for each sequence and the Stokes parameters summed-up at the end of the process, one would have to propagate the uncertainties on $Q$ and $U$ and combine them all together. All the results and their interpretation will be discussed in Section~\ref{pol_analysis}.

\subsubsection{SN 2021bnw}

Polarimetry was obtained on SN 2021bnw at 1 phase, about +89 days after maximum light. A measurement was obtained almost one year after maximum light (+347 days), when the SN was expected to be at least 3 to 4 magnitudes fainter than its host galaxy. This is suggested by the linear fit on the last point of the light curve of SN 2021bnw shown in Figure~\ref{fig:LCandPOL}, top. The measurement at +347 days relative to peak brightness is therefore expected to reflect the polarization properties of the galaxy without any contribution from the supernovae.

\subsubsection{SN 2021fpl}

Polarimetry of SN 2021fpl was obtained during 4 phases. A first measurement was obtained at an epoch close to maximum light (+2 days) and 3 additional measurements were obtained during the decreasing phase (+23 days, +38 days and +48 days), before the radioactive decaying or shock-heated 
diffusion phase. 

\section{Analysis} \label{analysis}

\subsection{Light Curves Modelling} \label{lc_modelling}

There are several mechanisms which can be used to power an SLSNe light curve: nickel decay, CSM interaction, and a central magnetar.  The large amount of nickel needed for SLSN light curves can only be produced in a pair instability SN, which requires an extremely massive progenitor and usually results of rise times of over 100 days \citep{Kasen2011}, which neither SN 2021bnw or SN 2021fpl show; pair instability SNe are also not expected to show significant polarization due to their quasi-spherical nature.  Models using CSM interaction sometimes require unphysically high explosion energies to match both the kinetic and radiated energies inferred from observations \citep[e.g.][]{chen2022_arxiv2, Kangas2022} and analytic models tend to produce different results than numerical models \citep{moriya2018}, leading to some inconsistencies in the interpretation of observations.  As such, we only present light curve fits using the magnetar model.

The multi-band light curve of each SN is fitted using the "slsn" magnetar spin-down model \citep{2017ApJ...850...55N} in the Modular Open-Source Fitter for Transients (\texttt{MOSFiT}) code, which uses a Markov Chain Monte Carlo (MCMC) algorithm to perform Bayesian parameter estimation for supernova light curves \citep{2018ApJS..236....6G}.  The Dynesty sampler \citep{2020MNRAS.493.3132S, 2019S&C....29..891H}, which utilizes dynamic nested sampling is used in the process. The uncertainty presented is only the statistical uncertainty in the fits, and does not include systematic uncertainty inherent in the simplified one-zone \texttt{MOSFiT} model.

\begin{figure*}
\begin{center}
\centering
\includegraphics[width=0.9\textwidth,angle=0]{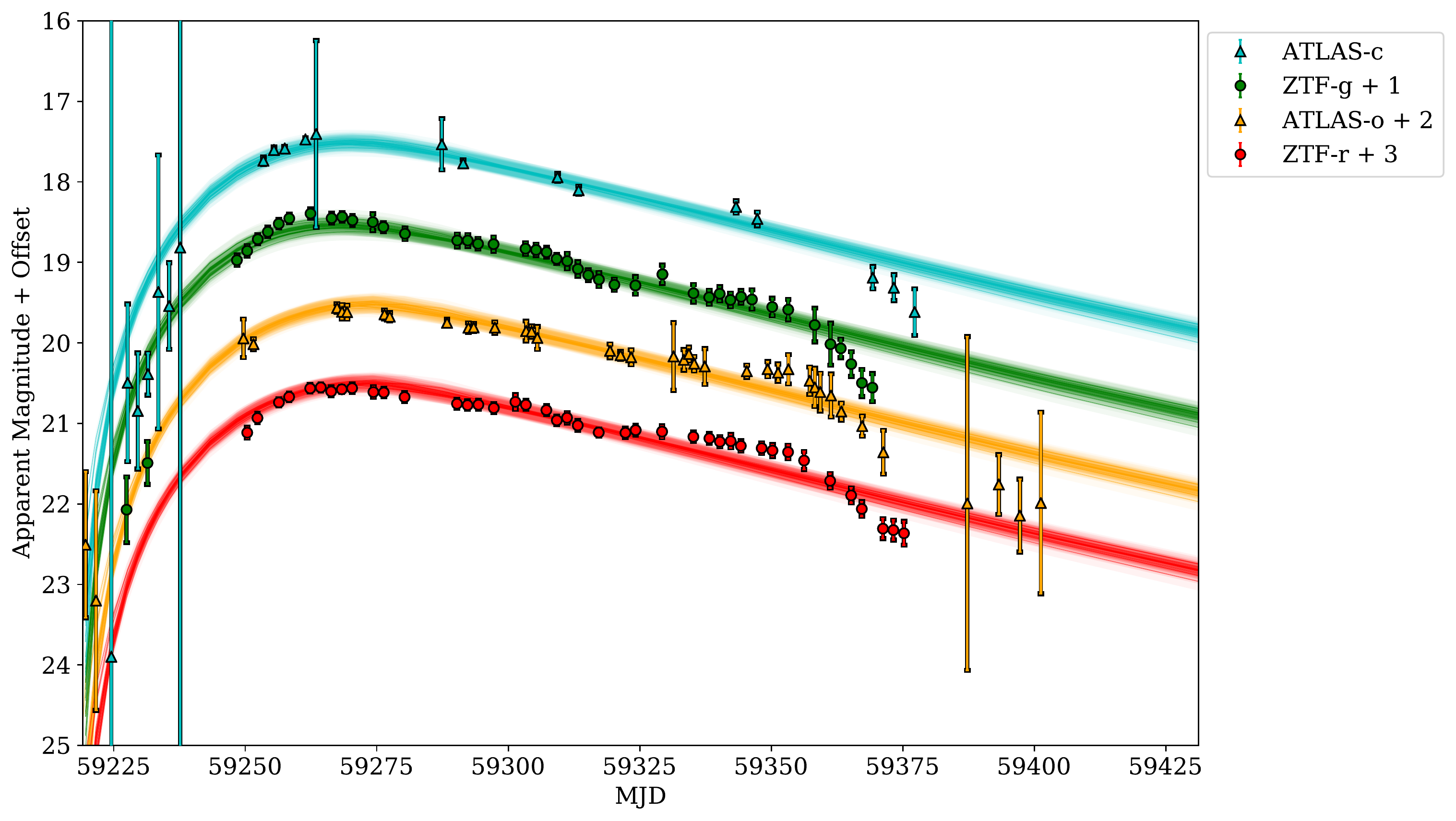}
\caption{Multi-band light curve of SN 2021bnw inferred from the magnetar-model, with each band offset for clarity.  The filled area shows the range of most likely models generated by \texttt{MOSFiT}. See Section \ref{lc_modelling} for details.}
\label{fig:mosfitfull_2021bnw}
\end{center}
\end{figure*}

\begin{figure*}
\begin{center}
\centering
\includegraphics[width=0.9\textwidth,angle=0]{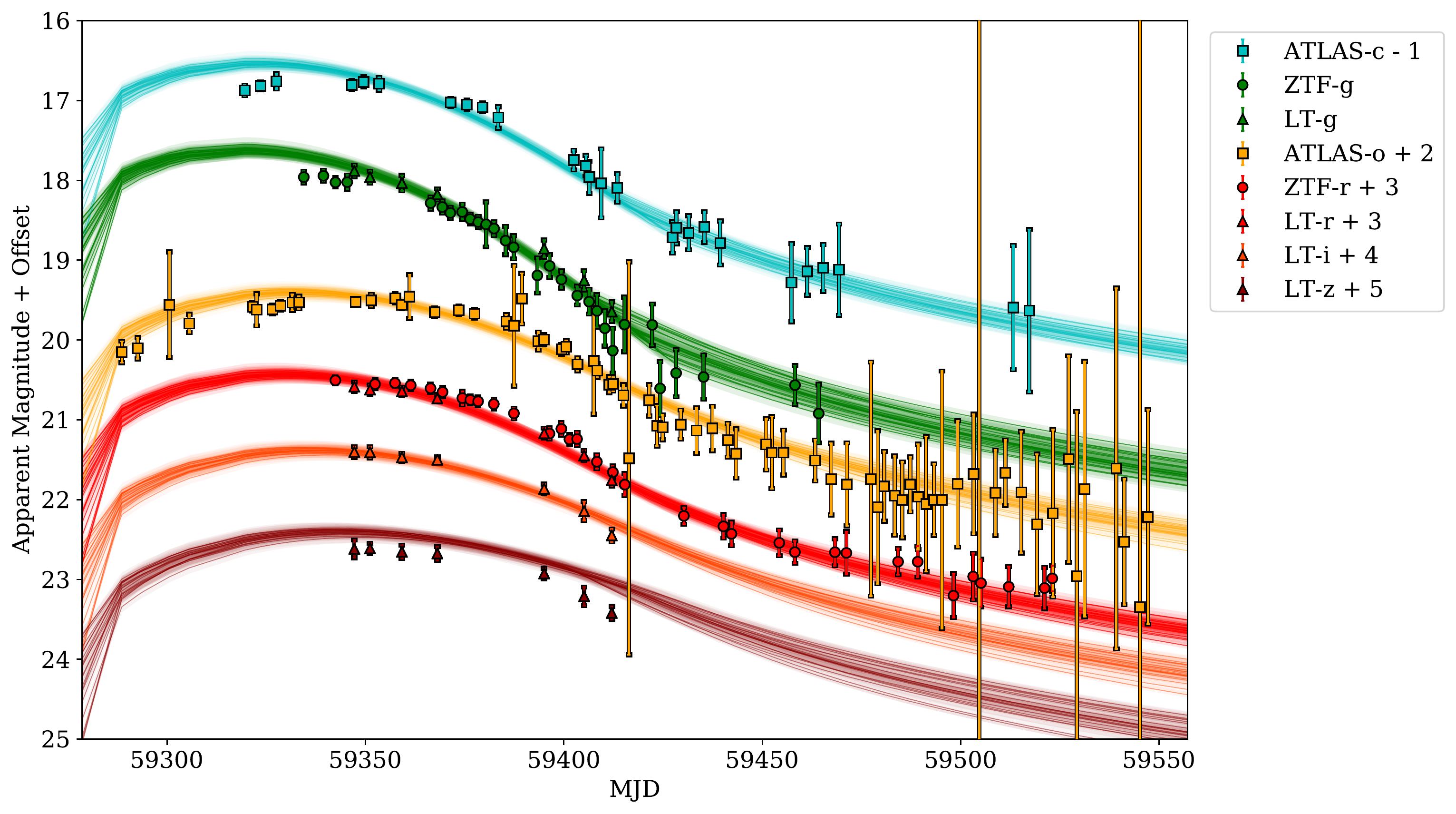}
\caption{Multi-band light curve of SN 2021fpl inferred from the magnetar-model, with each band offset for clarity.  The filled area shows the range of most likely models generated by \texttt{MOSFiT}. See Section \ref{lc_modelling} for details.}
\label{fig:mosfitfull_2021fpl}
\end{center}
\end{figure*}

\begin{table*}
	\centering
	\caption{Median and 1$\sigma$ best fit parameters for magnetar models obtained on SN 2021bnw and SN 2021fpl as discussed in Section \ref{lc_modelling}}
	\label{tab:mosfitparams}
	\begin{tabular}{cccccc} 
          Parameter Symbol & Definition & Prior & Best Fit Value & Best Fit Value &  Units\\
           &  &  & SN 2021bnw & SN 2021fpl  & \\
          \hline
          $B_\perp$ & Magnetar Magnetic Field Strength & [0.1,10] & $0.40^{+0.12}_{-0.11}$ & $0.78^{+0.14}_{-0.17}$ & 10$^{14}$ G\\
          $M_{\rm NS}$ & Neutron Star Mass      & [1.0,2.0]       & $1.58^{+0.26}_{-0.28}$&$1.71^{+0.19}_{-0.26}$ & $M_\odot$\\
          $P_{\rm spin}$ & Magnetar Spin Period & [1,10]          & $4.56^{+0.59}_{-0.64}$ & $1.43^{+0.30}_{-0.24}$ & ms \\
          $\log (\kappa_{\gamma})$ & Ejecta Gamma-Ray Opacity     & [-4,4]          & $-1.23^{+0.10}_{-0.14}$& $0.93^{1.84}_{-1.44}$ & cm$^2$ g$^{-1}$\\
          $M_{\rm ej}$ & Ejecta Mass     & [0.1,100]          & $3.59^{+1.06}_{-0.75}$& $21.59^{+8.22}_{-4.67}$ & $M_\odot$\\
          $T_{\rm min}$ &  Photospheric Plateau Temperature       & [3,10]          & $9.78^{+0.14}_{-0.23}$& $5.47^{+0.54}_{-0.28}$ & 10$^3$ K\\
          $v_{\rm ej}$ &  Ejecta Velocity       & [1,20]          & $7.87^{+0.50}_{-0.54}$& $6.94^{+0.28}_{-0.28}$ & 10$^3$ km/s\\
          \hline
	\end{tabular}
\end{table*}

\begin{figure*}
\begin{center}
\centering
\includegraphics[width=1.\textwidth,angle=0]{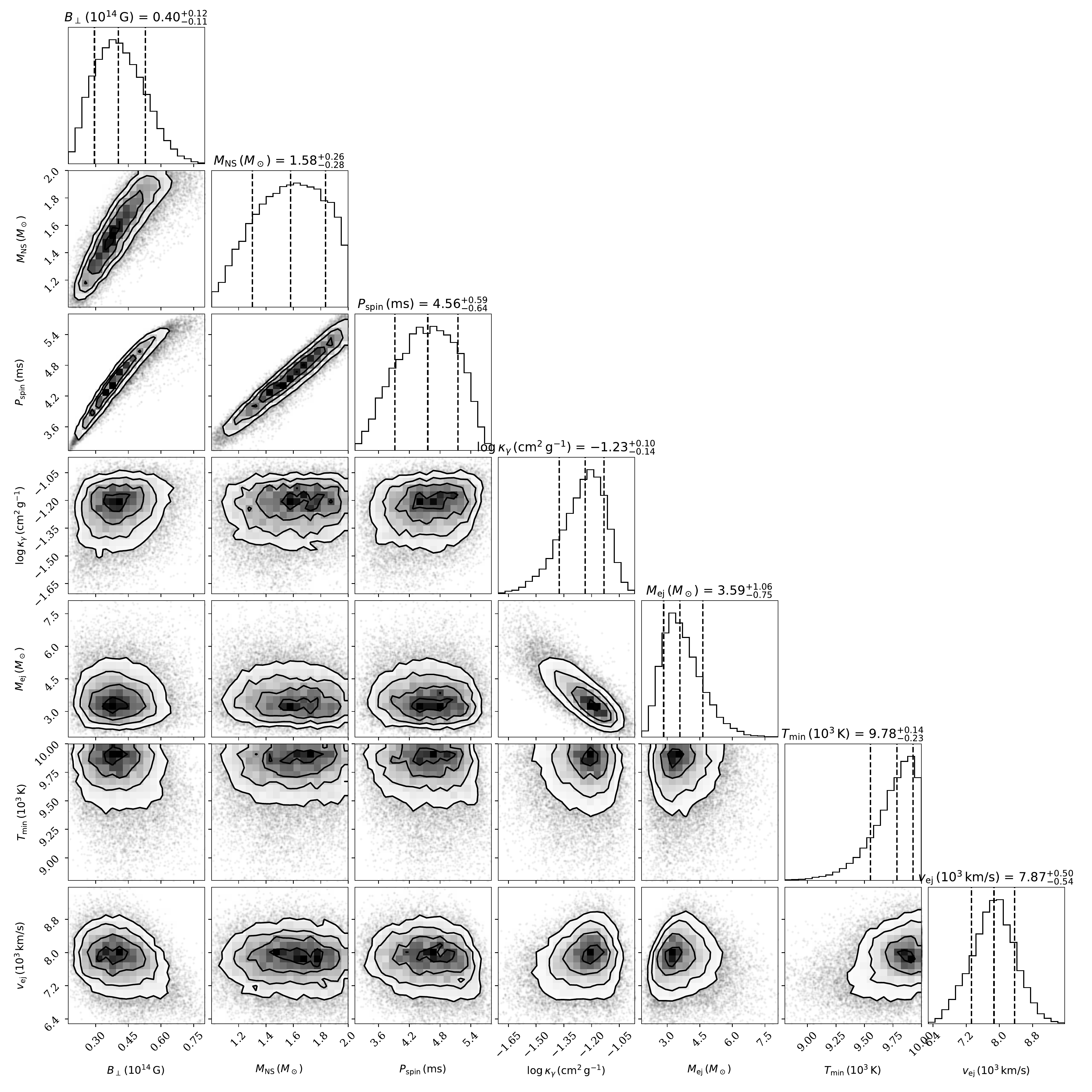}
\caption{1D and 2D posterior distributions of the
magnetar model parameters obtained from SN 2021bnw light curve analysis with \texttt{MOSFiT}. Median and 1$\sigma$ values are
marked and labeled - these are used as the best fit values (discussed in text).} 
\label{fig:mosfitpost_2021bnw}
\end{center}
\end{figure*}

\begin{figure*}
\begin{center}
\centering
\includegraphics[width=1.\textwidth,angle=0]{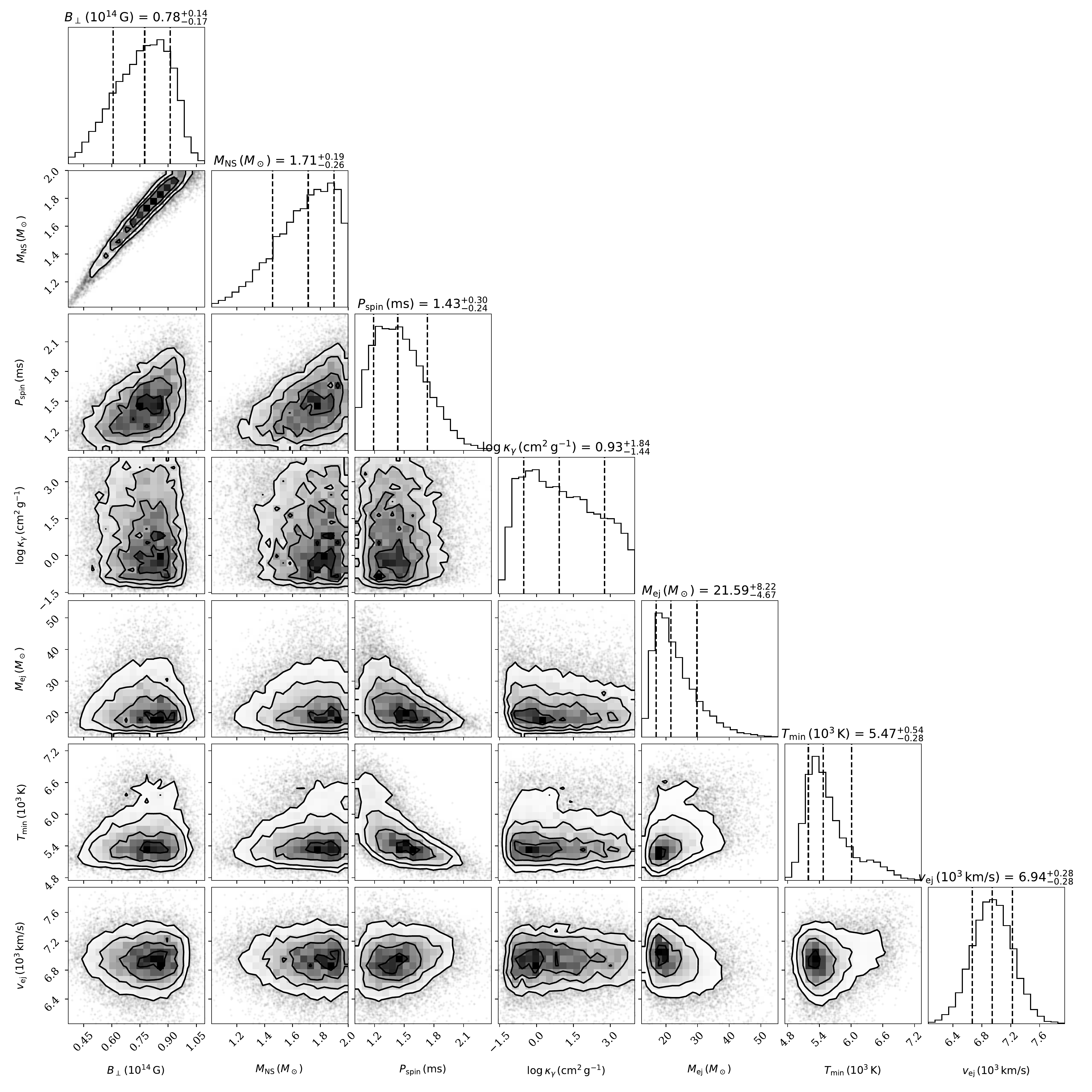}
\caption{1D and 2D posterior distributions of the
magnetar model parameters obtained from SN 2021fpl light curve analysis with \texttt{MOSFiT}. Median and 1$\sigma$ values are
marked and labeled - these are used as the best fit values (discussed in text).} 
\label{fig:mosfitpost_2021fpl}
\end{center}
\end{figure*}

The magnetar-model fits of SN 2021bnw and SN 2021fpl light curves are shown in Figure \ref{fig:mosfitfull_2021bnw} and in Figure \ref{fig:mosfitfull_2021fpl}, respectively. The most physically relevant parameters are listed in Table \ref{tab:mosfitparams}, with their posteriors shown in Figure \ref{fig:mosfitpost_2021bnw} for SN 2021bnw, and in Figure \ref{fig:mosfitpost_2021fpl} for SN 2021fpl.  The fits for both supernovae are quite good, although there are some small undulations followed by a large decrease in the luminosity of SN 2021bnw at around 100 days post-peak that is not captured by the model. The physical parameters we find for SN 2021bnw (SN 2021fpl) are $B_\perp \approx 4 \times 10^{13}$ ($8 \times 10^{13}$) G, M$_{\rm NS} \approx$ 1.6 (1.7) $M_\odot$, $P_{\rm spin} \approx 4.6$ (1.4) ms, $\log (\kappa_\gamma) \approx -1.2$ (0.9) cm$^2$ g$^{-1}$, $M_{\rm ej} \approx 3.6$ (22) $M_\odot$, $T_{\rm min} \approx 9800$ (5500) K, and $v_{\rm ej} \approx$ 7900 (6900) km/s; where $B_\perp$ is the magnetar magnetic field strength, M$_{\rm NS}$ is the neutron star mass, $P_{\rm spin}$ is the magnetar spin period, $\kappa_\gamma$ is the gamma-ray opacity of the ejecta \citep{Wang2015}, $M_{\rm ej}$ is the ejecta mass, $T_{\rm min}$ is the photospheric plateau temperature, and $v_{\rm ej}$ is the ejecta velocity.  These best fit parameters and uncertainties are the median and 1$\sigma$ values from the one-dimensional posterior for each of the parameters.  Although the parameters for these supernovae are quite different, they are both representative of the extremes of the SLSN population \citep{2017ApJ...850...55N}, with one extreme having high spin periods and low ejecta masses (like SN 2021bnw), and the other having low spin periods and high ejecta masses (like SN 2021fpl). The total kinetic energy of the ejecta for SN 2021bnw (SN 2021fpl) is calculated to be $E_{\rm K} \approx 1.3 \times 10^{51}$ ($6.2 \times 10^{51}$) ergs. 
The mass of the progenitor stars, $M_* = M_{\rm NS} + M_{\rm ej} \approx 5$ (23) $M_\odot$, which are consistent with the $3.6-40$ $M_\odot$ range inferred in the mass distribution found by \cite{2020ApJ...897..114B}. 

We also find correlations between $B_\perp$, $M_{\rm NS}$, and $P_{\rm spin}$ in the 2D posterior distributions of these parameters, as in \cite{Poidevin2022}, although the $P_{\rm spin}$ correlation in SN 2021fpl is not as strong as in SN 2020znr or SN 2021bnw.  The cause of these correlations is still unknown.

The gamma-ray opacity is important when modelling a late-time decline in luminosity due to gamma-ray leakage.  The leakage timescale is \citep{Wang2015}

\begin{equation}
    t_{\rm leak} \approx 80 \text{ days } \left(\frac{\kappa_\gamma}{0.1 \text{ cm$^2$ g$^{-1}$}}\right)^{1/2} \left(\frac{M_{\rm ej}}{M_\odot}\right)^{1/2} \left(\frac{v_{\rm ej}}{10^4 \text{ km s$^{-1}$}}\right)^{-1}.
    \label{eqn:leaktime}
\end{equation}
For SN 2021bnw, $t_{\rm leak}$ $\sim$ 150 days, which is about when the light curve starts to decline.  We also calculate the time when the optical depth of the ejecta $\tau_{\rm ej} =$ 1, at \citep{2017ApJ...850...55N}

\begin{equation}
    t_{\rm neb} = (3\kappa M_{\rm ej}/4\pi v_{\rm ej}^2)^{1/2},
    \label{eqn:tneb}
\end{equation}
where $\kappa$ is the opacity of the ejecta ($\kappa$ $\approx 0.15$ cm$^2$ g$^{-1}$ for both SNe).  For SN 2021bnw, $t_{\rm neb} \sim$ 240 days, which is much later than its light curve decline.  Even though the leakage time is correctly estimated, the model still cannot reproduce the observed luminosity decrease.  This is likely due to a more complicated physical process than what is described by the model \citep{Vurm2021}, such as CSM interaction, dust formation \citep{2019MNRAS.484.5468O}, or molecular cooling \citep{Liljegren2022}.  The value of $\kappa_\gamma$ is not well constrained for SN 2021fpl, since there is no observed luminosity decrease that looks consistent with gamma-ray leakage.

The ejecta velocities estimated by \texttt{MOSFiT} are not the same as the photospheric velocity, which can be estimated from spectra, since the photospheric velocity should decrease as the photosphere recedes further into the ejecta, while the ejecta velocity should remain relatively constant post-peak, with small increases and decreases due to acceleration from the pulsar wind nebula (PWN) and collision with circumstellar material (CSM).  However, the photospheric velocity at the light curve maximum can be used as a proxy for the ejecta velocity \citep{konyvestoth21}.  The \texttt{MOSFiT} magnetar model also includes a temperature floor for the photosphere \citep{2017ApJ...850...55N}, which is motivated by late-time observations of SLSNe \citep[e.g.][]{Inserra2013, Nicholl2017} -  this plateau temperature $T_{\rm min}$ can also be checked against spectral models.

\subsection{Spectrum modelling} \label{syn++}

We utilized the code named  {\tt SYN++} \citep{thomas11} to model the available spectra of SN~2021bnw and SN~2021fpl. This code uses some local parameters that fit the lines of the individual ions and global parameters referring to the entire model spectrum.
The global parameters are: 
\begin{itemize}
    \item {{$a_0$}: a constant multiplier to the overall model spectrum}
    \item {{$v_{\rm phot}$}: the velocity at the photosphere. The uncertainty of this parameter is $\sim$ 1000 km s$^{-1}$. For detailed explanation, see Section 3.3.2. in \citet{silverman15}.  }
    \item {{$T_{\rm phot}$}: the temperature at the photosphere. The uncertainty of $T_{\rm phot}$ is $\sim$2000 K, discussed in \citet{Konyves-Toth2022}}  
\end{itemize} 
and the local parameters are:
\begin{itemize}
    \item {{$\tau$}: the optical depth for the reference line of each ion}
    \item {{$v_{\rm min}$}: the inner velocity of the line forming region}
    \item {{$v_{\rm max}$}: the outer velocity of the line forming region}
    \item {{$aux$}: the scale height of the optical depth above the photosphere given in km s$^{-1}$. This parameter accounts for the width of the spectral features, that is in relation with the width of the line-forming region in the atmosphere. The larger the $aux$ parameter, the broader the feature.}
    \item{{$T_{\rm exc}$:}  excitation temperature of each element/ion, assuming  Local Thermodynamic Equilibrium (LTE)}.
\end{itemize} 

The best-fit models obtained for the spectra of SN~2021bnw are shown in Figure \ref{fig:21bnw_modeling}, while the same in case of SN~2021fpl are plotted in Figure \ref{fig:21fpl_modeling}. In these figures the observed spectra are shown with black line, while green color codes the best-fit models built in \texttt{SYN++}. The single-ion contributions to the overall model spectra are denoted with turquoise lines shifted vertically for clarification. The local parameter values of the best-fit models can be found in Table \ref{tab:sn21bnw_lokparams} and Table \ref{tab:sn21fpl_lokparams} in the Appendix.  

\subsubsection{SN~2021bnw} 
\begin{figure}
\begin{center}
\vspace*{2mm}
\centering
\hspace*{0.cm}
\includegraphics[width=8cm,angle=0]{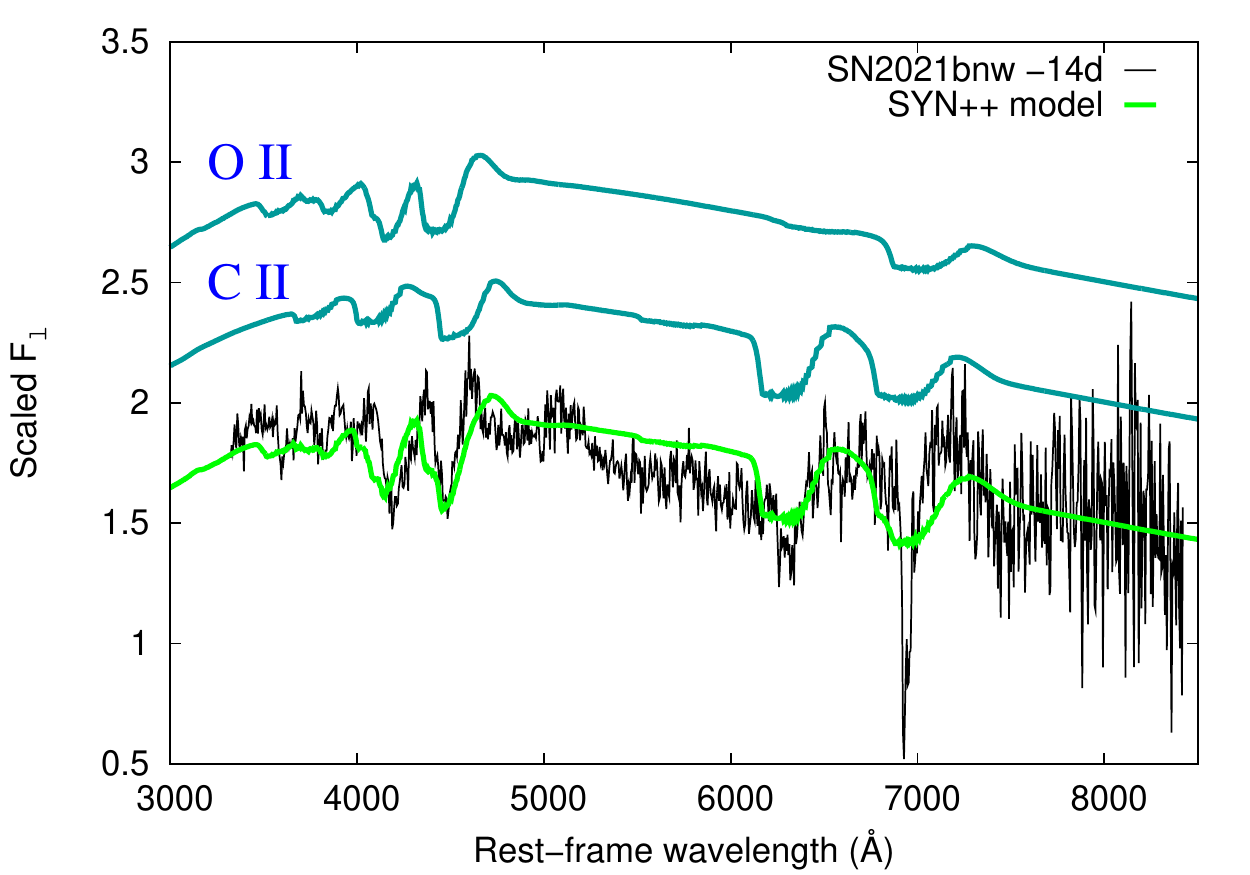}
\includegraphics[width=8cm,angle=0]{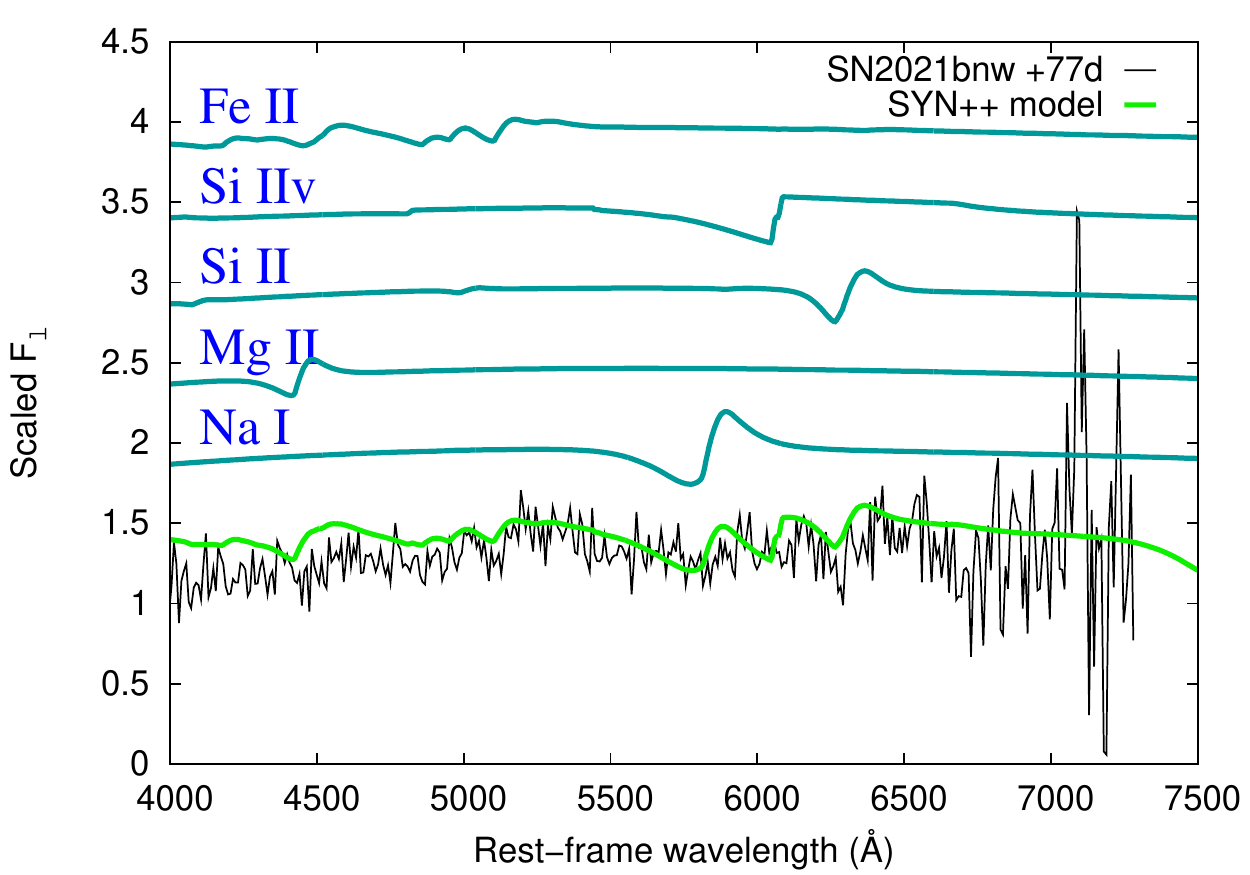}
\vspace*{0.7cm}
\caption{
{Top panel: Spectrum at -14 days relative to peak brightness of SN~2021bnw (black line) plotted together its best-fit model built using SYN++ (green line). Single ion contributions to the overall model spectrum (turquoise lines) are also plotted. 
Bottom panel: Spectrum at +77 days relative to peak brightness of SN~2021bnw together with its best-fit model. The color-coding is the same as in the left panel. 
}
}
\label{fig:21bnw_modeling}
\end{center}
\end{figure}

In case of SN~2021bnw two spectra were modeled, both from the photospheric phase taken at -14 days and +77 days rest-frame phase relative to the moment of the maximum light. The date of maximum was estimated from the \texttt{MOSFiT} fitting as MJD 59265. As shown in the left panel of Figure \ref{fig:21bnw_modeling}, the pre-maximum spectrum of SN~2021bnw shows some W-shaped O II features between 4000 and 5000 \AA, making this object a "Type W" SLSN-I using the classification scheme of \citet{konyvestoth21}. According to the best-fit \texttt{SYN++} - model, the velocity of the photosphere is 20~000 km s$^{-1}$, while the photospheric temperature is 14~000 K. These are not extraordinary, but high values compared to other SLSNe. 
The best-fit model of the second spectrum taken at +77 days phase (see the right panel of Figure \ref{fig:21bnw_modeling}) contains the features of Na I, Mg II, Si II, high velocity Si II (Si II v) and Fe II. By this phase, the O II and C II lines have disappeared and the photospheric velocity swiftly decreased to 4000 km s$^{-1}$ and the temperature of the photosphere dropped to 6500 K, making SN~2021bnw a spectroscopically fast evolving event. 

The estimated ejecta velocity is around 8000 km s$^{-1}$.  If this is the photospheric velocity at peak, then the photospheric velocity evolution in this system was very fast at pre-peak, decreasing by a factor of $\sim$ 2.5 in two weeks, followed by a fairly slow evolution post-peak, decreasing by a factor of $\sim$ 2 in $\sim$ 2.5 months.  The estimated plateau temperature is almost $\sim$ 10 000 K, which is much higher than the photospheric temperature inferred from the +77 day spectrum. This apparent discrepancy may come from model uncertainties within \texttt{MOSFiT}.

\subsubsection{SN~2021fpl}

\begin{figure}
\begin{center}
\vspace*{2mm}
\centering
\hspace*{0.cm}
\includegraphics[width=8cm,angle=0]{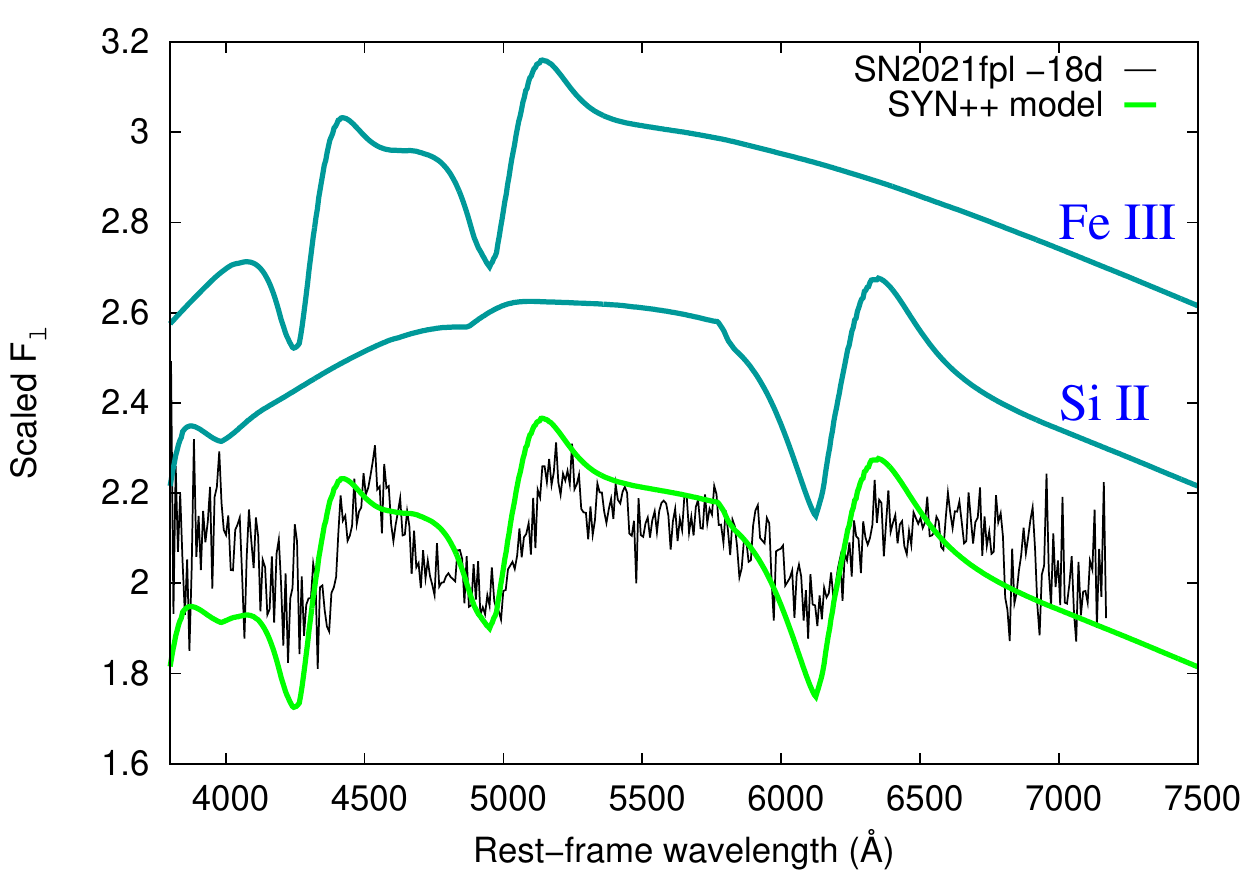}
\includegraphics[width=8cm,angle=0]{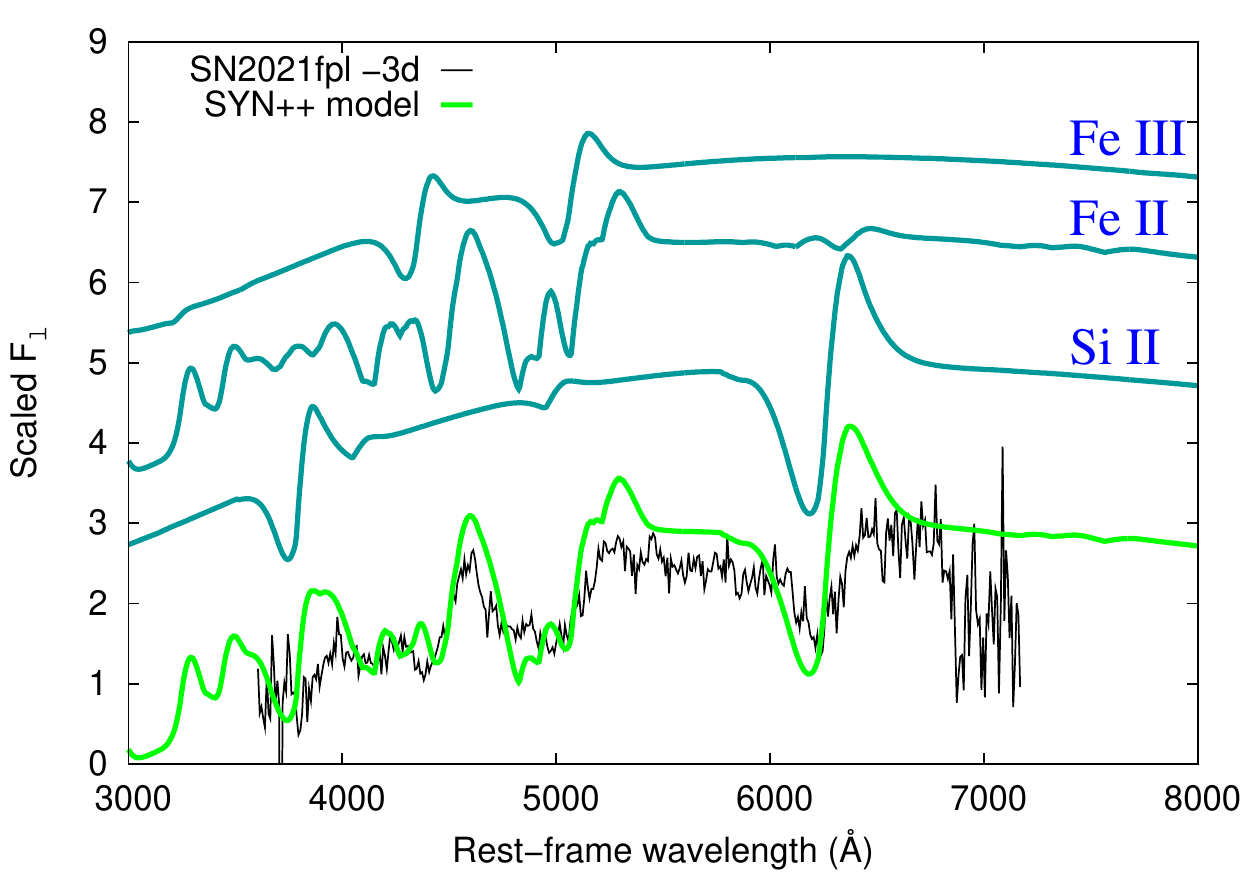}
\includegraphics[width=8cm,angle=0]{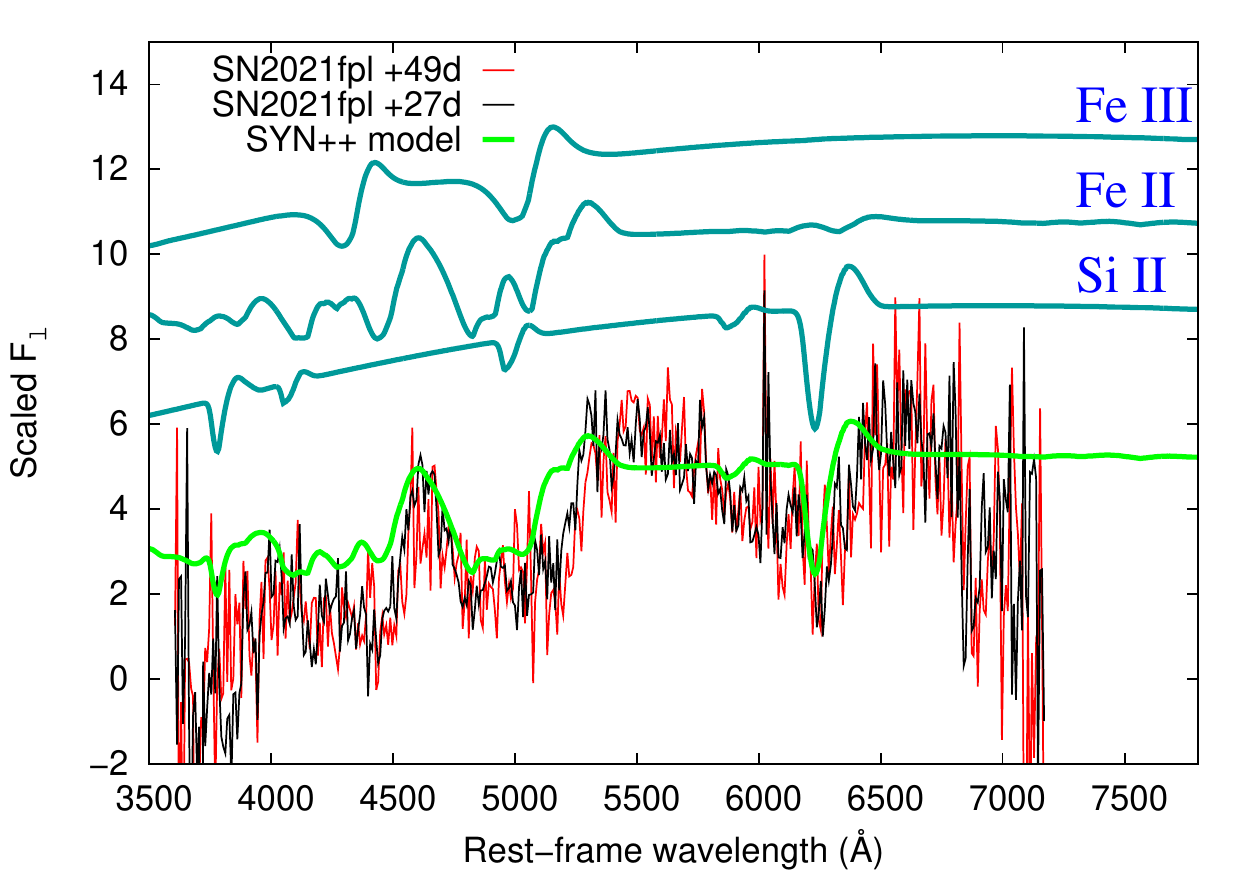}
\vspace*{0.7cm}
\caption{
Top panel: modelling the -18 days phase SN~2021fpl with \texttt{SYN++}. Middle panel: modelling of the -3 days phase spectrum. Bottom panel: modelling of the +27 days (black) and the +49 days (red) phase spectrum. The color coding is the same as in \ref{fig:21bnw_modeling}.
} 
\label{fig:21fpl_modeling}
\end{center}
\end{figure}

We carried out the spectrum modelling of the -18 days, -3 days, +27 and +49 days phase spectrum of SN~2021fpl, as can be seen in Figure \ref{fig:21fpl_modeling}. We estimated the date of the maximum as MJD 59350 using the \texttt{MOSFiT} fitting similarly to SN~2021bnw.
As the top left panel of Figure \ref{fig:21fpl_modeling} shows, the earliest phase spectrum of SN~2021fpl is differing from the -14d phase spectrum of SN~2010bnw, as the former does not show the W shaped O II features between 4000 and 5000 \AA, and rather can be fitted using Si II and Fe II lines. This fact makes SN~2021fpl a "Type 15bn" SLSN-I by  \citet{konyvestoth21}. The photospheric velocity ($v_{\rm phot} = 11~000$ km s$^{-1}$) and temperature ($T_{\rm phot} = 6800$) K are also much lower compared to the similar phase spectrum of SN~2021bnw, strengthening that SN~2021fpl resembles to SN~2015bn by its pre-maximum spectrum evolution. 

 The -3d phase spectrum of SN~2021fpl is modeled using Si II, Fe II and Fe III lines (see the top right panel of Figure \ref{fig:21fpl_modeling}). By this time, the $v_{\rm phot}$ diminished to 6000 km s$^{-1}$, while the best-fit $T_{\rm phot}$ is 5700 K, a result which is broadly consistent with the \texttt{MOSFiT} modelling. The best-fit model of the +27d phase spectrum (see the bottom panel of  Figure \ref{fig:21fpl_modeling}) shows similar lines and has the same photospheric velocity as the -3d phase and the photospheric temperature is similar as well (5200 K). This suggests that SN~2021fpl shows a spectroscopically slow evolution, like SN~2015bn. In Figure \ref{fig:21fpl_modeling}, the +49 days phase observed spectrum (red line) of SN~2021fpl is plotted together with the +27d phase spectrum (black line). It is seen that they are quite similar to each other, thus they share the best-fit \texttt{SYN++} model.

\subsection{Polarimetry Analysis} \label{pol_analysis}

The levels of polarization obtained on polarized and unpolarized calibration stars, on SN 2021bnw, on SN 2021fpl, and on several objects of interest in their field of view, are displayed in Table~\ref{tab:pol_results}. The steps followed to get estimates of the intrinsic polarization degree of SN 2021bnw and of SN 2021fpl (see last column in Table~\ref{tab:pol_results}) are given in the following sections. A visual summary of all these results can be seen on the $Q - U$ plots displayed in Figure~\ref{fig:QU_plots} 
for each polarimetry epoch (see Appendix~\ref{QU_plots_appendix}).
The polarization was corrected for bias following the equation given in \cite{1997ApJ...476L..27W}:
\begin{equation} 
    P = (P_{\rm obs} - \sigma_{P}^{2}/P_{\rm obs}) \times h(P_{\rm obs} - \sigma_{P}),
    \label{eqn:pdeb}
\end{equation}
where $h$ is the Heaviside function, $P_{\rm obs}$ is the observed polarization and $\sigma_{P}$ is the 1$\sigma$ error. 

\begin{table*}
	\centering
	\caption{V-band polarimetry results on SN 2021bnw and its host, and on SN 2021fpl. $^{\rm(a)}$: Stokes
          parameters, $\overline{Q}$ and $\overline{U}$, directly
          obtained from the ALFOSC data frames Extraordinary and 
          Ordinary images without applying any further corrections.
          $^{\rm(b)}$: instrumental polarization estimates.
          $^{\rm(c)}$: instrumental polarization corrected. 
          Assuming HD127769, HD251204, BD +64 106 and BD$+$32 3739 
          have a polarization angle of $147 ^{\circ}$ 
          the zero polarization angle, $ZPA$, estimates are 
          $83.9 ^{\circ}$ and
          $87.7 ^{\circ}$, respectively. 
          $^{\rm(d)}$: $IP$ and $ZPA$ corrected. 
          $^{\rm(e)}$: $IP$ and $ZPA$ corrected +
          Milky Way Interstellar polarization
          corrected with Stokes parameters obtained 
          on star 2$^{\rm(f)}$, 
          which is at the largest heliocentric distance, 
          then with weighted average Stokes 
          parameters obtained on stars STAR 1, STAR 2 and STAR 3 $^{\rm(g)}$  (See text for details). $^{\rm(bc)}$:bias corrected.
          }
	\label{tab:pol_results}
	\begin{tabular}{llccccccccc} 
          \hline
          Date
          &Source
          & $\overline{Q}^{\rm (a)}$
          & $\overline{U}^{\rm(a)}$
          & $P [\%]^{\rm(b)} $
          & $P [\%]^{\rm(c)}$
          & $\theta [^{\circ}]^{\rm(c)}$
          & $P [\%]^{\rm(d)} $
          & $\theta [^{\circ}]^{\rm(d)}$
          & $P [\%]^{\rm(e)} $\\
         \hline
2021-05-17 & GD 319 & 0.03 & -0.10 & \textbf{ 0.10 $\pm$ 0.09 }& ... & ... & ... & ... & ...  \\
 ... & HD127769 & 0.45 & -1.41 & ... & \textbf{ 1.38 $\pm$ 0.11 }& \textbf{ 143.85 $\pm$ 2.37 }& ... & ... & ... \\
 ... & SN 2021bnw & 0.04 & 0.03 & ... &....&...& 0.05 $\pm$ 0.10 $^{\rm(bc)}$& 132.98 $\pm$ 20.77  & ...  \\
2022-01-29 & HD94851 & -0.04 & -0.08 & \textbf{ 0.09 $\pm$ 0.08 }& ... & ... & ... & ... & ...  \\
 ... & HD251204 & -3.10 & 3.73 & ... & \textbf{ 4.88 $\pm$ 0.18 }& \textbf{ 64.38 $\pm$ 1.04 }& ... & ... & ... \\
 ... & BD +64 106 & 5.46 & 1.74 & ... & \textbf{ 5.80 $\pm$ 0.11 }& \textbf{ 9.16 $\pm$ 0.53 }& ... & ... & ... \\
 ... & SN 2021bnw Host & 0.00 & 0.00 & ... &....&...& 0.09 $\pm$ 0.21$^{\rm(bc)}$ & -12.25 $\pm$ 23.15  & ...  \\
          \hline          
2021-05-16 & BD$+$32 3739 & 0.08 & 0.09 & \textbf{ 0.12 $\pm$ 0.09 }& ... & ... & ... & ... & ...  \\
 ... & Hiltner960 & 1.99 & -5.51 & ... & \textbf{ 5.92 $\pm$ 0.09 }& \textbf{ 144.40 $\pm$ 0.44 }& ... & ... & ... \\
 ... & STAR 1 & -0.84 & -0.06 & ... &....&...& 0.94 $\pm$ 0.15 & 4.96 $\pm$ 4.57  & ...  \\
 ... & STAR 2 & -0.90 & 0.25 & ... &....&...& 1.00 $\pm$ 0.22 & 175.91 $\pm$ 6.36  & ...  \\
 ... & STAR 3 & -0.78 & -0.01 & ... &....&...& 0.87 $\pm$ 0.40 & 3.86 $\pm$ 13.26  & ...  \\
 ... & ISP & 0.00 & 0.00 & ... &....&...& \textbf{ 0.94 $\pm$ 0.12 }& \textbf{ 1.63 $\pm$ 3.63 }& ...  \\
 ... & SN 2021fpl $^{\rm(f)}$& -0.16 & -0.30 & ... &....&...& 0.46 $\pm$ 0.23 & 29.60 $\pm$ 14.44 & \textbf{1.10 $\pm$ 0.25$^{\rm(bc)}$  } \\
2021-06-13 & BD$+$32 3739 & 0.06 & 0.03 & \textbf{ 0.07 $\pm$ 0.06 }& ... & ... & ... & ... & ...  \\
 ... & Hiltner960 & 1.92 & -5.30 & ... & \textbf{ 5.65 $\pm$ 0.06 }& \textbf{ 144.58 $\pm$ 0.31 }& ... & ... & ... \\
 ... & STAR 1 & -1.02 & -0.28 & ... &....&...& 1.12 $\pm$ 0.12 & 8.39 $\pm$ 3.17  & ...  \\
 ... & STAR 2 & -0.90 & -0.14 & ... &....&...& 0.98 $\pm$ 0.15 & 5.25 $\pm$ 4.35  & ...  \\
 ... & STAR 3 & -0.55 & -0.16 & ... &....&...& 0.64 $\pm$ 0.15 & 8.96 $\pm$ 6.84  & ...  \\
 ... & ISP & 0.00 & 0.00 & ... &....&...& \textbf{ 0.93 $\pm$ 0.08 }& \textbf{ 7.46 $\pm$ 2.50 }& ...  \\
 ... & SN 2021fpl $^{\rm(f)}$& 0.19 & -0.06 & ... &....&...& 0.16 $\pm$ 0.22 & 71.80 $\pm$ 39.64 & \textbf{0.70 $\pm$ 0.23$^{\rm(bc)}$  } \\
2021-06-28 & BD$+$32 3739 & 0.06 & -0.04 & \textbf{ 0.07 $\pm$ 0.03 }& ... & ... & ... & ... & ...  \\
 ... & Hiltner960 & 2.06 & -5.41 & ... & \textbf{ 5.73 $\pm$ 0.06 }& \textbf{ 145.23 $\pm$ 0.31 }& ... & ... & ... \\
 ... & STAR 1 & -0.78 & -0.03 & ... &....&...& 0.83 $\pm$ 0.08 & 179.08 $\pm$ 2.64  & ...  \\
 ... & STAR 2 & -0.58 & 0.05 & ... &....&...& 0.64 $\pm$ 0.10 & 175.39 $\pm$ 4.31  & ...  \\
 ... & STAR 3 & -0.52 & 0.11 & ... &....&...& 0.60 $\pm$ 0.09 & 172.16 $\pm$ 4.51  & ...  \\
 ... & ISP & 0.00 & 0.00 & ... &....&...& \textbf{ 0.70 $\pm$ 0.05 }& \textbf{ 176.21 $\pm$ 2.08 }& ...  \\
 ... & SN 2021fpl $^{\rm(f)}$& 0.06 & 0.10 & ... &....&...& 0.15 $\pm$ 0.13 & 132.89 $\pm$ 23.98 & \textbf{0.63 $\pm$ 0.13$^{\rm(bc)}$  } \\
2021-07-08 & BD$+$32 3739 & 0.04 & 0.00 & \textbf{ 0.04 $\pm$ 0.06 }& ... & ... & ... & ... & ...  \\
 ... & Hiltner960 & 1.97 & -5.45 & ... & \textbf{ 5.79 $\pm$ 0.08 }& \textbf{ 144.74 $\pm$ 0.41 }& ... & ... & ... \\
 ... & STAR 1 & -0.89 & -0.04 & ... &....&...& 0.93 $\pm$ 0.09 & 1.31 $\pm$ 2.89  & ...  \\
 ... & STAR 2 & -0.69 & -0.18 & ... &....&...& 0.75 $\pm$ 0.12 & 7.18 $\pm$ 4.71  & ...  \\
 ... & STAR 3 & -0.71 & 0.13 & ... &....&...& 0.76 $\pm$ 0.14 & 175.20 $\pm$ 5.12  & ...  \\
 ... & ISP & 0.00 & 0.00 & ... &....&...& \textbf{ 0.82 $\pm$ 0.07 }& \textbf{ 1.38 $\pm$ 2.30 }& ...  \\
 ... & SN 2021fpl $^{\rm(f)}$& -0.08 & -0.05 & ... &....&...& 0.14 $\pm$ 0.17 & 11.70 $\pm$ 35.70 & \textbf{0.87 $\pm$ 0.17$^{\rm(bc)}$  } \\
 \hdashline
 4 epochs & STAR 1 & ... & ... & ... &....&...& 0.95 $\pm$ 0.11 & 3.80 $\pm$ 3.46  & ...  \\
 ... & STAR 2 & ... & ... & ... &....&...& 0.83 $\pm$ 0.15 & 1.05 $\pm$ 5.35  & ...  \\
 ... & STAR 3 & ... & ... & ... &....&...& 0.70 $\pm$ 0.23 & 0.26 $\pm$ 9.45  & ...  \\
 ... & ISP & ... & ... & ... &....&...& \textbf{ 0.85 $\pm$ 0.30 }& \textbf{ 2.27 $\pm$ 10.09 }& ...  \\
 2021-05-16 & SN 2021fpl $^{\rm(f)}$& -0.16 & -0.30 & ... &....&...& 0.46 $\pm$ 0.23 & 29.60 $\pm$ 14.44 & \textbf{0.49 $\pm$ 0.38$^{\rm(bc)}$  } \\
 2021-06-13 & SN 2021fpl $^{\rm(f)}$& 0.19 & -0.06 & ... &....&...& 0.16 $\pm$ 0.22 & 71.80 $\pm$ 39.64 & \textbf{0.84 $\pm$ 0.37$^{\rm(bc)}$ } \\
 2021-06-28 & SN 2021fpl $^{\rm(f)}$& 0.06 & 0.10 & ... &....&...& 0.15 $\pm$ 0.13 & 132.89 $\pm$ 23.98 & \textbf{0.77 $\pm$ 0.33$^{\rm(bc)}$}\\
 2021-07-08 & SN 2021fpl $^{\rm(f)}$& -0.08 & -0.05 & ... &....&...& 0.14 $\pm$ 0.17 & 11.70 $\pm$ 35.70 & \textbf{0.57 $\pm$ 0.34$^{\rm(bc)}$ } \\
 \hdashline
 4 epochs & ISP (STAR 1) & ... & ... & ... &....&...& \textbf{0.95 $\pm$ 0.11} & \textbf{3.80 $\pm$ 3.46}  & ...  \\
 2021-05-16 & SN 2021fpl $^{\rm(f)}$& -0.16 & -0.30 & ... &....&...& 0.46 $\pm$ 0.23 & 29.60 $\pm$ 14.44 & \textbf{0.67 $\pm$ 0.26$^{\rm(bc)}$} \\
 2021-06-13 & SN 2021fpl $^{\rm(f)}$& 0.19 & -0.06 & ... &....&...& 0.16 $\pm$ 0.22 & 71.80 $\pm$ 39.64 & \textbf{1.01 $\pm$ 0.25$^{\rm(bc)}$} \\
 2021-06-28 & SN 2021fpl $^{\rm(f)}$& 0.06 & 0.10 & ... &....&...& 0.15 $\pm$ 0.13 & 132.89 $\pm$ 23.98 & \textbf{0.96 $\pm$ 0.17$^{\rm(bc)}$} \\
 2021-07-08 & SN 2021fpl $^{\rm(f)}$& -0.08 & -0.05 & ... &....&...& 0.14 $\pm$ 0.17 & 11.70 $\pm$ 35.70 & \textbf{0.77 $\pm$ 0.20$^{\rm(bc)}$} \\
           \hline         
	\end{tabular}
\end{table*}

\subsubsection{SN 2021bnw} \label{pol_analysis_2021bnw}

The degree of polarization obtained after bias correction along the line-of-sight of SN 2021bnw about 3 months after maximum light (+89 days), is $p=0.05 \pm 0.10 \%$. This measurement, displayed in column 8 in Table \ref{tab:pol_results}, was obtained after instrumental polarization ($IP$), and zero polarization angle ($ZPA$) corrections only, without applying any correction to remove the polarization component foreground to the SN. Getting estimates of the Galactic polarization contribution is difficult for that source since there is no star available in the field of view covered by ALFOSC (see Figure~\ref{fig:images}, top). An alternative is to look to starlight polarization of Galactic stars compiled in the catalog provided by \citep[][]{heiles2000}. A few stars have been observed at distances about 6$^{\circ}$ from the position of SN 2021bnw. Relevant data retrieved from the catalog are given in Table~\ref{tab:heiles_2021bnw}. All measurements show very low levels of polarization, with $p \leq 0.2 \%$. The most relevant measurement is $p= 0.166 \pm 0.025 \%$ from HD 91316 (see bold line) which heliocentric distance of 791.5 parsecs is the highest. If this measurement is representative of the level of polarization along the LOS of SN 2021bnw, this means that one could expect that the overall level of polarisarion of SN 2021bnw and its host is consistent with a null-polarization degree.

Since a contribution to the degree of polarization foreground to SN 2021bnw could also come from its host galaxy, we observed SN 2021bnw and its host a second time about one year after maximum light (+347 days), when the SN was more than 4 magnitudes fainter than its host (see Figure~\ref{fig:LCandPOL}, top). After bias corection, the $IP$ and $ZPA$ corrected measured degree of polarization, given in column 8 in Table \ref{tab:pol_results}, is $p=0.09 \pm 0.21 \%$. This measurement is consistent with the first epoch measurement, and at the same level than the Galactic interstellar polarization estimated from the \citet{heiles2000} catalog in the vicinity of the system. Since the flux from the SN was negligible at this epoch, this means that the intrinsic degree of polarization of SN 2021bnw host galaxy is negligible and can be considered consistent with $0 \%$ at both epochs. This result suggests that an intrinsic null-polarization degree was measured on SN 2021bnw during the first epoch measurement (+89 days). Assuming the galaxy is fully unpolarized, we also estimate that the total flux of the galaxy would dilute the polarization level measured on SN 2021bnw during the first polarimetry epoch at the level of 30$\%$ in the V-band. This means that the intrinsic polarization level of SN 2021bnw would be 1.3 higher than its estimated value, but still consistent with a null-polarization measurement.

\subsubsection{SN 2021fpl} \label{pol_analysis_2021fpl}

The degree of polarization obtained on SN 2021fpl at the 4 epochs are displayed in bold in the last column of Table~\ref{tab:pol_results}. 
The first bias corrected 4 epochs estimates ($P=1.10 \pm 0.25 \%$, $P=0.70 \pm 0.23 \%$,
$P=0.63 \pm 0.13 \%$ and $P=0.87 \pm 0.17 \%$) have been obtained 
following one of the methods used to derive the intrinsic level 
of polarization on SN 2020znr in \citet{Poidevin2022} 
(in the following we call it METHOD 1). 
At each epoch the $Q$ and $U$ 
Stokes parameters measured on field stars, STAR 1, STAR 2, STAR 3, 
were $IP$ and $ZPA$ corrected using the unpolarized and polarized 
stars measurements. These measurements obtained on the 3 field stars 
were then weighted averaged to get a final estimate of the interstellar
polarization (ISP) from the Milky Way. 

A look to columns 8 and 9 in Table \ref{tab:pol_results} shows that, 
for each of the field star, the estimates obtained at the 4 epochs are 
almost all consistent with each other. The coordinates of these stars 
are displayed in Table~\ref{tab:gaia_edr3_dist}. Using TOPCAT \citep[][]{2005ASPC..347...29T}, 
we cross-matched the stars coordinates with the GAIA Early Data Release 3
(EDR3) catalog in order to obtain estimates of the distances to each star.
The median of the geometric distance posterior, $rgeo$, and the median of
the photogeometric posterior, $rpgeo$, extracted from the catalog are
displayed in column 4 and 5, respectively, in
Table~\ref{tab:gaia_edr3_dist}. The distance estimate to STAR 2 is of order
800 pc, while the one to STAR 3 is of about 3600 pc, and the one to STAR 1
is of order 5300 pc or higher. This seems to corroborate with the high
level of polarization of order $P=0.95 \pm 0.11\%$ observed on STAR 1 when
the 4 epochs measurements are weighted averaged, while similar averaged
measurements obtained on STAR 2 and STAR 3, are of $P=0.83\pm 0.15 \%$, and
of $P=0.70\pm 0.23 \%$, respectively. These values are displayed in 
columns 8 and 9 in Table~\ref{tab:pol_results}, under field 
Date $= 4$ epochs. Since these 3 estimates are almost consistent
with each other within their uncertainties, they were weighted averaged to
get a first estimate of the global 4 epochs ISP 
contribution, $P(\rm{ISP}) = 0.85 \pm 0.31\%$), and 
$\theta_{\rm{ISP}} = 2.27 \pm 10.09 ^{\circ}$ (METHOD 2). 
The relatively large uncertainty on $P$ ($\sigma_{P} =0.31 \%$) coming
from this combination propagates on the intrinsic estimates of $P$ 
obtained on SN 2021fpl at each epoch 
($P=0.49 \pm 0.38 \%$, $P=0.84 \pm 0.37 \%$,
$P=0.77 \pm 0.33 \%$ and $P=0.57 \pm 0.34 \%$), and leads to less accurate measurements than with METHOD 1.   

Finally, since STAR 1 is the field star at the higher distance from 
the sun, we consider
the 4 epochs weighted measurements obtained on that star, as another good
proxy of the averaged ISP obtained at the 4 epochs (METHOD 3). The
bias corected intrinsic level of polarization of SN 2021fpl obtained with that method 
are given in the last columns of the 4 last rows in
Table~\ref{tab:pol_results}. Similarly to
the results obtained with METHOD 1, these estimates show $> 3 \sigma$
measurements, leading to the possibility  that the photosphere of 
SN 2021fpl, observed close to and after maximum light, was not symmetrical.
A result that was also observed on SN 2015bn, and on SN 2017egm. 



\begin{table}
	\centering
	\caption{Gaia EDR3 distances to the field stars, STAR 1, STAR 2 and STAR 3, displayed in Figure~\ref{fig:images}. Parameter, $rgeo$, is the geometric distance, while parameter, $rpgeo$, is the photogeometric distance \citep[see][for details]{bailer-jones2021}}
	\label{tab:gaia_edr3_dist}
	\begin{tabular}{ccccccc} 
          \hline
          Star name &  RA (J2000) &  Dec (J2000)  & $rgeo$ & $rpgeo$ \\
           &   $[^{\circ}]$&  $[^{\circ}]$ & [pc] &  [pc] \\
          \hline
          STAR 1 & 303.5720 & -18.1781 & 6643.95166 & 5319.87744 \\ 
          STAR 2 & 303.5723 & -18.1644 & 816.467285 & 776.057312 \\ 
          STAR 3 & 303.5768 & -18.1864 & 3749.81885 & 3465.27148 \\ 
          \hline
	\end{tabular}
\end{table}


\begin{table*}
	\centering
	\caption{Starlight polarization from the \citet{heiles2000}
          agglomeration file catalog in the vicinity of SN2021bnw.}
	\label{tab:heiles_2021bnw}
	\begin{tabular}{cccccccccccc} 
          \hline
          Star name &  RA (J2000) &  Dec (J2000)  &GLON     &GLAT &  $P$&  $\sigma_{P}$ &  $\theta $ &  $\sigma_{\theta}$ &  V  &  Heliocentric  & Distance to \\
           &   &   &     & &  &   &  &   &   &   Distance &  SN2021bnw \\  
           &   $[^{\circ}]$&  $[^{\circ}]$ & $[^{\circ}]$    & $[^{\circ}]$&  [\%] &   $[\%]$ &  $[^{\circ}]$ &  $[^{\circ}]$ &  [mag] &  [pc] & $[^{\circ}]$ \\
          \hline
97179.0 &  167.844135 &       8.2670 & 246.8084 & 59.7332 & 0.090 &             0.069 &                 53.8 &                          21.0 &      9.0 &           436.5 &                          6.04 \\
91636.0 &  158.759130 &       8.6504 & 236.2949 & 52.8599 & 0.080 &             0.120 &                 73.0 &                          36.9 &      5.6 &            65.0 &                          5.99 \\
\textbf{91316.0} &  \textbf{158.202810} &       \textbf{9.3065} & \textbf{234.8871} & \textbf{52.7675} & \textbf{0.166} &             \textbf{0.025} &                 \textbf{98.1} &                           \textbf{4.3} &      \textbf{3.8} &           \textbf{791.5} &                          \textbf{6.09} \\
97907.0 &  168.966240 &      13.3075 & 239.9347 & 63.7440 & 0.050 &             0.120 &                 50.0 &                          50.2 &      5.3 &            87.1 &                          5.55 \\
97633.0 &  168.560175 &      15.4297 & 235.3722 & 64.5853 & 0.010 &             0.120 &                 93.0 &                          80.5 &      3.3 &            25.0 &                          5.80 \\
    \hline
	\end{tabular}
\end{table*}

\section{Discussion} \label{discussion}


\begin{table}
	\centering
	\caption{Possible categorization of some SLSNe that have been probed with linear polarimetry. 
	$^{\rm(a)}$: \citet{konyvestoth21}. 
	$^{\rm(b)}$: This Work. 
	$^{\rm(c)}$: \citet{leloudas2017_erratum}.
	$^{\rm(d)}$: \citet{2017ApJ...850...55N}.
	$^{\rm(e)}$: \citet{inserra2016, leloudas2017}. 
	$^{\rm(f)}$: \citet{2017ApJ...845L...8N}.
	$^{\rm(g)}$: \citet{saito2020}. 
	$^{\rm(h)}$: \citet{2020MNRAS.497..318L}
	$^{\rm(i)}$: \citet{lee2019}.
	$^{\rm(j)}$: \citet{pursiainen2022}.
	$^{\rm(k)}$: \citet{2018AA...620A..67A}
	$^{\rm(l)}$: \citet{Poidevin2022}.
	$^{\rm(m)}$: \citet{lee2020}.}
	\label{tab:comparisons}
	\begin{tabular}{llll} 
          \hline
          SLSN &  W / 15bn$^{\rm(a)}$ &  Fast / Slow &  $P$ $^{\rm(b)}$  \\
               &  Type &   evolving event &  [\%]   \\
          \hline
          LSQ14mo & W$^{\rm(a)}$    & Slow $^{\rm(a)}$             & null$^{\rm(c)}$ \\
          2015bn  & 15bn$^{\rm(a)}$ & Slow$^{\rm(a,d)}$ & increase with time$^{\rm(e)}$ \\
          2017egm & W$^{\rm(f)}$    & Fast$^{\rm(f)}$ & increase with time$^{\rm(g)}$ \\
          2018bsz & W$^{\rm(j)}$    & Slow$^{\rm(a,k)}$ & change with time$^{\rm(j)}$ \\
          2020znr & W$^{\rm(l)}$    & Slow$^{\rm(l)}$ & null$^{\rm(l)}$\\
          2020ank$^{\rm(b)}$ & W & Fast$^{\rm(b)}$ & null$^{\rm(m)}$\\
          2021bnw & W$^{\rm(b)}$ & Fast$^{\rm(b)}$ & null$^{\rm(b)}$ \\
          2021fpl & 15bn$^{\rm(b)}$ & Slow$^{\rm(b)}$ & non-zero $^{\rm(b)}$ \\ 
          \hline
	\end{tabular}
\end{table}

From the previous analysis, SN 2021bnw can be classified as a fast evolver, of spectral type W, showing no evidence of intrinsic polarization over time (-14 and +77 days in the observer frame). On the other hand, SN 2021fpl is a slow evolver, of spectral type 15bn, showing hints of intrinsic polarization after maximum light ($\approx$ +23, +38 and +48 days in the observer frame). A summary of this information and similar information compiled on other type I SLSN observed with polarimetry is given in Table~\ref{tab:comparisons}. The sample is obviously quite low for one to derive any strong conclusion but it is interesting to note that, while polarimetry is not available before maximum light on SN 2021fpl, it is showing an intrinsic level of polarization after maximum light, as does, SN 2015bn which is also classified as a slow evolver. On the other hand, since no polarimetry was obtained earlier before maximum light, a contribution from SN 2021fpl's host galaxy can not be discarded at this stage.

To better understand SLSN-I polarization properties, in the following sections we first make a summary of the variation of the intrinsic (or at least ISP corrected) linear polarization of SLSNe as a function of their empirical diffusion timescale estimates. We then have a look at the several spectral features observed with linear polarimetry on this sample of objects. Indeed a similar filter will capture different absorption features as a function of the redshifts and phases of the transients. Such a study may help to better understand the polarization sample currently available. Then, we explore the \texttt{MOSFiT} space parameters to search for any correlation between SLSN-I polarization properties and the parameters inferred from the modelling of the light curves.

\subsection{SLSNe light curves polarimetry sampling}

\begin{figure}
\begin{center}
\vspace*{2mm}
\centering
\hspace*{0.cm}
\includegraphics[width=85mm,angle=0]{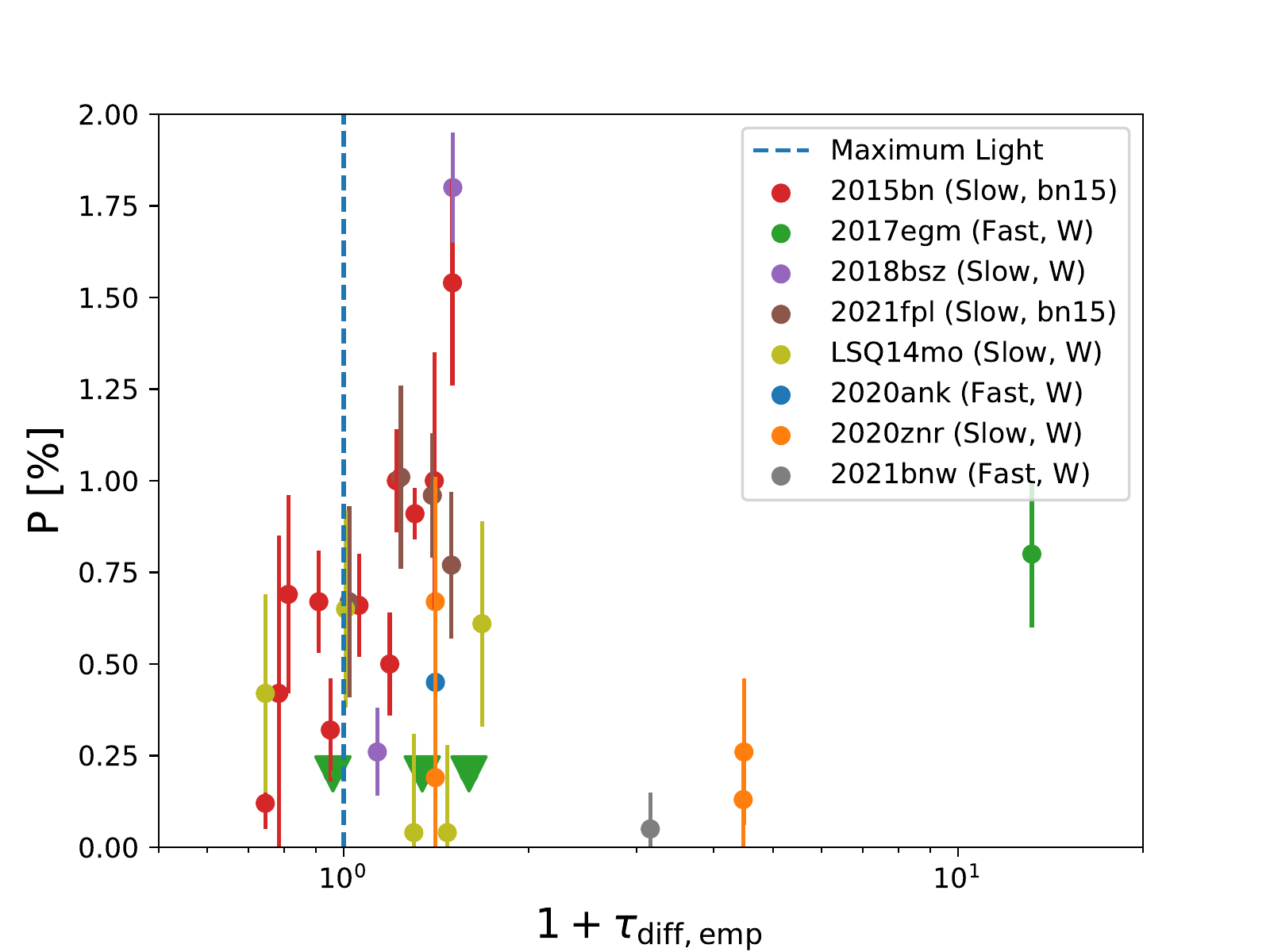}
\includegraphics[width=85mm,angle=0]{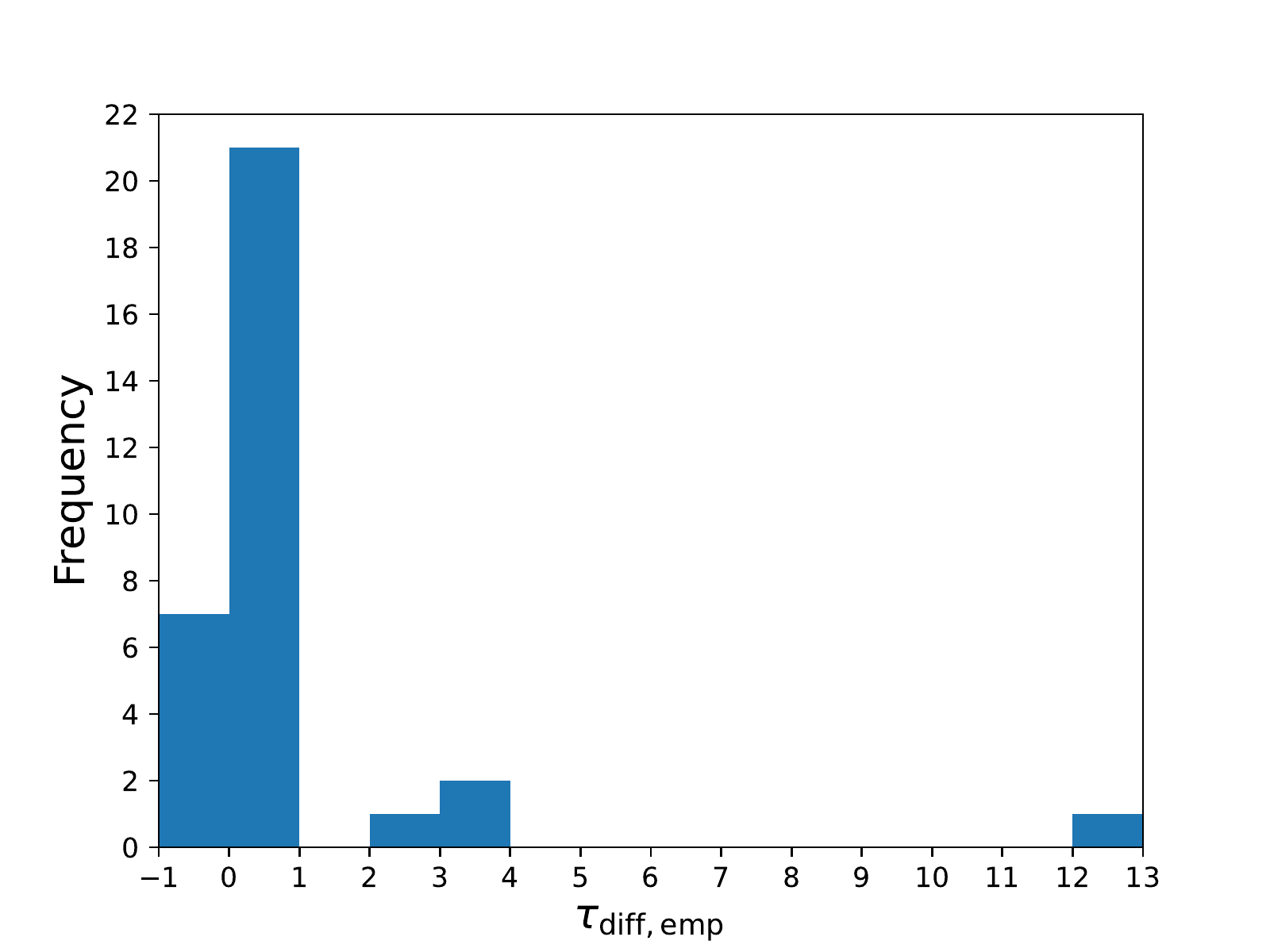}
\vspace*{0.cm}
\caption{Top: Distribution of the degree of polarization as a function of the diffusion time scale from the sample of SLSN-I for which intrinsic polarization measurements are available. Bottom: histogram of the diffusion time scales when polarimetry was obtained.}
\label{fig:polarimetry_SLSNe_vs_phases}
\end{center}
\end{figure}


ISP corrected or intrinsic polarization measurements have been obtained on SN 2015bn \citep[][]{inserra2016, leloudas2017}, SN 2017egm \citep[][]{saito2020}, SN 2018bsz \citep[][]{pursiainen2022}, SN 2021fpl (this work), SN LSQ14mo \citep[see][with the second reference giving revised ISP corrected polarimetry of the data presented in the first reference.]{leloudas2015, leloudas2017_erratum}, SN 2020ank \citep[][]{lee2020}, SN 2020znr \citep[][]{Poidevin2022} and SN 2021bnw (this work). The polarimetry of SN 2017egm is similar to linear polarimetry obtained from spectropolarimetry after some integration in the wavelength range $\approx$ 4600--9400 \,{\AA}, rest frame \citep[see][]{saito2020}. Linear polarimetry was also conducted by \citet{maund2019} on that source but the measurements were lacking sensitivy therefore they are not included in this analysis. Similarily, the results obtained on SN 2018bsz are from spectropolarimetry after some integration in the wavelength range $\approx$ 3200--9100 \,{\AA}, rest frame. Spectropolarimetry is also available on SN 2015bn \citep[see][]{inserra2016}. The data obtained at two epochs integrated over the ALFOSC V-band filter give results consistent with those obtained with linear polarimetry by \citet{leloudas2017}, and are included in this analysis. All these spectropolarimetry data are considered and discussed in more details in Section~\ref{specpoldisc}. 

Figure~\ref{fig:polarimetry_SLSNe_vs_phases}, top, shows the distribution of the polarimetry obtained on SN 2015bn, SN 2017egm, SN 2018bsz, SN 2021fpl, SN LSQ14mo, SN 2020ank, SN 2020znr and SN 2021bnw as a function of $\tau_{\rm diff, emp}+1$ (in log scale), where $\tau_{\rm diff, emp}$ is the empirical diffusion time scale.
The empirical diffusion timescales were obtained by normalizing the SLSNe phases when polarimetry was measured by the rising timescale estimates retrieved from light curve analysis (see Table~\ref{tab:mosfitparamsallsne}). Maximum light is shown by the vertical dashed-line in the plot. The data spread in the empirical diffusion time scale range -0.3--12.2. This is illustrated in the histogram displayed in Figure~\ref{fig:polarimetry_SLSNe_vs_phases}, bottom.
Seven intrinsic polarization measurements are available before maximum light ($\tau_{\rm diff, emp} < 0$). A few of them obtained on SN 2015bn are more than 3$\sigma$ detections but all of them are showing $P<0.75\%$. Departure from this limit are seen after maximum light and at empirical diffusion time scales between 0 and 1 on the transients SN 2015bn (Slow evolver, bn15 type), SN 2018bsz (Slow evolver, W type) and SN 2021fpl (Slow evolver, 15bn type), while the remaining sources (all W type) are all showing low level of polarization ($P<0.75\%$).
Above $\tau_{\rm diff, emp} > 1$, in the empirical diffusion time scale range $2 < \tau_{\rm diff, emp} < 4$, the measurements are very scarce with null-polarization detections on SN 2021bnw (Fast evolver, W type) and 
on SN 2020znr (Slow evolver, W type). Then another high polarization detection is available on SN 2017egm (Fast evolver, W type) at a high empirical diffusion time scale, $\tau_{\rm diff, emp} \approx 12.2$. 

With a total of 32 measurements, empirical diffusion time scales are mainly sampled by polarimetry in the range, $-1 < \tau_{\rm diff, emp} < 1$ (28  measurements). In that domain, independently of the spectral type, only some of the slow evolvers are showing an increase of polarization and one has to probe empirical diffusion time scales $\tau_{\rm diff, emp} > 3$ to detect an increase of polarization on a fast evolver (SN 2017egm). Obviously, higher polarimetry cadence samples on a larger sample of SLSN-I and on the wide range of empirical diffusion time scales discussed here ($-1 < \tau_{\rm diff, emp}$ < 13) may lead to different conclusions. 
If such a sample was available, its interpretation could also be subject to the limit of the spectral ranges probed by linear polarimetry, when this technique is applied.
For a given pass-band the spectral domain probed in the rest-frame of the SLSN will vary with the redshift and may lead to higher or lower probability to detect polarization variations. We discuss these aspects in the following section.    


\begin{table*}
	\centering
	\caption{Magnetar model parameters from \texttt{MOSFiT} for some SLSNe that have been probed with linear polarimetry.  The first three have shown an increase or change in polarization over time or a non negligible level of polarization, while the rest have shown only null detections.  .
	$^{\rm(a)}$: If not mentioned otherwise the rise time is the estimated explosion date to g- or V-band peak, and has an uncertainty of around 10\%.
	$^{\rm(b)}$:  Rise time estimates from \citet{konyvestoth21}.
	$^{\rm(c)}$:  Parameters will likely be estimated in Roy et al. (in prep).
	$^{\rm(d)}$: Fit with \texttt{MINIM} \citep{Chatzopoulos2013}. 
	}
	\label{tab:mosfitparamsallsne}
	\begin{tabular}{lccccccc} 
	\hline
          SN Name & $P_{\rm spin}$ (ms) & $B_\perp$ (10$^{14}$ G) & $M_{\rm ej}$ ($M_\odot$) & $v_{\rm ej}$ (10$^3$ km/s) & $M_{\rm NS}$ ($M_\odot$) & Rise Time$^{\rm(a)}$ (days) & Source\\
          \hline
          SN 2015bn & ${2.16}_{-0.17}^{+0.29}$	& ${0.31}_{-0.05}^{+0.07}$ & ${11.73}_{-1.34}^{+0.83}$ &	${5.46}_{-0.14}^{+0.16}$ & 	${1.78}_{-0.23}^{+0.28}$ & 90.88$^{\rm(b)}$ & \cite{2017ApJ...850...55N} \\
          SN 2017egm & ${4.38}_{-0.67}^{+0.44}$	& ${0.79}_{-0.22}^{+0.20}$ & ${3.72}_{-0.90}^{+1.65}$ &	${6.35}_{-0.22}^{+0.22}$ & 	${1.67}_{-0.33}^{+0.22}$ & 16 & \cite{2017ApJ...845L...8N} \\
          SN 2018bsz$^{\rm(c)}$ & & & & & & 76.17$^{\rm(b)}$ & \cite{2018AA...620A..67A}  \\
          SN 2021fpl & ${1.43}_{-0.24}^{+0.30}$ & ${0.78}_{-0.17}^{+0.14}$ & ${21.59}_{-4.67}^{+8.22}$ & ${6.94}_{-0.28}^{+0.28}$ & ${1.71}_{-0.26}^{+0.19}$ & 96 & This Work\\
          \hline 
          SN LSQ14mo & ${4.97}_{-0.71}^{+0.65}$ & ${1.01}_{-0.30}^{+0.27}$ & ${2.10}_{-0.36}^{+0.42}$ & ${10.74}_{-0.41}^{+0.52}$ & ${1.85}_{-0.27}^{+0.22}$ & 34.22$^{\rm(b)}$ & \cite{2017ApJ...850...55N}\\
          SN 2020ank$^{\rm(d)}$ & ${2.23}\pm 0.51$ & ${2.09}\pm 0.07$ & $3.58\pm 0.04$ & $12.27\pm 0.91$ & N/A & 36 & \cite{Kumar2021}\\
          SN 2020znr & ${2.80}_{-0.39}^{+0.26}$ & ${0.52}_{-0.13}^{+0.10}$ & ${21.37}_{-1.43}^{+1.53}$ & ${5.56}_{-0.13}^{+0.13}$ & ${1.68}_{-0.31}^{+0.21}$ & 90 & \cite{Poidevin2022} \\
          SN 2021bnw & ${4.56}_{-0.64}^{+0.59}$ & ${0.40}_{-0.11}^{+0.12}$ & ${3.59}_{-0.75}^{+1.06}$ & ${7.87}_{-0.54}^{+0.50}$ & ${1.58}_{-0.28}^{+0.26}$ & 41 & This Work\\
          \hline
	\end{tabular}
\end{table*}


\begin{table*}
	\centering
	\caption{Polarimetry summary table. $^{\rm(a)}$: as from \citet{inserra2016} after integration of the VLT spectropolarimetry data in the ALFOSC V-band filter. $^{\rm(b)}$:e.g. \citet{konyvestoth21}, $^{\rm(c)}$: \citet{nicholl2013}, $^{\rm(d)}$: \citet{chen2017}, $^{\rm(e)}$: This work.}
	\label{tab:pol_spectral_features}
	\begin{tabular}{llllccccccl} 
          \hline
SN & Redshift & MJD (max)  & MJD (pol) & pol. phase & $\tau_{\rm diff, emp}$& $P$ & $\sigma_{P}$ & $\lambda_{\rm min}$ & $\lambda_{\rm max}$ & Spectral  \\
 &  & [days] & [days] & [days]& & $[\%]$ & $[\%]$ & [{\AA}] & [{\AA}] & Features  \\
\hline  \\
SN 2021fpl & 0.115 &    59265.0 &    59267.0 &        1.8 &       0.02 & 0.67 & 0.26 &    4292.38 &    5782.53 & Fe III, Fe II, Si II $\nu$ $^{\rm(e)}$  \\
... & ... & ... &    59288.0 &       20.6 &       0.24 & 1.01 & 0.25 & ... & ... &  \\
... & ... & ... &    59303.0 &       34.1 &       0.39 & 0.96 & 0.17 & ... & ... &  \\
... & ... & ... &    59313.0 &       43.0 &       0.50 & 0.77 & 0.2 & ... & ... &  \\
SN 2015bn & 0.1136 &    57102.0 &    57080.1 &      -19.6 &      -0.22 & 0.42 & 0.43 &    4297.77 &    5789.80 & Fe III, Fe II$^{\rm(b)}$ \\
... & ... & ... &    57083.0 &      -17.0 &      -0.19 & 0.69 & 0.27 & ... & ... & ... \\
... & ... & ... &    57093.0 &       -8.1 &      -0.09 & 0.67 & 0.14 & ... & ... & ... \\
... & ... & ... &    57097.1 &       -4.4 &      -0.05 & 0.32 & 0.14 & ... & ... & ... \\
... & ... & ... &    57108.0 &        5.4 &       0.06 & 0.66 & 0.14 & ... & ... & ... \\
... & ... & ... &    57121.1 &       17.1 &       0.19 & 0.5 & 0.14 & ... & ... & ... \\
... & ... & ... &    57124.1 &       19.8 &       0.22 & 1.0 & 0.14 & ... & ... & ... \\
... & ... & ... &    57142.9 &       36.8 &       0.40 & 1.0 & 0.35 & ... & ... & ... \\
... & ... & ... &    57153.0 &       45.8 &       0.50 & 1.54 & 0.28 & ... & ... & Fe II, Si II, Si II$\nu$ $^{\rm(b)}$ \\
... & ... & ... &    57076.2 &      -23.7$^{\rm(a)}$ &      -0.25 & 0.12 & 0.07 & ... & ... & Fe III, Fe II $^{\rm(b)}$ \\
... & ... & ... &    57133.0 &       27.5$^{\rm(a)}$ &       0.31 & 0.91 & 0.07 & ... & ... & ... \\
\hdashline
SN 2017egm & 0.030721 &    57925.8 &    57925.2 &       -0.6 &      -0.04 & 0.2 & ... &    4559.91 &    8828.77 & OII$^{\rm(b)}$ \\
... & ... & ... &    57931.2 &        5.2 &       0.34 & 0.2 & ... & ... & ... & ... \\
... & ... & ... &    57935.2 &        9.1 &       0.60 & 0.2 & ... & ... & ... & ... \\
... & ... & ... &    58116.5 &      185.0 &      12.18 & 0.8 & 0.2 & ... & ... &  Fe II, Na I, [Ca II] $^{\rm(c)}$ \\
SN 2020ank & 0.22 &    58894.3 &    58909.2 &       12.2 &       0.41 & 0.45 & 0.3 &    3922.95 &    5284.85 & OII$^{\rm(b)}$ \\
SN 2021bnw & 0.098 &    59265.0 &    59354.0 &       81.1 &       2.16 & 0.05 & 0.1 &    4358.83 &    5872.06 & Fe II, Na I $\nu$ $^{\rm(e)}$  \\
\hdashline
SN LSQ14mo & 0.253 &    56698.9 &    56690.2 &       -6.9 &      -0.25 & 0.42 & 0.27 &    3817.06 &    5371.12 & OII, (Si III$\nu$) $^{\rm(d)}$  \\
... & ... & ... &    56699.1 &        0.2 &       0.01 & 0.65 & 0.27 & ... & ... & ... \\
... & ... & ... &    56709.2 &        8.2 &       0.30 & 0.04 & 0.27 & ... & ... & Fe II, Mg II $^{\rm(d)}$ \\
... & ... & ... &    56715.1 &       12.9 &       0.47 & 0.04 & 0.24 & ... & ... & ... \\
... & ... & ... &    56722.1 &       18.5 &       0.68 & 0.61 & 0.28 & ... & ... & ... \\
SN 2020znr & 0.1 &    59233.0 &    59267.0 &       30.9 &       0.41 & 0.67 & 0.34 &    4350.91 &    5861.38 & Fe II$^{\rm(c)}$ \\
... & ... & ... &    59267.0 &       30.9 &       0.41 & 0.19 & 0.32 &    5046.99 &    7747.96 &  Na I, Si II, [Ca II] $^{\rm(c)}$\\
... & ... & ... &    59521.0 &      261.8 &       3.47 & 0.13 & 0.14 & ... & ... & ... \\
... & ... & ... &    59522.0 &      262.7 &       3.48 & 0.26 & 0.2 & ... & ... & ... \\
\hdashline
SN 2018bsz & 0.0267 &    58267.5 &    58278.0 &       10.2 &       0.13 & 0.26 & 0.12 &    3300.00 &    9325.00 &  See \cite{pursiainen2022}\\
... & ... & ... &    58307.0 &       38.5 &       0.51 & 1.8 & 0.15 & ... & ... & ... \\
           \hline
           \hline
	\end{tabular}
\end{table*}

\subsection{Spectral windows observed with polarimetry} \label{specpoldisc}

In this section we discuss and compare some of the spectral features integrated over the various filters used for linear polarimetry. 
The wavelength ranges covered by each filter and a summary of the polarimetry are given in Table~\ref{tab:pol_spectral_features}. The polarization values of SN 2020ank \citep{lee2020} and SN 2020znr \citep{Poidevin2022} were bias corrected using Equation ~\ref{eqn:pdeb}. The polarization values of SN 2015bn \citep{inserra2016,leloudas2017}, SN 2017egm \citep{saito2020}, SN LSQ14mo \citep{leloudas2015} and SN 2018bsz \citep{pursiainen2022} were already bias corrected and are given as so.
The spectra of SN 2015bn, SN 2017egm, SN PTF12dam and SN LSQ14mo used in this section were retrieved from the WISEReP\footnote{WISEReP, {\tt
    https://www.wiserep.org/}} \citep[see][]{yaron2012} repository. When spectra on SN 2020ank, SN 2020znr are not available at phases close to the epochs when polarimetry was obtained, the spectra of SN 2017egm and SN PFT12dam are used as spectral templates for comparisons and identification of some spectral features, respectively. If not already provided by our SYN++ analysis, this information was gathered in the literature. From one study to the other, some spectral features are sometimes described with different atomic elements. The list of the most typical features are indicated in the last column of Table~\ref{tab:pol_spectral_features}. A summary of this information is also provided on the several plots shown in Figure~\ref{fig:polarimetry_spectra_2021fpl} and Figure~\ref{fig:polarimetry_spectra_wtype}.

\subsubsection{15bn Type, Slow Evolver}

We first focus on the 15bn Type. The polarimetry sample only contains Slow evolvers with this spectral type. SN 2021fpl (z=0.115) and SN 2015bn (z=0.1136) have very similar redshift meaning that polarimetry obtained through the ALFOSC V-band are integrating the signals over very similar wavelength ranges, making spectroscopy and polarimetry comparisons more straightforward. 

\begin{figure*}
\begin{center}
\vspace*{2mm}
\centering
\hspace*{0.cm}
\includegraphics[width=140mm,angle=0]{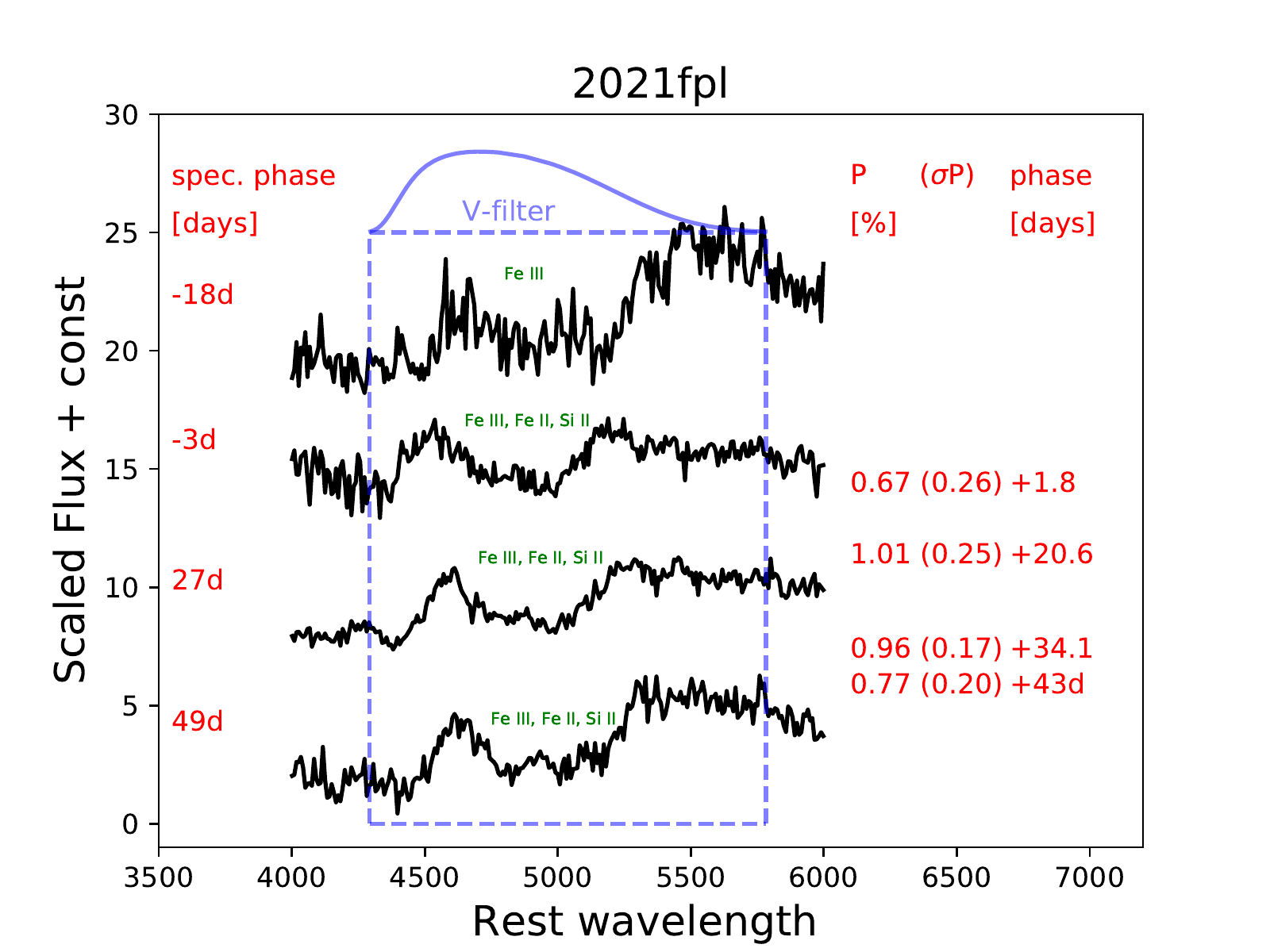}
\includegraphics[width=140mm,angle=0]{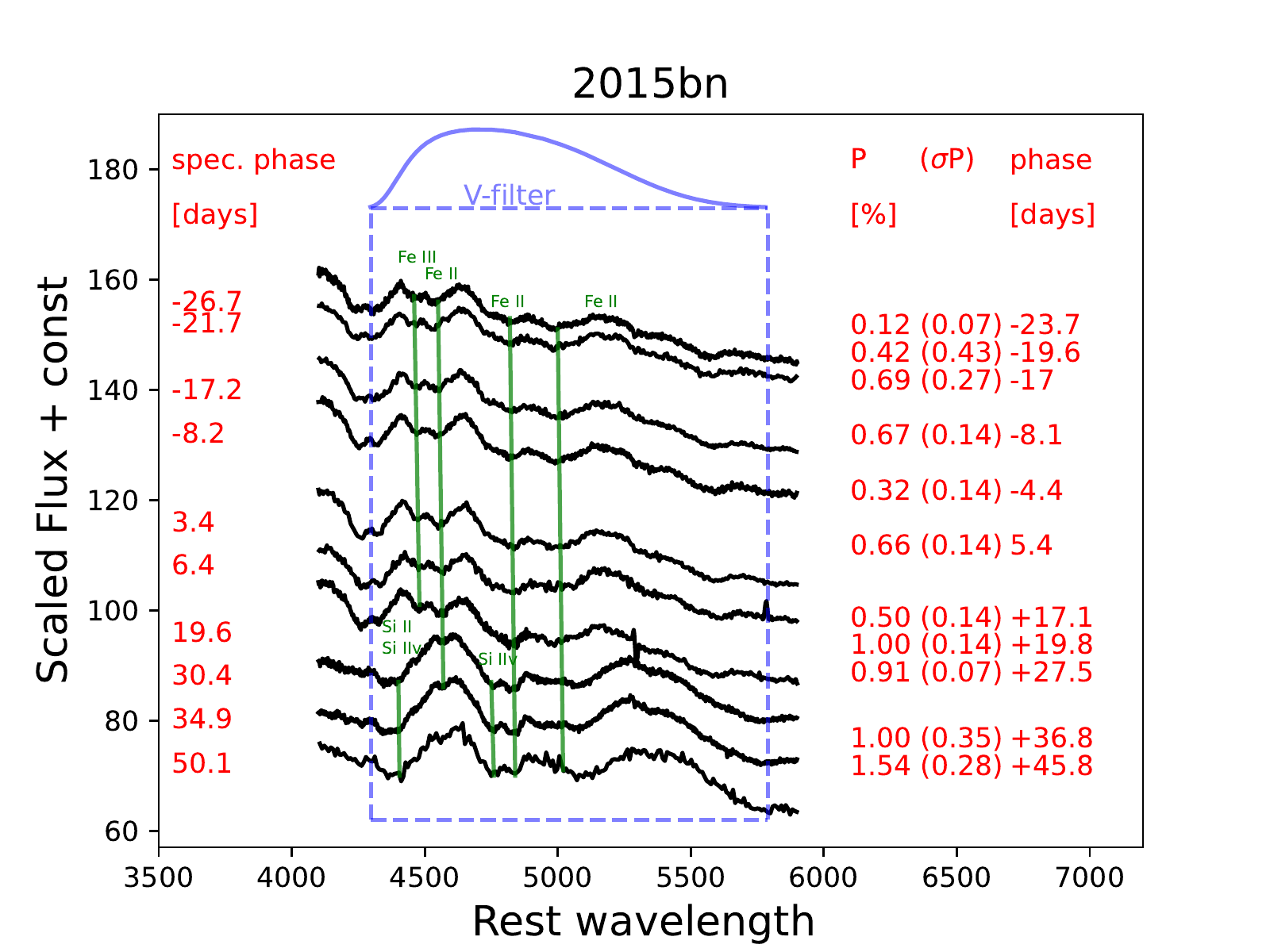}
\vspace*{0.cm}
\caption{Spectral features observed in the rest-framed NOT ALFOSC Bessel V filter. Top: SN 2021fpl. Bottom: SN 2015bn. The phases when spectroscopy and polarimetry were obtained are indicated in red. The levels of polarization are also given. The filter band-passes are shown in blue and with the blue boxes. The main absorption spectral features integrated over the filters are shown in green.}
\label{fig:polarimetry_spectra_2021fpl}
\end{center}
\end{figure*}

Figure~\ref{fig:polarimetry_spectra_2021fpl}, top, shows the spectral features integrated in the linear polarimetry ALFOSC V-band filter during the observations close to and after maximum light of SN 2021fpl. Spectroscopy obtained before maximum light (-18 days) is also shown. The main absorption spectral features captured by the polarimetry filter around and after maximum light are produced by Fe II and Fe III, with contributions from Si II, as from the \texttt{SYN++} analysis discussed in Section~\ref{syn++}. The Fe III absorption may be less strong at phase -18 days than at higher phases and Fe II and Si II are not required to fit that spectrum, inferring a spectral transition phase occurred between -18 days and -3 days. For comparison, SN 2015bn shows a significant spectral evolution between +7 days and +20 days, with very slow spectral evolution before +7 days and after +20 days \citep[see for example][]{2016ApJ...826...39N}. This is illustrated with the sequence of spectra displayed in Figure~\ref{fig:polarimetry_spectra_2021fpl}, bottom. Before a rest frame phase of +7 days SN 2015bn is very lowly polarized ($P \approx 0.5 \%$) while the level of polarization is higher after +20 days \citep[see results summarised in Table~\ref{tab:pol_spectral_features}, results displayed in the bottom plot in Figure~\ref{fig:polarimetry_spectra_2021fpl} and][]{leloudas2017}. The level of polarization of SN 2021fpl before maximum light is not known therefore we do not know if SN 2021fpl underwent a polarization transition phase as did SN 2015bn \citep[][]{inserra2016, leloudas2017}, during its spectral transition phase. On the other hand, despite their mild quality, the spectra, at phases -3 days, +27 days and +49 days, obtained on SN 2021fpl show similar spectral structures to those seen in the spectra obtained on SN 2015bn after $\approx$ + 20 days (see Figure~\ref{fig:polarimetry_spectra_2021fpl}, bottom). This gives support to the possibility that SN 2021fpl underwent a polarization transition phase earlier than SN 2015bn. 

As mentioned previously, SN 2015bn was observed with spectropolarimetry by \citet{inserra2016} at two epochs, one pre-peak at -23.7 days, and another 27.5 days after maximum in the rest-frame. The analysis of the variation of the polarization shows the presence of a dominant axis in the $Q-U$-plane with no departure from it which is interpreted as the axis of symmetry of the inner photosphere. This result was confirmed by \citet{leloudas2017} with the linear polarimetry data discussed before. The data obtained on SN 2021fpl also show the presence of a dominant axis over the period of about 40 days rest-frame during which polarimetry was acquired from about maximum light. The polarization angles obtained from the first to the fourth epochs given in the last 4 lines in Table~\ref{tab:pol_results} are $\theta = 79.4^{\circ} \pm 9.9$, $\theta = 90.8^{\circ} \pm 6.6$, $\theta = 98.0^{\circ} \pm 4.9$ and $\theta = 92.5^{\circ} \pm 7.1$, respectively.This is another interesting point suggesting that SN 2021fpl and SN 2015bn could have similar polarization properties and are part of the same class of objects. The intrinsic levels of polarization measured after SN 2015bn and SN 2021fpl underwent a spectroscopic transition are very similar to each other which also supports this hypothesis. If this is the correct this means that, similarly to SN 2015bn, the ellipticity of the photosphere of SN 2021fpl could have receded from an initially external layer with a prolate geometry of axis ratio $a/c \approx 0.9$ where $b=c$, keeping its geometry over time while the inner shell would have increased its asphericity from $\approx 0.9$ to $\approx 0.6$, as from the quantitative model proposed by \citet{inserra2016} to interpret the increase of intrinsic linear polarization of SN 2015bn.

\subsubsection{W Type, Fast Evolver}

\begin{figure*}
\begin{center}
\vspace*{2mm}
\centering
\hspace*{0.cm}
\includegraphics[width=140mm,angle=0]{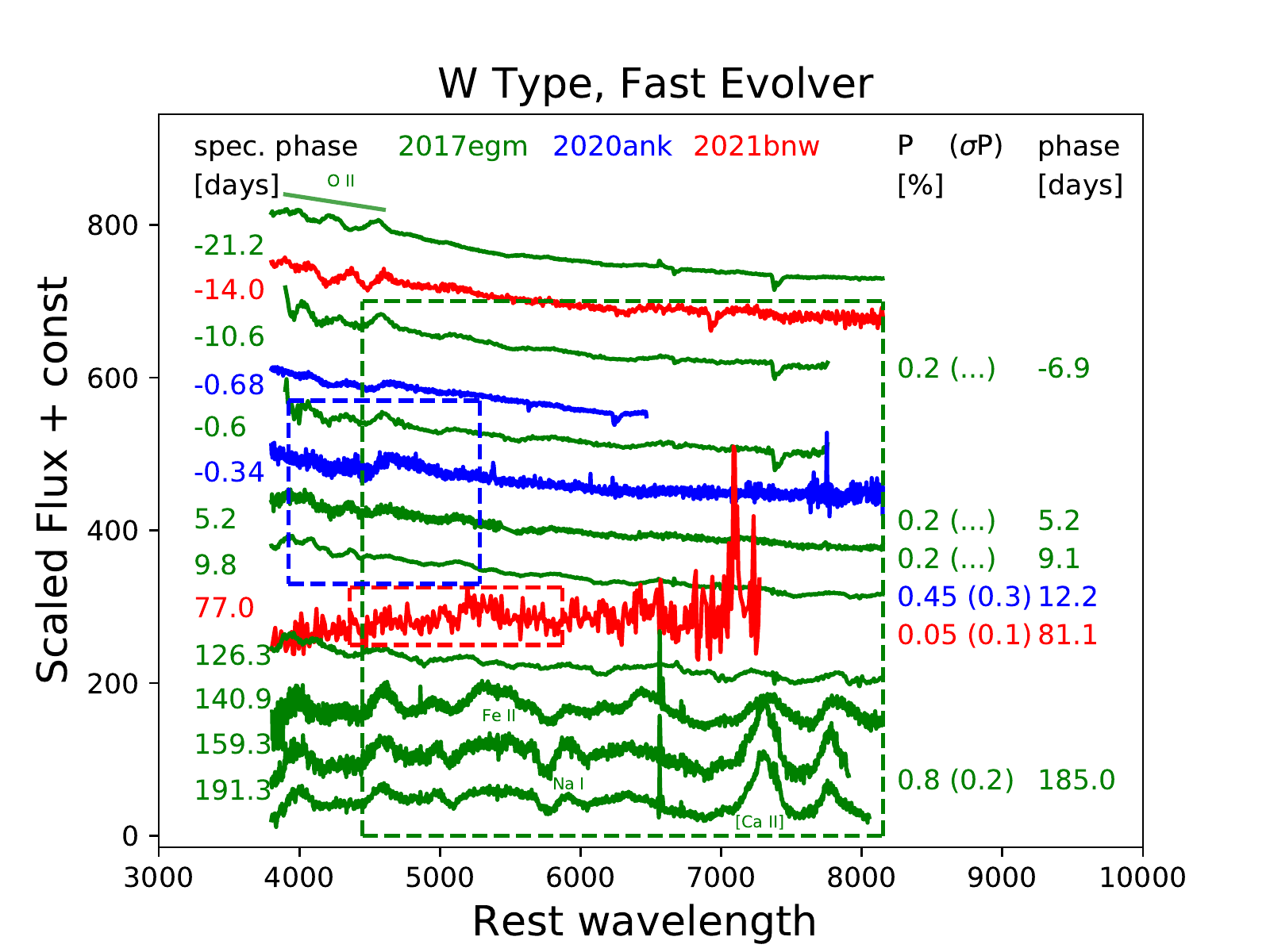}
\includegraphics[width=140mm,angle=0]{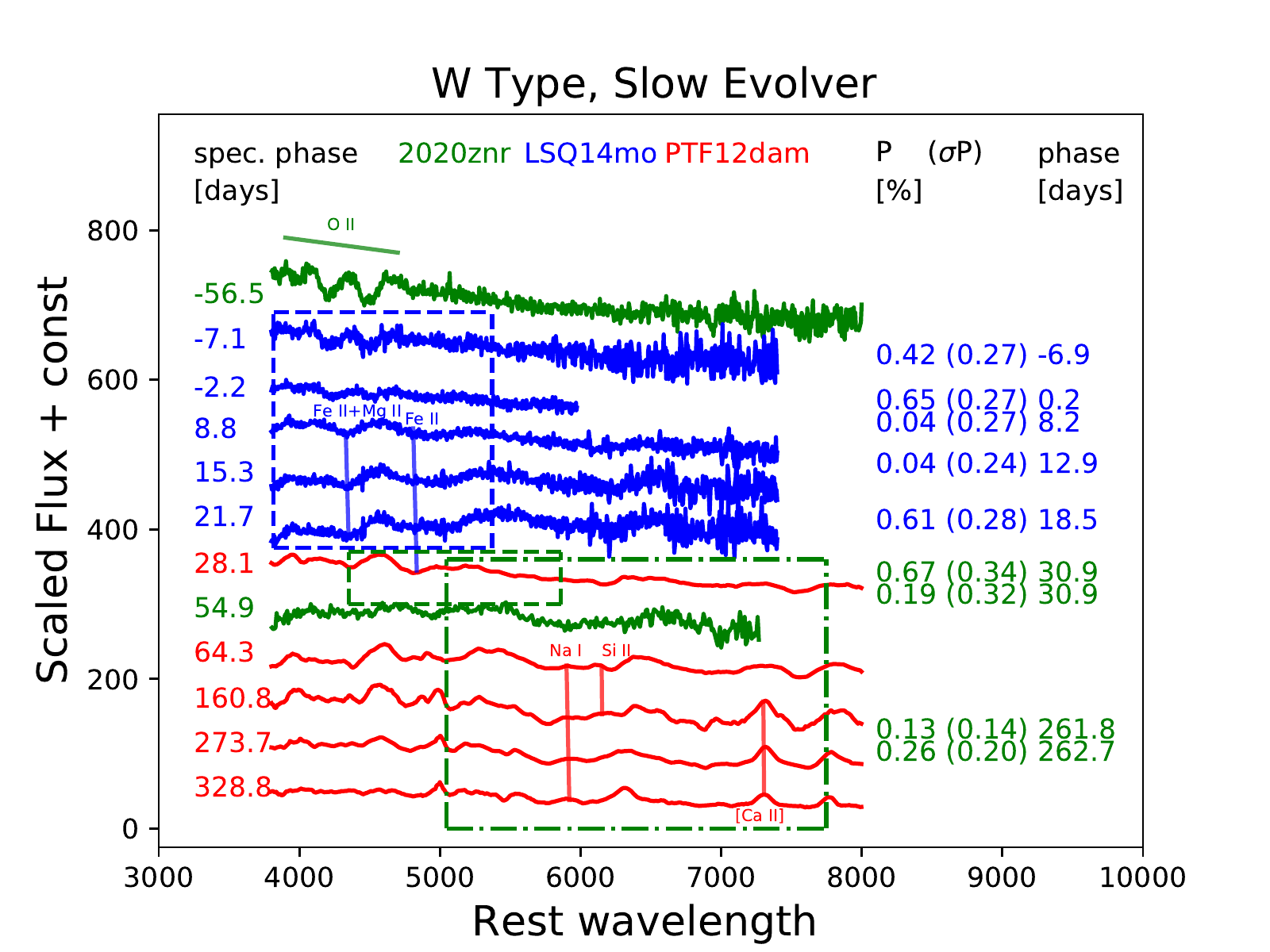}
\vspace*{0.cm}
\caption{Top plot: Spectra of the W type, Fast Evolvers SN 2017egm (green), SN 2020ank (blue) and SN 2021bnw (red). The rest-framed NOT ALFOSC Bessel V-filters are shown with the red box for SN 2021bnw, and with the blue box for SN 2020ank. The phases when spectroscopy and polarimetry were obtained are indicated in the plot, as well as the levels of polarization. Bottom plot: same as top plot but for the Slow evolvers SN 2020znr (green), LSQ14mo (blue) and PTF12dam (red). The rest-framed V-filters wavelength ranges are shown with the blue and with the small green dashed-line boxes. The wide green dashed-line box shows the wavelength range covered by the NOT ALFOSC Bessel R-filter that was used to observe SN 2020znr.}
\label{fig:polarimetry_spectra_wtype}
\end{center}
\end{figure*}

SN 2017egm was observed with spectropolarimetry at 4 epochs by \citet{saito2020}. 
A sequence of some of the spectra publicly available on WISEReP are shown in green in  Figure~\ref{fig:polarimetry_spectra_wtype}. The wavelength range over which \citet{saito2020} made simple average of the polarization was between 4450 and 8150 {\AA}. This is illustrated by the dashed-line green box in the Figure. Also plotted in the Figure are the publicly available spectra of SN 2020ank shown in blue, and the spectra of SN 2021bnw shown in red and discussed in this work. All together, these spectra show the spectral evolution one could roughly expect for each of these SLSN. The dashed-blue box shows the wavelength range of the V-band filter used for polarimetry on SN 2020ank by \citet{lee2020}. One could expect the spectral feature integrated over the filter about 12 days after maximum light was mainly the well known W-shape absorption feature produced by O II and C II, as from the analysis of the -14 days rest-frame spectrum of SN 2021bnw discussed in Section~\ref{syn++}. This feature is partly captured by the bandwidth used to derive the polarization level of SN 2017egm shown with the green dashed-box. All put together these results show a consistent picture suggesting that the photosphere of the W Type, Fast evolvers is still almost spherical until a few weeks after maximum light during which the W feature is still visible, i.e. the photospheres have prolate geometry of axis-ratio of order $0.9$ -- $0.95$ \citep[see][]{1991A&A...246..481H}. 

The filter used for polarimetry on SN 2021bnw about 80 days after maximum light is shown with the red dashed-box. The low level of polarization obtained in this filter suggests that the spectral features now produced by Fe II and Na I are associated to a photosphere still almost spherical. At later phase around +185 days SN 2017egm underwent an increase of polarization which interpretation is complex \citep[see discussion in][]{saito2020}. The increase in polarization could come from a loss of symmetry on the inner photosphere, but additional late epochs polarimetry would have been helpful to show if the change of polarization stands along a dominant axis as for SN 2015bn, or if it was produced by some interactions with a clumpy CSM. The Fe II and Na I already observed at earlier phase in the +77 days spectrum of SN 2021bnw are visible in the spectra of SN 2017egm close to that epoch and, for that reason, may be rejected as the cause of the change in polarization. 


\subsubsection{W Type, Slow Evolver}

SN 2018bsz has been observed with spectropolarimetry and the results are discussed by \citet{pursiainen2022}. The spectra of SN 2018bsz exhibit several features commonly seen in SLSNe but its spectral evolution looks uncommon among the W, Slow Evolver Type discussed here, with multi-component H$_{\alpha}$ profile appearing about +30 days after maximum light. The polarization results obtained by \citet{pursiainen2022} are compiled in Table~\ref{tab:pol_spectral_features} and show a measurement consistent with null-polarization after maximum light (+10.2 days) while a relatively high level of polarization $P=1.80 \pm 0.15$~$\%$ is found after maximum light (+38.4 days). In that case the ISP is unknown and there was a fairly significant change in polarization angle between the two epochs, for these reasons the increase of polarization is most likely explained by highly aspherical, possibly disk-like, CSM with several emitting regions. For all the reasons mentioned above, we did not add SN 2018bsz to the plot in Figure~\ref{fig:polarimetry_spectra_wtype}, bottom.
On the other hand a sequence of spectra including spectra of SN 2020znr (shown in green), of LSQ14mo (shown in blue) and of SN PTF12dam (shown in red) is show in that Figure. The spectra of SN PTF12dam are assumed to be representative of the W, Slow Evolver type at epochs when polarimetry was obtained on SN 2020znr, without a spectroscopy counterpart. 

In Figure~\ref{fig:polarimetry_spectra_wtype}, the blue-dashed box shows the bandwidth of the V-filter used to get polarimetry on SN LSQ14mo by \citet{leloudas2015}. The final results given by \citet{leloudas2017_erratum} are also given in the plot. As for Fast evolvers, the W-shape spectral feature associated to O II seems not to be polarised (-7.1 days). After that epoch the spectra show LSQ14mo underwent a spectral evolution with spectral features likely and mainly produced by Fe II and Mg II around 4300~{\AA} \citep[see][]{chen2017}. The lack of high signal-to-noise ratio polarization measurements obtained after maximum light on that source suggest none of these spectral features could give insight on a loss of symmetry on the inner ejecta. The same is found at later epoch (+30.9 days) with the polarimetry of SN 2020znr in the V-filter spectral window shown with the green-dashed box. The same conclusion is corroborated by the results obtained toward the R-filter window shown with the green dashed-dotted-box in the plot. At later epoch, the V-band filter was not used and the spectral window does not include the spectral features in the wavelength range 4000--5000~{\AA}. 

\subsubsection{Synthesis of the results}

The results discussed in the previous section suggest than none of the several spectral groups of SLSN-I show a substantial level of polarization before maximum light. If it happens, a change of polarization property seems rather to occur after a spectral transition of the photosphere. Regarding the 15bn group, the previous analysis suggest that SN 2021fpl underwent a spectral evolution similar to SN 2015bn, and possibly a polarization evolution as well, even though, no polarimetry is available at early phase on SN 2021fpl. On the contrary, it looks like the W Type SLSNe, SN 2020ank, SN 2021bnw, SN 2020znr and SN LSQ14mo, whether they are Slow or Fast evolvers, did not undergo a polarization transition, except maybe in the case of SN 2017egm.
In this spectral class, the increase of polarization observed on SN 2017egm (+185 days) lack additional polarimetry data epochs to know if the polarization would have evolved along a dominant axis. From the spectral analysis, though, it looks like 
the early and late spectra of SN 2017egm show H$_{\alpha}$ emission features likely coming from its host galaxy, and detected when the flux of the SLSN was not dominating the one of the galaxy. The shape of the H$_{\alpha}$ emission features is not suggesting interactions with a clumpy CSM and the increase of polarization could therefore be a probe of a loss of symmetry of the inner photosphere. From the identification or compilation of the spectral features observed by polarimetry and summarised in Table~\ref{tab:pol_spectral_features} it is difficult to associate any specific element to an increase of polarization. As mentioned by \cite{saito2020} some late-phase line profiles are sensitive to the element distribution while the continuum polarization are sensitive to the distribution of the free electron which produce polarization, therefore specific studies of these two effects would be needed to disentangle them.

\subsection{Light curve modelling parameter space}

\begin{figure*}
\begin{center}
\vspace*{2mm}
\centering
\hspace*{-1.cm}
\includegraphics[width=200mm,angle=0]{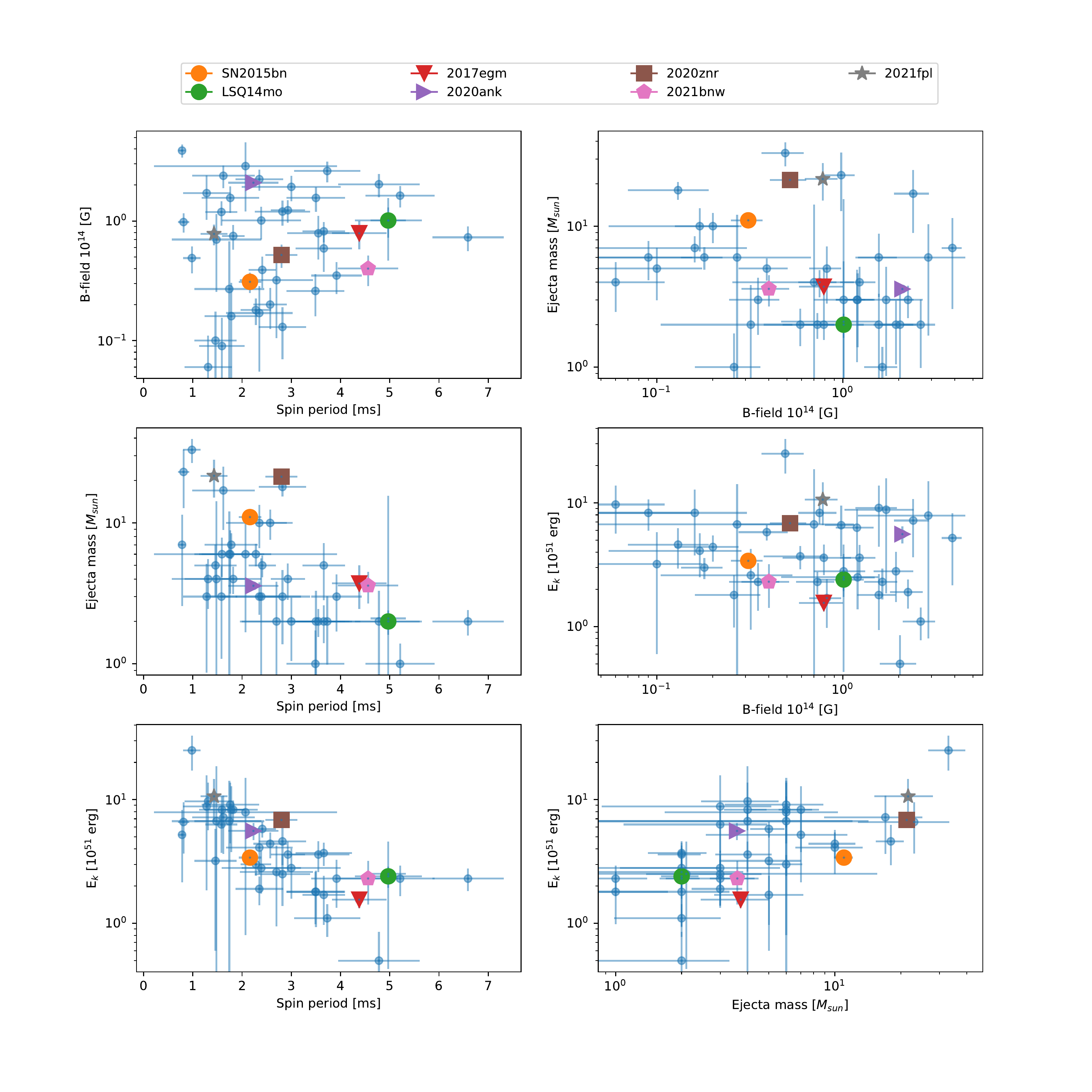}
\vspace*{0.cm}
\caption{Distribution of the \texttt{MOSFiT} best fit magnetar model parameters as from \citet{Nicholl2017}. Overplotted with different coloured symbols are the parameters 
obtained on SN 2015bn, SN 2017egm, SN LSQ14mo, SN 2020ank, 
SN 2020znr, SN 2021bnw and SN 2021fpl given in Table~\ref{tab:mosfitparamsallsne}, for which polarimetry data are available.}
\label{fig:nicholl2017_fig3_updated}
\end{center}
\end{figure*}

Based on the preceding analysis, and assuming that all SLSNe in the sample share a common progenitor that could be described by a magnetar model, we produce a comparison of the \texttt{MOSFiT} best fit parameters obtained with that model. Figure~\ref{fig:nicholl2017_fig3_updated} shows the 
distribution of the \texttt{MOSFiT} best fit parameters as from \citet{Nicholl2017} and retrieved from their work. 
Overplotted are the parameters obtained on SN 2015bn, SN 2017egm, SN LSQ14mo, SN 2020ank, SN 2020znr, SN 2021bnw and SN 2021fpl given in Table~\ref{tab:mosfitparamsallsne}.
In each plot, it is interesting to note that the \texttt{MOSFiT} best fit parameters of SN 2021fpl (grey star symbol) and SN 2015bn (orange disk symbol) are quite close to each other in the parameter space. The only other source, close to these two sources in each of the six plots, and on which polarimetry is available is the W Type SN 2020znr. The data at hand do not show that SN 2020znr underwent a polarization transition, may be because observations were obtained in the R-filter only (see plot in Figure~\ref{fig:polarimetry_spectra_wtype}, bottom). On the other hand, the parameters obtained for SN 2017egm share a distinct locus in the parameter space
with SN LSQ14mo and SN 2021bnw, which might imply that later polarization measurements on them (which would have been too faint to be observed) would have been interesting test beds of W-type polarization properties.   
Finally, we point out that the kinetic energy estimate of SN 2021fpl is slightly higher than the one of SN 2015bn while SN 2015bn has twice the spin period of SN 2021fpl and a lower magnetic field. This means that the magnetar from SN 2021fpl released its energy a lot faster than SN 2015bn, which might give us hints about the possibility that the polarization transition phase of SN 2021fpl could have occurred earlier than the one of SN 2015bn. If this is true, from its position in the plot of $E_{k}$ versus spin period, SN 2020ank would also be polarized, which is not.  However, the parameter inference for SN 2020ank was done using \texttt{MINIM} instead of \texttt{MOSFiT}, which may have systematic and model differences.
All these facts, put together, may indicate that 15bn Type SLSNe have different polarization properties than W Type SLSNe but surely a larger polarimetry sample with higher cadence, ideally in both V- and R-filters, or with spectropolarimetry, is needed to test that hypothesis.

\section{Conclusions}

Understanding polarization properties of H-poor SLSNe to retrieve information from their inner photospheres that could help to understand the nature of their progenitor is not an easy problem to deal with. Its looks reasonable to assume that before maximum light, SLSN-I photospheres are almost spherical \citep[This work,][]{inserra2016,2023arXiv230108111P} with low linear polarization mainly produced by Thomson scattering of light from free electrons, and this independently of the viewing angle of the inner system. According to the SYN++ analysis the photosphere temperatures of SN 2021bnw (W type) were higher compared to those of SN 2021fpl (15bn type) consistently with \citet{Konyves-Toth2022} when polarimetry was obtained on each of these sources. In such a case, the electron scattering opacities could have dominated over the line opacities. This could be one reason why measuring an increase of polarization after maximum light on 15bn type SLSNe-I is more likely to happen than on W type SLSNe-I, even though the 15bn type represents only about 33$\%$ of poor hydrogen SLSNe \citep{Konyves-Toth2022}.

In addition to this point, it is interesting to note that none of the light curves of SN 2021bnw and SN 2021fpl show any strong bumps or undulations after maximum light, contrary to what can be seen, for example, from the the light curves of SN 2018bsz, SN 2017egm and 2017gci \citep[see plot in Figure 5 of][]{2023arXiv230108111P}. After that stage, the light curves are rather smoothly decreasing. It looks therefore reasonable to assume that none of the polarization measurements obtained on these two SLSNe were possibly 'contaminated' by polarization components that would be produced by interactions of the photospheres with the CSM, which is an important point raised by \citet{2023arXiv230108111P}. The presence of a dominant axis in the $Q-U$ plane observed on SN 2021fpl also strengthen that hypothesis. 

Finally, element mixing and clumpiness could also be additional sources of complexity on the interpretation of the polarization results \citep{Tanaka2017}. At late times, once they reach the nebular phase, the inner ejecta of SLSNe becomes visible and constraints on the nucleosynthesis and core structure from spectral analysis and modeling point out to clumped ejecta \citep[][]{2017ApJ...835...13J}. SLSNe produced by stronger engines could imply more mixing and clumping in the ejecta, and possibily additional sources of asymmetry in the system, than for SLSNe produced by weaker engines \citep{Suzuki2021}. From this point of view, SN 2015bn, SN 2021fpl and SN 2020znr are all located in the upper left part of the spin period versus ejecta mass diagram (see Figure~\ref{fig:nicholl2017_fig3_updated}), and can therefore be considered as strong engines. For SN 2015bn and SN 2021fpl, the increase of polarization, or high level of polarization, detected after optical maximum light could be favored by higher mixing and clumping than from weaker engines. On the other hand, as it is mentioned by \citet{2023arXiv230108111P}, SN 2020znr (W type) was observed with polarimetry during the nebular phase but there was no detection of an increase of polarization on that source in that phase \citep[][]{Poidevin2022}. In this case it could be that the temperature of its photosphere was still high enough to prevent the detection of the effects of a high level of mixing, or that the mass of the ejecta was so high that effects of it's clumpy structure was still not detectable in the R-filter used for linear polarimetry, or both. It is worth noting that clumping and asymmetry can strongly affect other signals as well, such as nebular spectra \citep{2017ApJ...835...13J, Dessart2019, Omand2022} and non-thermal emission \citep{Suzuki2021} in radio \citep{Omand2018, Eftekhari2019, Law2019, Eftekhari2021, Murase2021} and x-rays \citep{Kotera2013, Metzger2014, Murase2015, Vurm2021}, which may lead to correlation between polarization and these signals.

Modeling of H-poor SLSN photosphere evolution and fuller analysis of a larger observation sample might be really helpful to explore all these scenarios. 

\section{Summary}

In this work we present new photometric, spectrocopic and imaging polarimetry data combined with publicly available data to study some of the physical properties of the two H-poor SLSN SN 2021bnw and SN 2021fpl. The spectroscopy data were obtained with LT SPRAT. The polarimetry data were obtained on the NOT with ALFOSC in the V-band filter. The photometry data were obtained on SN 2021fpl with LT IO:O. 
From the analysis of these data we find that:
\begin{itemize}
    \item For each SLSN, the best-fit parameters obtained from the magnetar model with MOSFiT do not depart from the range of parameter obtained on other SLSNe discussed in the literature.
    \item An analysis of the spectra with \texttt{SYN++} shows that SN 2021bnw is a W Type, Fast evolver, while SN 2021fpl is a 15bn Type, Slow evolver.
    \item An analysis of the polarimetry data shows no departure from null polarization for SN 2021bnw after maximum light (+81.1 days rest-frame), while $> 3\sigma$ polarization measurements in the range 0.8--1~$\%$ are obtained for SN 2021fpl at four epochs close to and after maximum light (+1.8, +20.6, +34.1 and +43.0 days).
\end{itemize}

These results are compared with results obtained on SN 2020ank, SN 2020znr, SN 2017egm, SN 2015bn, LSQ14mo, and SN 2018bsz, a sample of SNSLe for which constrained polarization measurements are available in the literature. From these comparisons we find that:

\begin{itemize}
    \item The majority of the polarimetry was obtained at diffuse timescale ranging between -1 and 1. 
    \item SN 2021fpl underwent a spectroscopic transition as bit earlier than SN 2015, during which it could also have undergone a polarization transition.  
    \item None of the photospheres of the W Type SLSNe, whether they are Slow or Fast evolvers, show a clear evidence of a departure from symmetry at empirical diffusion timescale lower than 2. The only exception being SN 2017egm which showed an increase of polarization at a empirical diffusion timescale of about 12.  
    \item Measurements at higher empirical diffusion timescale may be needed to see any departure from symmetry of the W Type group.
\end{itemize}


\section*{Acknowledgements}

The authors would like to thank David Morate for his kind support on the NOT, and David Young for the ATLAS python public code he developed to stack and bin ATLAS forced photometry public data.  

F.P. acknowledges support from the Spanish State Research Agency (AEI) under grant number PID2019-105552RB-C43. I.P.-F. acknowledges support from the Spanish State Research Agency (AEI) under grant numbers ESP2017-86852-C4-2-R and PID2019-105552RB- C43. R.K.T. has been supported by the NKFIH/OTKA FK-134432 grant of the National Research, Development and Innovation Office of Hungary (NKFIH) and by the ÚNKP-22-4 New National Excellence Program of the Ministry for Culture and
Innovation from the source of the National Research, Development and Innovation Fund.”

Based on observations made with the Nordic Optical Telescope (NOT), owned in collaboration by the University of Turku and Aarhus University, and operated jointly by Aarhus University, the University of Turku and the University of Oslo, representing Denmark, Finland and Norway, the University of Iceland and Stockholm University at the Observatorio del Roque de los Muchachos, La Palma, Spain, of the Instituto de Astrofisica de Canarias. The data presented here were obtained in part with ALFOSC, which is provided by the Instituto de Astrofisica de Andalucia (IAA) under a joint agreement with the University of Copenhagen and NOT. Some of the data were obtained during CAT service observation Spanish time. ALFOSC polarimetry imaging data of SN2020ank and calibration data were retrieved from the NOT public archive. 

The Liverpool Telescope is operated on the island of La Palma by Liverpool John Moores University in the Spanish Observatorio del Roque de los Muchachos of the Instituto de Astrofisica de Canarias with financial support from the UK Science and Technology Facilities Council.

This work is based in part on observations obtained with the Samuel Oschin 48-inch Telescope at the Palomar Observatory as part of the Zwicky Transient Facility project. ZTF is supported by the NSF under grant AST-1440341 and a collaboration including Caltech, IPAC, the Weizmann Institute for Science, the Oskar Klein Center at Stockholm University, the University of Maryland, the University of Washington, Deutsches Elektronen-Synchrotron and Humboldt University, Los Alamos National Laboratories, the TANGO Consortium of Taiwan, the University of Wisconsin at Milwaukee, and the Lawrence Berkeley National Laboratory. Operations are conducted by the Caltech Optical Observatories (COO), the Infrared Processing and Analysis Center (IPAC), and the University of Washington (UW).

This work has made use of data from the Asteroid Terrestrial- impact Last Alert System (ATLAS) project. The Asteroid Terrestrial- impact Last Alert System (ATLAS) project is primarily funded to search for near earth asteroids through NASA grants NN12AR55G, 80NSSC18K0284, and 80NSSC18K1575; byproducts of the NEO search include images and catalogs from the survey area. This work was partially funded by Kepler/K2 grant J1944/80NSSC19K0112 and HST GO-15889, and STFC grants ST/T000198/1 and ST/S006109/1. The ATLAS science products have been made possible through the contributions of the University of Hawaii Institute for Astronomy, the Queen s University Belfast, the Space Telescope Science Institute, the South African Astronomical Observatory, and The Millennium Institute of Astrophysics (MAS), Chile.

Lasair is supported by the UKRI Science and Technology Facilities Council and is a collaboration between the University of Edinburgh (grant ST/N002512/1) and Queen’s University Belfast (grant ST/N002520/1) within the LSST:UK Science Consortium.

This research has made use of ``Aladin sky atlas'' developed at CDS, Strasbourg Observatory, France 2000A\&AS..143...33B and 2014ASPC..485..277B.

SNID is Copyright (C) 1999-2007 St\'{e}phane Blondin and John L. Tonry, and is available under the GNU General Public License.

This work made use of the python public code \texttt{astropy/photutils: 1.0.2} release developed by \citet{2021zndo...4453725B}.

This research made use of the Transient Name Server (TNS) which is the official IAU mechanism for reporting new astronomical transients such as supernova candidates, As of January 1, 2016.

This research made use of DESI LS DR9 data. The Legacy Surveys consist of three individual and complementary projects: the Dark Energy Camera Legacy Survey (DECaLS; Proposal ID \#2014B-0404; PIs: David Schlegel and Arjun Dey), the Beijing-Arizona Sky Survey (BASS; NOAO Prop. ID \#2015A-0801; PIs: Zhou Xu and Xiaohui Fan), and the Mayall z-band Legacy Survey (MzLS; Prop. ID \#2016A-0453; PI: Arjun Dey). DECaLS, BASS and MzLS together include data obtained, respectively, at the Blanco telescope, Cerro Tololo Inter-American Observatory, NSF’s NOIRLab; the Bok telescope, Steward Observatory, University of Arizona; and the Mayall telescope, Kitt Peak National Observatory, NOIRLab. The Legacy Surveys project is honored to be permitted to conduct astronomical research on Iolkam Du’ag (Kitt Peak), a mountain with particular significance to the Tohono O’odham Nation.

NOIRLab is operated by the Association of Universities for Research in Astronomy (AURA) under a cooperative agreement with the National Science Foundation.

This project used data obtained with the Dark Energy Camera (DECam), which was constructed by the Dark Energy Survey (DES) collaboration. Funding for the DES Projects has been provided by the U.S. Department of Energy, the U.S. National Science Foundation, the Ministry of Science and Education of Spain, the Science and Technology Facilities Council of the United Kingdom, the Higher Education Funding Council for England, the National Center for Supercomputing Applications at the University of Illinois at Urbana-Champaign, the Kavli Institute of Cosmological Physics at the University of Chicago, Center for Cosmology and Astro-Particle Physics at the Ohio State University, the Mitchell Institute for Fundamental Physics and Astronomy at Texas A$\&$M University, Financiadora de Estudos e Projetos, Fundacao Carlos Chagas Filho de Amparo, Financiadora de Estudos e Projetos, Fundacao Carlos Chagas Filho de Amparo a Pesquisa do Estado do Rio de Janeiro, Conselho Nacional de Desenvolvimento Cientifico e Tecnologico and the Ministerio da Ciencia, Tecnologia e Inovacao, the Deutsche Forschungsgemeinschaft and the Collaborating Institutions in the Dark Energy Survey. The Collaborating Institutions are Argonne National Laboratory, the University of California at Santa Cruz, the University of Cambridge, Centro de Investigaciones Energeticas, Medioambientales y Tecnologicas-Madrid, the University of Chicago, University College London, the DES-Brazil Consortium, the University of Edinburgh, the Eidgenossische Technische Hochschule (ETH) Zurich, Fermi National Accelerator Laboratory, the University of Illinois at Urbana-Champaign, the Institut de Ciencies de l’Espai (IEEC/CSIC), the Institut de Fisica d’Altes Energies, Lawrence Berkeley National Laboratory, the Ludwig Maximilians Universitat Munchen and the associated Excellence Cluster Universe, the University of Michigan, NSF’s NOIRLab, the University of Nottingham, the Ohio State University, the University of Pennsylvania, the University of Portsmouth, SLAC National Accelerator Laboratory, Stanford University, the University of Sussex, and Texas A$\&$M University.

The Legacy Surveys imaging of the DESI footprint is supported by the Director, Office of Science, Office of High Energy Physics of the U.S. Department of Energy under Contract No. DE-AC02-05CH1123, by the National Energy Research Scientific Computing Center, a DOE Office of Science User Facility under the same contract; and by the U.S. National Science Foundation, Division of Astronomical Sciences under Contract No. AST-0950945 to NOAO.

\section*{Data Availability}


For science reproducibility purposes the spectra presented in this work are available via WISEReP. The photometry data displayed in Table~\ref{tab:photometry_2021bnw} to Table~\ref{tab:photometry_2021bnw_3} and Table~\ref{tab:photometry_2021fpl} to Table~\ref{tab:photometry_2021fpl_4} will be available online. 





\bibliographystyle{mnras}
\bibliography{sn2021bnw} 




\newpage 
\clearpage

\appendix

\section{LT IO:O Observation logs} \label{ltiooobslog}

The log of the photometry obtained on SN 2021fpl with LT IO$:$O is displayed in Table~\ref{tab:ioolog} (see Section~\ref{ltioo}).

\begin{table}
	\centering
	\caption{Log of the photometry obtained on SN 2021fpl with the LT IO$:$O}
	\label{tab:ioolog}
	\begin{tabular}{llcc} 
          \hline
          UT Time & Exp. Time & Filter & Seeing \\
           &    [s] & & [$\arcsec$] \\
          \hline
          2021-05-13 04:51:47    & 180 & SDSS-U & 1.9\\
          2021-05-13 04:55:17     & 60 & SDSS-G & 1.6\\
          2021-05-13 04:56:49     & 60 & SDSS-R & 1.7\\
          2021-05-13 04:58:19    & 60 & SDSS-I & 1.3\\
          2021-05-13 04:59:49    & 120 & SDSS-Z & 1.5\\
          \hline
          2021-05-17 03:50:52     & 180 & SDSS-U & 2.8 \\
          2021-05-17 03:54:22     & 60 & SDSS-G & 2.0 \\
          2021-05-17 03:55:55     & 60 & SDSS-R & 1.7 \\
          2021-05-17 03:57:25     & 60 & SDSS-I & 1.9 \\
          2021-05-17 03:58:55     & 120 & SDSS-Z & 1.7 \\
          \hline
          2021-05-25 05:00:13     & 180 & SDSS-U & 2.3\\
          2021-05-25 05:03:43      & 60 & SDSS-G & 2.3\\
          2021-05-25 05:05:16     & 60 & SDSS-R & 1.8\\
          2021-05-25 05:06:47   & 60 & SDSS-I & 1.7\\
          2021-05-25 05:08:18    & 120 & SDSS-Z & 1.5\\
          \hline
          2021-06-03 03:10:09     & 180 & SDSS-U & 1.8 \\
          2021-06-03 03:13:38      & 60 & SDSS-G & 1.4 \\
          2021-06-03 03:15:11     & 60 & SDSS-R & 1.5\\
          2021-06-03 03:16:42   & 60 & SDSS-I & 1.4\\
          2021-06-03 03:18:12    & 120 & SDSS-Z & 1.4\\
          \hline
          2021-06-09 02:28:37     & 180 & SDSS-U & 7.9\\
          2021-06-09 02:32:06      & 60 & SDSS-G & 5.4\\
          2021-06-09 02:33:39      & 60 & SDSS-R & ...\\
          2021-06-09 02:35:09   & 60 & SDSS-I & 5.2\\
          2021-06-09 02:36:39    & 120 & SDSS-Z & 3.8\\
          \hline
          2021-06-15 03:50:25     & 180 & SDSS-U & 1.4\\
          2021-06-15 03:53:54   & 60 & SDSS-G & 1.2\\
          2021-06-15 03:55:26    & 60 & SDSS-R & 1.1\\
          2021-06-15 03:56:57  & 60 & SDSS-I & 1.3\\
          2021-06-15 03:58:27    & 120 & SDSS-Z & 1.2\\
          \hline
          2021-06-30 01:02:01     & 180 & SDSS-U & 1.6\\
          2021-06-30 01:05:30      & 60 & SDSS-G & 1.4\\
          2021-06-30 01:07:02     & 60 & SDSS-R & 1.3\\
          2021-06-30 01:08:33   & 60 & SDSS-I & 1.2\\
          2021-06-30 01:10:04    & 120 & SDSS-Z & 1.4\\
          \hline
          2021-07-10 02:43:23     & 180 & SDSS-U & 3.4\\
          2021-07-10 02:46:53      & 60 & SDSS-G & 2.5\\
          2021-07-10 02:48:24     & 60 & SDSS-R & 2.4\\
          2021-07-10 02:49:54   & 60 & SDSS-I & 3.3\\
          2021-07-10 02:51:25    & 120 & SDSS-Z & 2.7 \\
          \hline
          2021-07-17 01:54:31     & 180 & SDSS-U & 1.8\\
          2021-07-17 01:57:59      & 60 & SDSS-G & 1.5\\
          2021-07-17 01:59:31     & 60 & SDSS-R & 1.5\\
          2021-07-17 02:01:01    & 60 & SDSS-I & 1.2\\
          2021-07-17 02:02:31    & 120 & SDSS-Z & 1.2\\
          \hline
	\end{tabular}
\end{table}

\section{Spectroscopy Observation logs} \label{specobslog}

Information about the spectroscopy obtained on SN 2021bnw and 2021fpl is given in Table~\ref{tab:speclog} (see Section~\ref{dataspec}).

\begin{table*}
	\centering
	\caption{Spectra obtained on SN 2021bnw and SN 2021fpl}
	\label{tab:speclog}
	\begin{tabular}{lllllll} 
          \hline
          Source & UT Time & Telescope/Instrument & Exp. Time & Observer/s & reducer/s & Group   \\
          & &  &  [s] && &  \\
          \hline
          SN 2021bnw & 2021-02-04 06:59:55  & ESO-NTT / EFOSC2-NTT & 600 & M. Magee, J. Terwel & S. Prentice, L. Harvey & ePESSTO+ \\
          SN 2021bnw & 2021-05-16 23:39:45  & LT / SPRAT & 3$\times$ 300 & I. P\'{e}rez-Fournon & I. P\'{e}rez-Fournon & This work \\
          SN 2021fpl & 2021-04-26 05:17:50  & LT / SPRAT & ? & S. Prentice & M. Deckers & TCD \\
          SN 2021fpl & 2021-05-13 04:27:30  & LT / SPRAT & 3$\times$ 240 & I. P\'{e}rez-Fournon & I. P\'{e}rez-Fournon & This work \\
          SN 2021fpl & 2021-06-15 03:33:31  & LT / SPRAT & 3$\times$ 300 & I. P\'{e}rez-Fournon & I. P\'{e}rez-Fournon & This work \\
          SN 2021fpl & 2021-07-10 02:23:54  & LT / SPRAT & 3$\times$ 350 & I. P\'{e}rez-Fournon & I. P\'{e}rez-Fournon & This work \\
          \hline
          \hline
	\end{tabular}
\end{table*}

\section{ALFOSC polarimetry Observation logs} \label{polobslog}

The observation logs of the imaging polarimetry obtained on SN 2021bnw, on its host and on SN 2021fpl are displayed in Table~\ref{tab:pollog} (see Section~\ref{datapol}).

\begin{table}
	\centering
	\caption{Observations log of the imaging polarimetry observations.}
	\label{tab:pollog}
	\begin{tabular}{lllcc} 
          \hline
          UT Time & Object & Exp. Time & Filter & Seeing \\
           &  &  [s] & & [$\arcsec$] \\
          \hline
          2021-05-17 22:11:48 &  SN 2021bnw     & $ 1 \times ( 4 \times 450 $) & V &1.9\\
          2021-05-17 22:53:56 &  HD127769    & $ 2 \times ( 4\times 1) $ & V &1.5\\
          2021-05-17 22:57:30 &  GD319       & $ 2 \times ( 4 \times 5) $ & V &1.3\\
          2021-05-18 03:18:03 &  SN 2021fpl     & $ 5 \times ( 4 \times 45) $ & V &1.0\\
          2021-05-18 04:00:19 &  BD+32 3739     & $ 1 \times (4 \times 2 )$ & V &0.8\\
          2021-05-18 04:03:33 &  Hiltner960     & $ 1 \times (4 \times 5) $ & V &0.8\\
          \hline
          2021-06-14 04:34:33 &  SN 2021fpl     & $ 7 \times ( 4 \times 45) $ & V &0.7\\
          2021-06-14 05:01:40 &  BD+32 3739     & $1 \times ( 4 \times 2 )$ & V &0.7\\
          2021-06-14 05:04:25 &  Hiltner960     & $1 \times ( 4 \times 5 )$ & V &0.6\\
          \hline
          2021-06-29 02:24:13 &  SN 2021fpl     & $ 5 \times ( 4 \times 90) $ & V &0.7\\
          2021-06-29 02:59:49 &  BD+32 3739     & $ 3 \times  (4 \times 2) $ & V &0.7\\
          2021-06-29 03:03:41 &  Hiltner960     & $ 2 \times  (4 \times 5) $ & V &0.9\\
          \hline
          2021-07-09 03:11:58 &  SN 2021fpl     & $ 6 \times ( 4 \times 90) $ & V &1.6\\
          2021-07-09 03:55:12 &  BD+32 3739     & $ 3 \times  (4 \times 2) $ & V &1.3\\
          2021-07-09 03:59:23 &  Hiltner960     & $ 2 \times  (4 \times 5) $ & V &1.3\\          \hline
          2022-01-29 20:24:46 &  BD+64 106     & $ 3 \times ( 4 \times 1) $ & V &1.2\\
          2022-01-30 01:36:31 & HD25120 & $ 4 \times ( 4 \times 1) $ & V &0.9\\
          2022-01-30 01:42:57 & HD94851 & $ 4 \times ( 4 \times 1) $ & V &1.1\\
          2022-01-30 03:43:06 & SN 2021bnw Host     & $1 \times (  4 \times 900 )$ & V &0.9\\
          \hline
	\end{tabular}
\end{table}

\section{Photometry Table} \label{photometry_table}

The photometry from SN 2021bnw and SN 2021fpl used along this work are compiled in Table~\ref{tab:photometry_2021bnw} to Table~\ref{tab:photometry_2021bnw_3} and Table~\ref{tab:photometry_2021fpl} to Table~\ref{tab:photometry_2021fpl_4}, respectively  (see Section~\ref{dataphot} and Section~\ref{lc_modelling}).

\begin{table}
	\centering
	\caption{ZTF and ATLAS photometry of SN 2021bnw.}
	\label{tab:photometry_2021bnw}
	\begin{tabular}{lcccccc} 
          \hline
          MJD &
                Magnitude &  $\sigma_{\rm Magnitude}$& band & survey \\
          \hline
 59219.640000 &    20.508 &       0.907 &      r &     ATLAS \\
 59221.630000 &    21.202 &       1.363 &      r &     ATLAS \\
 59224.540000 &    23.900 &       9.375 &      r &     ATLAS \\
 59227.440637 &    21.072 &       0.408 &      r &       ZTF \\
 59227.640000 &    20.496 &       0.978 &      g &     ATLAS \\
 59229.570000 &    20.845 &       0.721 &      g &     ATLAS \\
 59231.419965 &    20.492 &       0.266 &      r &       ZTF \\
 59231.500000 &    20.389 &       0.261 &      g &     ATLAS \\
 59233.480000 &    19.368 &       1.702 &      c &     ATLAS \\
 59235.540000 &    19.543 &       0.539 &      c &     ATLAS \\
 59237.640000 &    18.816 &      37.824 &      c &     ATLAS \\
 59248.414120 &    17.973 &       0.064 &      r &       ZTF \\
 59249.660000 &    17.944 &       0.238 &      r &     ATLAS \\
 59250.370023 &    17.857 &       0.064 &      g &       ZTF \\
 59250.377361 &    18.115 &       0.067 &      g &       ZTF \\
 59251.570000 &    18.016 &       0.068 &      g &     ATLAS \\
 59252.318773 &    17.934 &       0.058 &      g &       ZTF \\
 59252.379005 &    17.713 &       0.056 &  sdssg &       ZTF \\
 59253.450000 &    17.737 &       0.044 &      c &     ATLAS \\
 59254.300023 &    17.620 &       0.058 &      r &       ZTF \\
 59255.460000 &    17.605 &       0.037 &      c &     ATLAS \\
 59256.369352 &    17.520 &       0.051 &  sdssg &       ZTF \\
 59256.399120 &    17.740 &       0.045 &      g &       ZTF \\
 59257.510000 &    17.587 &       0.025 &      g &     ATLAS \\
 59258.341076 &    17.671 &       0.046 &  sdssr &       ZTF \\
 59258.366447 &    17.454 &       0.051 &      g &       ZTF \\
 59261.440000 &    17.474 &       0.024 &      r &     ATLAS \\
 59262.349213 &    17.565 &       0.046 &      g &       ZTF \\
 59262.410694 &    17.394 &       0.057 &      r &       ZTF \\
 59263.480000 &    17.406 &       1.162 &      g &     ATLAS \\
 59264.360023 &    17.552 &       0.043 &      r &       ZTF \\
 59266.332535 &    17.603 &       0.056 &      g &       ZTF \\
 59266.417014 &    17.452 &       0.062 &  sdssg &       ZTF \\
 59267.460000 &    17.566 &       0.051 &      r &     ATLAS \\
 59268.354398 &    17.577 &       0.038 &      g &       ZTF \\
 59268.410000 &    17.614 &       0.089 &      g &     ATLAS \\
 59268.431817 &    17.434 &       0.050 &      r &       ZTF \\
 59269.350000 &    17.618 &       0.087 &      r &     ATLAS \\
 59270.309595 &    17.563 &       0.050 &      g &       ZTF \\
 59270.382465 &    17.477 &       0.061 &      r &       ZTF \\
 59274.239607 &    17.500 &       0.104 &  sdssg &       ZTF \\
 59274.359919 &    17.611 &       0.066 &      r &       ZTF \\
 59276.256076 &    17.561 &       0.056 &      g &       ZTF \\
 59276.337083 &    17.616 &       0.054 &      g &       ZTF \\
 59276.480000 &    17.646 &       0.056 &      o &     ATLAS \\
 59277.470000 &    17.675 &       0.052 &      r &     ATLAS \\
 59280.286493 &    17.674 &       0.054 &      r &       ZTF \\
 59280.286493 &    17.674 &       0.054 &  sdssr &       ZTF \\
 59280.359826 &    17.644 &       0.070 &      g &       ZTF \\
 59287.380000 &    17.534 &       0.319 &      g &     ATLAS \\
 59288.360000 &    17.751 &       0.043 &      o &     ATLAS \\
 59290.255706 &    17.756 &       0.056 &      g &       ZTF \\
 59290.295637 &    17.728 &       0.079 &  sdssg &       ZTF \\
 59291.440000 &    17.770 &       0.043 &      g &     ATLAS \\
 59292.254606 &    17.774 &       0.052 &  sdssr &       ZTF \\
 59292.294757 &    17.728 &       0.072 &      r &       ZTF \\
 59292.430000 &    17.812 &       0.060 &      o &     ATLAS \\
 59293.420000 &    17.808 &       0.049 &      r &     ATLAS \\
 59294.253889 &    17.770 &       0.066 &      g &       ZTF \\
 59294.319282 &    17.768 &       0.053 &      r &       ZTF \\
 59297.205984 &    17.774 &       0.088 &  sdssg &       ZTF \\
            \hline
	\end{tabular}
\end{table}

\begin{table}
	\centering
	\caption{ZTF and ATLAS photometry of SN 2021bnw.}
	\label{tab:photometry_2021bnw_2}
	\begin{tabular}{lcccccc} 
          \hline
          MJD &
                Magnitude &  $\sigma_{\rm Magnitude}$& band & survey \\
          \hline
 59297.256701 &    17.809 &       0.054 &      r &       ZTF \\
 59297.450000 &    17.810 &       0.070 &      o &     ATLAS \\
 59301.356435 &    17.733 &       0.091 &      g &       ZTF \\
 59303.278657 &    17.830 &       0.072 &      r &       ZTF \\
 59303.351562 &    17.769 &       0.056 &      g &       ZTF \\
 59303.400000 &    17.853 &       0.123 &      g &     ATLAS \\
 59304.430000 &    17.859 &       0.070 &      r &     ATLAS \\
 59305.290150 &    17.846 &       0.070 &  sdssg &       ZTF \\
 59305.290150 &    17.846 &       0.070 &  sdssg &       ZTF \\
 59305.520000 &    17.938 &       0.142 &      o &     ATLAS \\
 59307.247315 &    17.835 &       0.056 &      g &       ZTF \\
 59307.279155 &    17.873 &       0.060 &      r &       ZTF \\
 59309.192778 &    17.956 &       0.053 &  sdssg &       ZTF \\
 59309.233009 &    17.962 &       0.051 &      g &       ZTF \\
 59309.360000 &    17.943 &       0.048 &      r &     ATLAS \\
 59311.170463 &    17.930 &       0.064 &      r &       ZTF \\
 59311.211875 &    17.984 &       0.095 &  sdssg &       ZTF \\
 59313.212685 &    18.025 &       0.060 &  sdssr &       ZTF \\
 59313.233762 &    18.082 &       0.090 &      g &       ZTF \\
 59313.380000 &    18.104 &       0.049 &      g &     ATLAS \\
 59315.211956 &    18.159 &       0.066 &      r &       ZTF \\
 59317.192662 &    18.113 &       0.043 &  sdssr &       ZTF \\
 59317.192662 &    18.113 &       0.043 &      r &       ZTF \\
 59317.233762 &    18.213 &       0.086 &      r &       ZTF \\
 59317.233762 &    18.213 &       0.086 &  sdssg &       ZTF \\
 59319.350000 &    18.100 &       0.085 &      o &     ATLAS \\
 59320.173715 &    18.278 &       0.068 &      g &       ZTF \\
 59321.350000 &    18.154 &       0.048 &      r &     ATLAS \\
 59322.270417 &    18.117 &       0.057 &      r &       ZTF \\
 59323.360000 &    18.179 &       0.092 &      o &     ATLAS \\
 59324.171412 &    18.286 &       0.109 &      r &       ZTF \\
 59324.171412 &    18.286 &       0.109 &      g &       ZTF \\
 59324.171412 &    18.286 &       0.109 &      g &       ZTF \\
 59324.212095 &    18.084 &       0.060 &  sdssr &       ZTF \\
 59324.212095 &    18.084 &       0.060 &      g &       ZTF \\
 59324.212095 &    18.084 &       0.060 &  sdssr &       ZTF \\
 59329.232905 &    18.103 &       0.073 &      r &       ZTF \\
 59329.232905 &    18.103 &       0.073 &  sdssr &       ZTF \\
 59329.232905 &    18.103 &       0.073 &      g &       ZTF \\
 59329.312373 &    18.148 &       0.113 &      r &       ZTF \\
 59329.312373 &    18.148 &       0.113 &      r &       ZTF \\
 59329.312373 &    18.148 &       0.113 &  sdssg &       ZTF \\
 59331.420000 &    18.169 &       0.421 &      g &     ATLAS \\
 59333.390000 &    18.211 &       0.127 &      o &     ATLAS \\
 59334.340000 &    18.138 &       0.084 &      o &     ATLAS \\
 59335.191123 &    18.167 &       0.059 &      g &       ZTF \\
 59335.211921 &    18.384 &       0.108 &      r &       ZTF \\
 59335.340000 &    18.259 &       0.088 &      r &     ATLAS \\
 59337.360000 &    18.292 &       0.222 &      g &     ATLAS \\
 59338.192245 &    18.434 &       0.086 &  sdssg &       ZTF \\
 59338.237176 &    18.186 &       0.060 &      g &       ZTF \\
 59340.192211 &    18.390 &       0.089 &      r &       ZTF \\
 59340.233472 &    18.228 &       0.063 &  sdssr &       ZTF \\
 59342.235521 &    18.467 &       0.084 &      r &       ZTF \\
 59342.290637 &    18.220 &       0.083 &      g &       ZTF \\
 59343.300000 &    18.311 &       0.076 &      c &     ATLAS \\
 59344.208750 &    18.279 &       0.064 &  sdssr &       ZTF \\
 59344.269143 &    18.428 &       0.077 &      g &       ZTF \\
 59345.350000 &    18.353 &       0.076 &      g &     ATLAS \\
 59346.331991 &    18.461 &       0.119 &  sdssg &       ZTF \\
 59347.350000 &    18.460 &       0.085 &      c &     ATLAS \\
            \hline
	\end{tabular}
\end{table}

\begin{table}
	\centering
	\caption{ZTF and ATLAS photometry of SN 2021bnw.}
	\label{tab:photometry_2021bnw_3}
	\begin{tabular}{lcccccc} 
          \hline
          MJD &
                Magnitude &  $\sigma_{\rm Magnitude}$& band & survey \\
          \hline
 59348.175706 &    18.308 &       0.066 &      g &       ZTF \\
 59349.310000 &    18.326 &       0.094 &      r &     ATLAS \\
 59350.177998 &    18.554 &       0.108 &      r &       ZTF \\
 59350.229606 &    18.338 &       0.073 &  sdssr &       ZTF \\
 59351.290000 &    18.370 &       0.107 &      g &     ATLAS \\
 59353.192778 &    18.356 &       0.084 &      r &       ZTF \\
 59353.221829 &    18.587 &       0.128 &      g &       ZTF \\
 59353.260000 &    18.327 &       0.182 &      r &     ATLAS \\
 59356.235822 &    18.461 &       0.112 &  sdssr &       ZTF \\
 59357.310000 &    18.471 &       0.170 &      g &     ATLAS \\
 59358.257118 &    18.778 &       0.209 &      g &       ZTF \\
 59358.280000 &    18.552 &       0.236 &      o &     ATLAS \\
 59359.310000 &    18.611 &       0.235 &      o &     ATLAS \\
 59361.194583 &    18.713 &       0.090 &  sdssr &       ZTF \\
 59361.255590 &    19.017 &       0.261 &  sdssg &       ZTF \\
 59361.350000 &    18.653 &       0.268 &      o &     ATLAS \\
 59363.214005 &    19.069 &       0.115 &  sdssg &       ZTF \\
 59363.310000 &    18.853 &       0.106 &      r &     ATLAS \\
 59365.181620 &    18.894 &       0.093 &  sdssr &       ZTF \\
 59365.204595 &    19.263 &       0.154 &      g &       ZTF \\
 59367.208229 &    19.062 &       0.091 &  sdssr &       ZTF \\
 59367.233565 &    19.498 &       0.169 &  sdssg &       ZTF \\
 59367.290000 &    19.034 &       0.125 &      o &     ATLAS \\
 59369.224005 &    19.556 &       0.174 &      g &       ZTF \\
 59369.300000 &    19.191 &       0.139 &      r &     ATLAS \\
 59371.225613 &    19.308 &       0.125 &  sdssr &       ZTF \\
 59371.310000 &    19.359 &       0.272 &      o &     ATLAS \\
 59373.217245 &    19.325 &       0.125 &  sdssr &       ZTF \\
 59373.300000 &    19.315 &       0.160 &      r &     ATLAS \\
 59375.226968 &    19.365 &       0.148 &  sdssr &       ZTF \\
 59377.260000 &    19.617 &       0.289 &      c &     ATLAS \\
 59387.270000 &    19.994 &       2.072 &      o &     ATLAS \\
 59393.260000 &    19.759 &       0.372 &      o &     ATLAS \\
 59397.260000 &    20.146 &       0.453 &      o &     ATLAS \\
 59401.250000 &    19.989 &       1.128 &      o &     ATLAS \\
           \hline
	\end{tabular}
\end{table}

\begin{table}
	\centering
	\caption{ZTF and ATLAS photometry of SN 2021fpl.}
	\label{tab:photometry_2021fpl}
	\begin{tabular}{lcccccc} 
          \hline
          MJD &
                Magnitude &  $\sigma_{\rm Magnitude}$& band & survey \\
          \hline
59288.630000 &    18.151 &       0.135 &     o &     ATLAS \\
59292.630000 &    18.105 &       0.134 &     o &     ATLAS \\
59300.620000 &    17.559 &       0.661 &     o &     ATLAS \\
59305.620000 &    17.793 &       0.115 &     o &     ATLAS \\
59319.620000 &    17.872 &       0.065 &     c &     ATLAS \\
59321.600000 &    17.586 &       0.054 &     o &     ATLAS \\
59322.610000 &    17.622 &       0.202 &     o &     ATLAS \\
59323.590000 &    17.817 &       0.052 &     c &     ATLAS \\
59326.590000 &    17.619 &       0.054 &     o &     ATLAS \\
59327.580000 &    17.759 &       0.096 &     c &     ATLAS \\
59328.610000 &    17.572 &       0.054 &     o &     ATLAS \\
59331.600000 &    17.531 &       0.100 &     o &     ATLAS \\
59333.270000 &    17.530 &       0.080 &     o &     ATLAS \\
59334.480243 &    17.959 &       0.072 & sdssg &       ZTF \\
59339.425671 &    17.944 &       0.069 & sdssg &       ZTF \\
59342.440185 &    17.506 &       0.044 & sdssr &       ZTF \\
59342.461933 &    18.025 &       0.059 & sdssg &       ZTF \\
59345.421933 &    18.023 &       0.088 & sdssg &       ZTF \\
59346.590000 &    17.806 &       0.056 &     c &     ATLAS \\
59347.205057 &    17.883 &       0.073 & sdssg &        LT \\
59347.206127 &    17.586 &       0.060 & sdssr &        LT \\
59347.207162 &    17.405 &       0.068 & sdssi &        LT \\
59347.208211 &    17.618 &       0.112 & sdssz &        LT \\
59347.540000 &    17.522 &       0.043 &     o &     ATLAS \\
59349.560000 &    17.769 &       0.064 &     c &     ATLAS \\
59351.162752 &    17.960 &       0.074 & sdssg &        LT \\
59351.163828 &    17.623 &       0.060 & sdssr &        LT \\
59351.164874 &    17.408 &       0.070 & sdssi &        LT \\
59351.165920 &    17.612 &       0.065 & sdssz &        LT \\
59351.480000 &    17.505 &       0.073 &     o &     ATLAS \\
59352.424387 &    17.553 &       0.059 & sdssr &       ZTF \\
59353.470000 &    17.791 &       0.079 &     c &     ATLAS \\
59357.461401 &    17.540 &       0.041 & sdssr &       ZTF \\
59357.540000 &    17.478 &       0.060 &     o &     ATLAS \\
59359.170000 &    17.561 &       0.049 &     o &     ATLAS \\
59359.210918 &    18.034 &       0.096 & sdssg &        LT \\
59359.211989 &    17.643 &       0.049 & sdssr &        LT \\
59359.213040 &    17.471 &       0.048 & sdssi &        LT \\
59359.214095 &    17.655 &       0.084 & sdssz &        LT \\
59361.110000 &    17.457 &       0.273 &     o &     ATLAS \\
59361.419190 &    17.568 &       0.055 & sdssr &       ZTF \\
59366.415613 &    17.606 &       0.053 & sdssr &       ZTF \\
59366.451863 &    18.285 &       0.076 & sdssg &       ZTF \\
59367.480000 &    17.653 &       0.055 &     o &     ATLAS \\
59368.134473 &    18.176 &       0.069 & sdssg &        LT \\
59368.135543 &    17.732 &       0.046 & sdssr &        LT \\
59368.136592 &    17.501 &       0.038 & sdssi &        LT \\
59368.137643 &    17.676 &       0.086 & sdssz &        LT \\
59369.376401 &    18.335 &       0.076 & sdssg &       ZTF \\
59369.441238 &    17.652 &       0.056 & sdssr &       ZTF \\
59371.397963 &    18.409 &       0.069 & sdssg &       ZTF \\
59371.490000 &    18.024 &       0.047 &     c &     ATLAS \\
59373.530000 &    17.627 &       0.050 &     o &     ATLAS \\
59374.398518 &    18.396 &       0.097 & sdssg &       ZTF \\
59374.438843 &    17.728 &       0.085 & sdssr &       ZTF \\
59375.540000 &    18.053 &       0.059 &     c &     ATLAS \\
59376.398866 &    18.485 &       0.061 & sdssg &       ZTF \\
59376.437940 &    17.751 &       0.048 & sdssr &       ZTF \\
59377.490000 &    17.671 &       0.056 &     o &     ATLAS \\
59378.396910 &    17.769 &       0.054 & sdssr &       ZTF \\
59378.419213 &    18.523 &       0.070 & sdssg &       ZTF \\
           \hline
	\end{tabular}
\end{table}

\begin{table}
	\centering
	\caption{ZTF and ATLAS photometry of SN 2021fpl.}
	\label{tab:photometry_2021fpl_2}
	\begin{tabular}{lcccccc} 
          \hline
          MJD &
                Magnitude &  $\sigma_{\rm Magnitude}$& band & survey \\
          \hline
59379.520000 &    18.086 &       0.055 &     c &     ATLAS \\
59380.405000 &    18.552 &       0.281 & sdssg &       ZTF \\
59382.357153 &    17.803 &       0.061 & sdssr &       ZTF \\
59382.414201 &    18.605 &       0.088 & sdssg &       ZTF \\
59383.480000 &    18.212 &       0.134 &     c &     ATLAS \\
59385.317766 &    18.755 &       0.177 & sdssg &       ZTF \\
59385.510000 &    17.769 &       0.084 &     o &     ATLAS \\
59387.357014 &    18.839 &       0.144 & sdssg &       ZTF \\
59387.421238 &    17.920 &       0.065 & sdssr &       ZTF \\
59387.460000 &    17.822 &       0.756 &     o &     ATLAS \\
59389.420000 &    17.482 &       0.319 &     o &     ATLAS \\
59393.335301 &    19.192 &       0.220 & sdssg &       ZTF \\
59393.560000 &    18.014 &       0.113 &     o &     ATLAS \\
59395.040000 &    17.997 &       0.063 &     o &     ATLAS \\
59395.045484 &    18.856 &       0.108 & sdssg &        LT \\
59395.046556 &    18.168 &       0.063 & sdssr &        LT \\
59395.047606 &    17.865 &       0.059 & sdssi &        LT \\
59395.048655 &    17.926 &       0.066 & sdssz &        LT \\
59396.378276 &    18.173 &       0.071 & sdssr &       ZTF \\
59396.419769 &    19.072 &       0.142 & sdssg &       ZTF \\
59399.378588 &    18.114 &       0.078 & sdssr &       ZTF \\
59399.398901 &    19.243 &       0.107 & sdssg &       ZTF \\
59399.460000 &    18.117 &       0.062 &     o &     ATLAS \\
59400.600000 &    18.085 &       0.078 &     o &     ATLAS \\
59401.419143 &    18.245 &       0.059 & sdssr &       ZTF \\
59402.520000 &    18.746 &       0.122 &     c &     ATLAS \\
59403.334491 &    18.241 &       0.082 & sdssr &       ZTF \\
59403.376840 &    19.444 &       0.119 & sdssg &       ZTF \\
59403.460000 &    18.305 &       0.083 &     o &     ATLAS \\
59405.115887 &    19.254 &       0.120 & sdssg &        LT \\
59405.116946 &    18.448 &       0.061 & sdssr &        LT \\
59405.117988 &    18.145 &       0.118 & sdssi &        LT \\
59405.119034 &    18.211 &       0.116 & sdssz &        LT \\
59405.560000 &    18.816 &       0.130 &     c &     ATLAS \\
59406.373866 &    19.520 &       0.156 & sdssg &       ZTF \\
59406.440000 &    18.961 &       0.202 &     c &     ATLAS \\
59407.540000 &    18.262 &       0.668 &     o &     ATLAS \\
59408.332465 &    18.526 &       0.087 & sdssr &       ZTF \\
59408.336262 &    19.636 &       0.204 & sdssg &       ZTF \\
59408.420000 &    18.384 &       0.086 &     o &     ATLAS \\
59409.440000 &    19.039 &       0.432 &     c &     ATLAS \\
59410.335984 &    19.854 &       0.223 & sdssg &       ZTF \\
59411.460000 &    18.560 &       0.101 &     o &     ATLAS \\
59412.081938 &    19.644 &       0.122 & sdssg &        LT \\
59412.082998 &    18.759 &       0.072 & sdssr &        LT \\
59412.084038 &    18.450 &       0.081 & sdssi &        LT \\
59412.085085 &    18.421 &       0.083 & sdssz &        LT \\
59412.335116 &    20.134 &       0.282 & sdssg &       ZTF \\
59412.357037 &    18.656 &       0.076 & sdssr &       ZTF \\
59412.430000 &    18.553 &       0.087 &     o &     ATLAS \\
59413.480000 &    19.096 &       0.180 &     c &     ATLAS \\
59415.020000 &    18.694 &       0.134 &     o &     ATLAS \\
59415.289873 &    19.807 &       0.343 & sdssg &       ZTF \\
59415.289873 &    19.807 &       0.343 & sdssg &       ZTF \\
59415.358368 &    18.812 &       0.137 & sdssr &       ZTF \\
59415.358368 &    18.812 &       0.137 & sdssr &       ZTF \\
59416.410000 &    19.484 &       2.462 &     o &     ATLAS \\
59421.510000 &    18.756 &       0.197 &     o &     ATLAS \\
59422.277384 &    19.811 &       0.261 & sdssg &       ZTF \\
59423.490000 &    19.076 &       0.254 &     o &     ATLAS \\
59424.272523 &    20.608 &       0.340 & sdssg &       ZTF \\
           \hline
	\end{tabular}
\end{table}

\begin{table}
	\centering
	\caption{ZTF and ATLAS photometry of SN 2021fpl.}
	\label{tab:photometry_2021fpl_3}
	\begin{tabular}{lcccccc} 
          \hline
          MJD &
                Magnitude &  $\sigma_{\rm Magnitude}$& band & survey \\
          \hline
59424.900000 &    19.094 &       0.156 &     o &     ATLAS \\
59427.390000 &    19.714 &       0.198 &     c &     ATLAS \\
59428.294340 &    20.415 &       0.297 & sdssg &       ZTF \\
59428.390000 &    19.594 &       0.201 &     c &     ATLAS \\
59429.450000 &    19.061 &       0.181 &     o &     ATLAS \\
59430.251817 &    19.204 &       0.105 & sdssr &       ZTF \\
59431.450000 &    19.658 &       0.214 &     c &     ATLAS \\
59433.450000 &    19.136 &       0.285 &     o &     ATLAS \\
59435.214213 &    20.464 &       0.277 & sdssg &       ZTF \\
59435.420000 &    19.586 &       0.192 &     c &     ATLAS \\
59437.430000 &    19.109 &       0.278 &     o &     ATLAS \\
59439.420000 &    19.785 &       0.276 &     c &     ATLAS \\
59440.230671 &    19.335 &       0.148 & sdssr &       ZTF \\
59440.230671 &    19.335 &       0.148 & sdssr &       ZTF \\
59441.350000 &    19.257 &       0.213 &     o &     ATLAS \\
59442.232188 &    19.429 &       0.148 & sdssr &       ZTF \\
59443.410000 &    19.423 &       0.309 &     o &     ATLAS \\
59450.960000 &    19.307 &       0.321 &     o &     ATLAS \\
59452.430000 &    19.410 &       0.452 &     o &     ATLAS \\
59454.292998 &    19.541 &       0.161 & sdssr &       ZTF \\
59455.360000 &    19.410 &       0.283 &     o &     ATLAS \\
59457.380000 &    20.282 &       0.491 &     c &     ATLAS \\
59458.230266 &    19.658 &       0.141 & sdssr &       ZTF \\
59458.251817 &    20.566 &       0.245 & sdssg &       ZTF \\
59461.350000 &    20.141 &       0.298 &     c &     ATLAS \\
59463.360000 &    19.510 &       0.258 &     o &     ATLAS \\
59464.231701 &    20.922 &       0.369 & sdssg &       ZTF \\
59465.350000 &    20.098 &       0.292 &     c &     ATLAS \\
59467.360000 &    19.743 &       0.450 &     o &     ATLAS \\
59468.319954 &    19.659 &       0.170 & sdssr &       ZTF \\
59469.350000 &    20.120 &       0.571 &     c &     ATLAS \\
59471.209850 &    19.668 &       0.267 & sdssr &       ZTF \\
59471.330000 &    19.810 &       0.520 &     o &     ATLAS \\
59477.380000 &    19.743 &       1.467 &     o &     ATLAS \\
59479.100000 &    20.096 &       0.959 &     o &     ATLAS \\
59480.750000 &    19.834 &       0.436 &     o &     ATLAS \\
59483.330000 &    19.951 &       0.497 &     o &     ATLAS \\
59484.230718 &    19.778 &       0.164 & sdssr &       ZTF \\
59485.320000 &    20.003 &       0.476 &     o &     ATLAS \\
59487.320000 &    19.810 &       0.345 &     o &     ATLAS \\
59489.172743 &    19.778 &       0.194 & sdssr &       ZTF \\
59489.290000 &    19.965 &       0.660 &     o &     ATLAS \\
59491.310000 &    20.055 &       0.846 &     o &     ATLAS \\
59493.280000 &    20.002 &       0.448 &     o &     ATLAS \\
59495.290000 &    20.002 &       1.612 &     o &     ATLAS \\
59498.185764 &    20.203 &       0.276 & sdssr &       ZTF \\
59499.250000 &    19.804 &       0.790 &     o &     ATLAS \\
59503.148148 &    19.964 &       0.292 & sdssr &       ZTF \\
59503.270000 &    19.680 &       0.750 &     o &     ATLAS \\
59504.690000 &    24.142 &      96.156 &     o &     ATLAS \\
59505.114792 &    20.046 &       0.299 & sdssr &       ZTF \\
59508.770000 &    19.915 &       0.535 &     o &     ATLAS \\
59511.250000 &    19.663 &       0.410 &     o &     ATLAS \\
59512.111655 &    20.091 &       0.252 & sdssr &       ZTF \\
59513.250000 &    20.594 &       0.779 &     c &     ATLAS \\
59515.280000 &    19.909 &       0.761 &     o &     ATLAS \\
59517.310000 &    20.634 &       1.019 &     c &     ATLAS \\
59519.240000 &    20.308 &       0.880 &     o &     ATLAS \\
59521.149178 &    20.108 &       0.258 & sdssr &       ZTF \\
59523.085220 &    19.987 &       0.159 & sdssr &       ZTF \\
59523.240000 &    20.172 &       1.048 &     o &     ATLAS \\
           \hline
	\end{tabular}
\end{table}

\begin{table}
	\centering
	\caption{ZTF and ATLAS photometry of SN 2021fpl.}
	\label{tab:photometry_2021fpl_4}
	\begin{tabular}{lcccccc} 
          \hline
          MJD &
                Magnitude &  $\sigma_{\rm Magnitude}$& band & survey \\
          \hline
59527.200000 &    19.491 &       1.293 &     o &     ATLAS \\
59529.230000 &    20.960 &       2.063 &     o &     ATLAS \\
59531.240000 &    19.868 &       1.601 &     o &     ATLAS \\
59539.230000 &    19.610 &       2.260 &     o &     ATLAS \\
59541.220000 &    20.530 &       0.787 &     o &     ATLAS \\
59545.190000 &    21.347 &      12.038 &     o &     ATLAS \\
59547.210000 &    20.216 &       1.345 &     o &     ATLAS \\
           \hline
	\end{tabular}
\end{table}

\section{Spectrum modelling}

The best-fit local parameter values of the \texttt{SYN++} modelling obtained on SN 2021bnw and SN 2021fpl are displayed in Table~\ref{tab:sn21bnw_lokparams} and Table~\ref{tab:sn21fpl_lokparams}, respectively (see Section~\ref{syn++}).

\begin{table}
\caption{Best-fit local parameter values of the  \texttt{SYN++} modelling of SN~2021bnw. }
\label{tab:sn21bnw_lokparams}
\centering
\begin{tabular}{lcccccc}
\hline 
Phase: -14d & & & & & & \\
\hline
\hline
Ions & O II & C II &  &  &  &  \\
\hline
$\log_\tau$ & -0.30 & -0.10 &  &  &  &  \\
$v_{\rm min}$ & 20.00 & 20.00 &  &  &  &  \\
$v_{\rm max}$ & 30.00 & 30.00 &  &  &  &  \\
$aux$ & 1.00 & 0.70 &  &  &  &  \\
$T_{\rm exc}$ & 12.00 & 14.00 &  &  &  &  \\
\hline
\hline
Phase: +77d &&&&&& \\
\hline
\hline
Ions & O I  & Na I & Si II & Mg II & Fe II & Si II $\nu$ \\
\hline
$\log_\tau$ & 0.42 & 0.20 & 0.10 & -0.30 & -0.40 & 0.30 \\
$v_{\rm min}$ & 4.00 & 4.00 & 4.00 & 4.00 & 4.00 & 15.00 \\
$v_{\rm max}$ & 50.00 & 50.00 & 50.00 & 50.00 & 50.00 & 50.00 \\
$aux$ & 4.00 & 5.00 & 2.00 & 4.00 & 5.00 & 10.00 \\
$T_{\rm exc}$ & 6.50 & 6.50 & 6.50 & 6.50 & 6.50 & 6.50 \\
\hline
\end{tabular}
\end{table}

\begin{table}
\caption{Best-fit local parameter values of the  \texttt{SYN++} modelling of SN~2021fpl. }
\label{tab:sn21fpl_lokparams}
\centering
\begin{tabular}{lcccc}
\hline 
Phase: -18d & & & &  \\
\hline
\hline
Ions & Si II & Fe III &  &  \\
\hline
$\log_\tau$ & -0.50 & -1.40 &  &  \\
$v_{\rm min}$ & 11.00 & 11.00 &  &  \\
$v_{\rm max}$ & 30.00 & 30.00 &  &  \\
$aux$ & 6.00 & 4.00 &  &  \\
$T_{\rm exc}$ & 6.50 & 5.70 &  &  \\
\hline
\hline
Phase: -3d & & & &  \\
\hline
Ions & Si II & Fe III  & Fe II \\
\hline
$\log_\tau$ & 0.50 & -1.00 &  0.50 & \\
$v_{\rm min}$ & 6.00 & 6.00  & 6.00 & \\
$v_{\rm max}$ & 30.00 & 30.00  & 30.00 & \\
$aux$ & 4.00 & 4.00 &  3.00 & \\
$T_{\rm exc}$ & 5.70 & 5.70  & 5.70 & \\
\hline
\hline
Phase: +27d and +49d & & &  & \\
\hline
Ions & Si II & Fe III &  Fe II v \\
\hline
$\log_\tau$ & 2.00 & -1.00    & -3.50 &\\
$v_{\rm min}$ & 6.00 & 6.00   &15.00 & \\
$v_{\rm max}$ & 30.00 & 30.00   &30.00& \\
$aux$ & 0.50 & 4.00   &2.00& \\
$T_{\rm exc}$ & 5.00 & 5.00  & 12.00 & \\
\hline
\end{tabular}
\end{table}

\section{Q-U plots} \label{QU_plots_appendix}

\begin{figure*}
\begin{center}
\vspace*{2mm}
\centering
\hspace*{0.cm}
\includegraphics[width=80mm,angle=0]{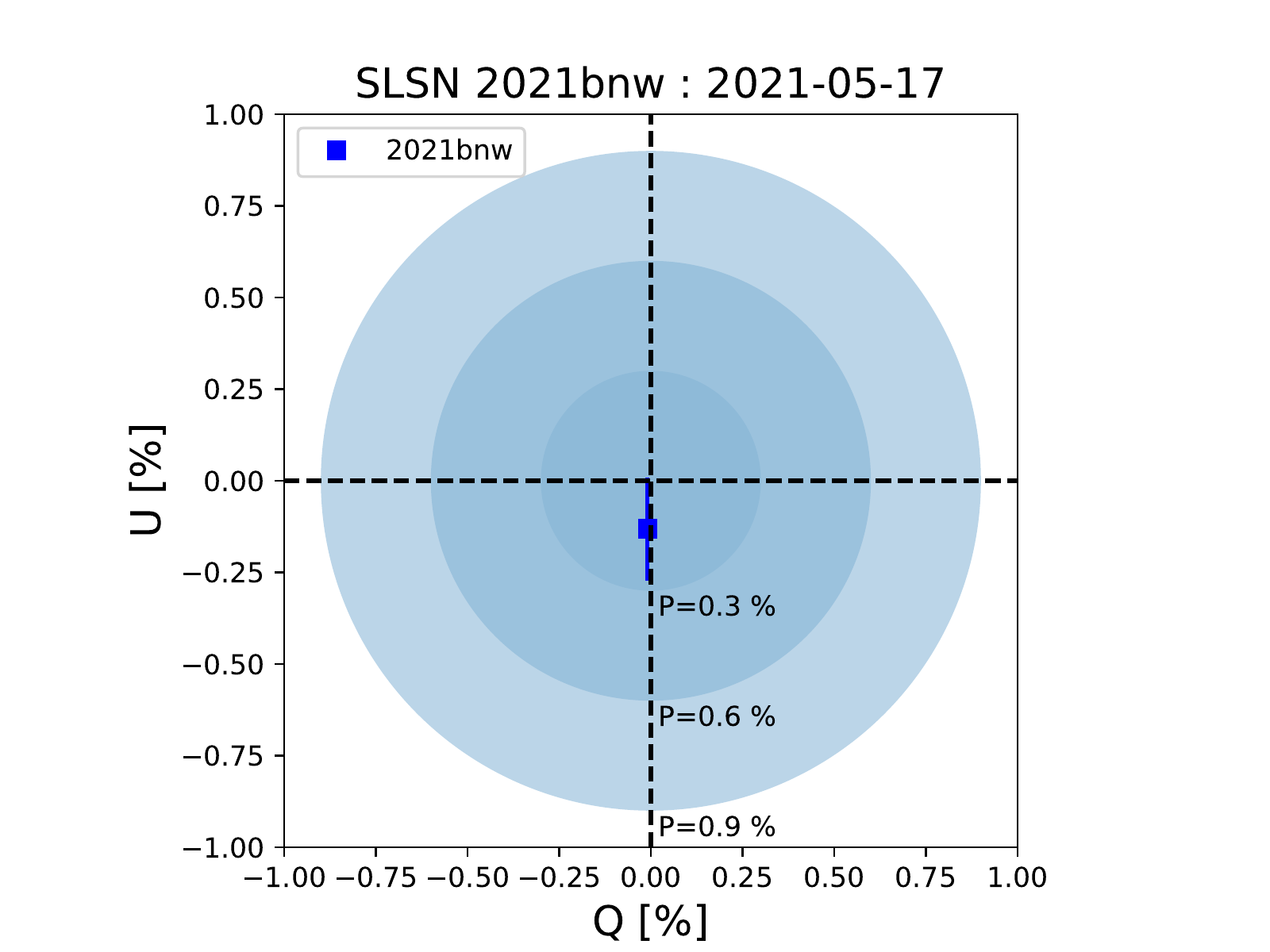}
\includegraphics[width=80mm,angle=0]{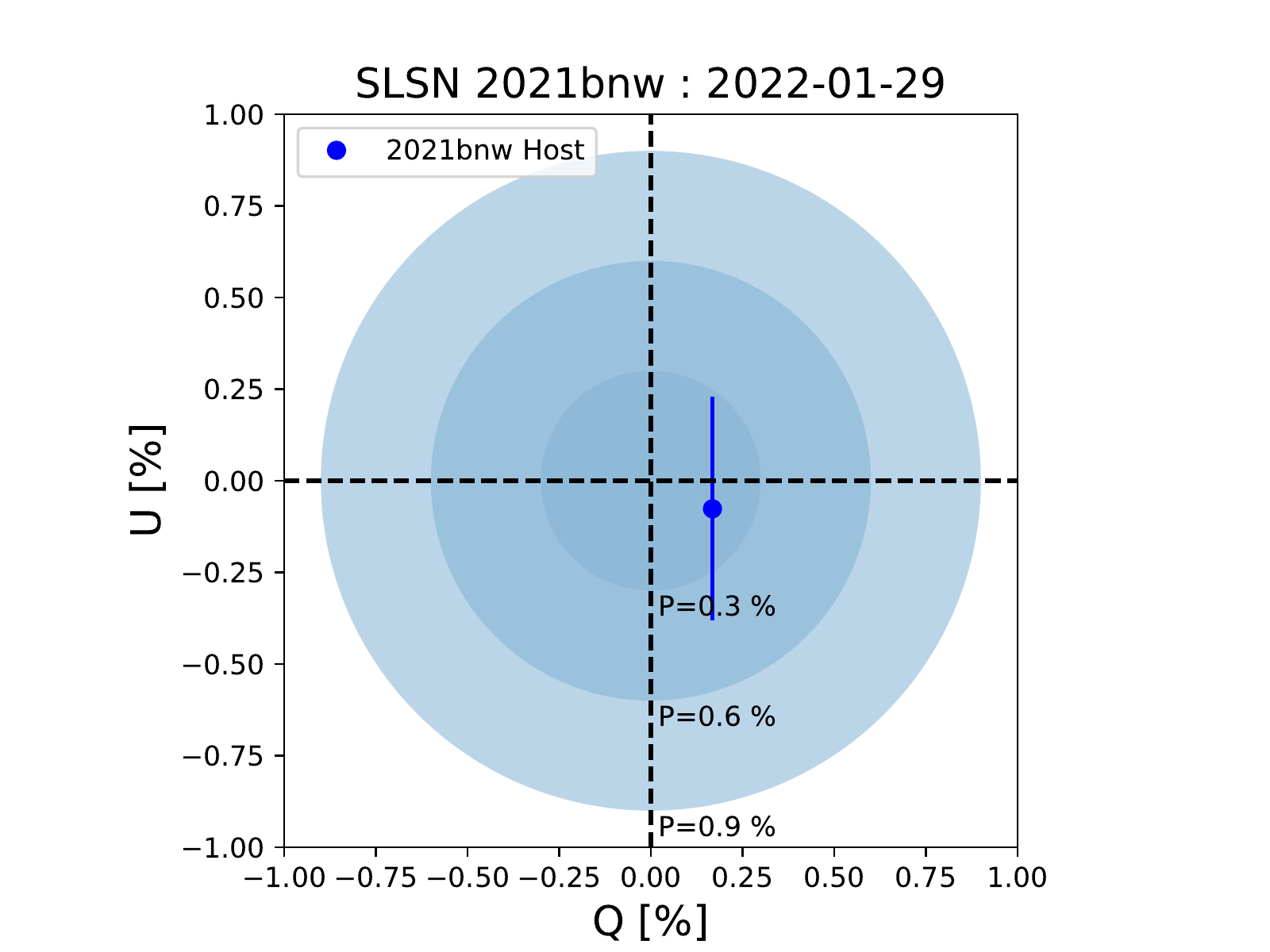}
\includegraphics[width=80mm,angle=0]{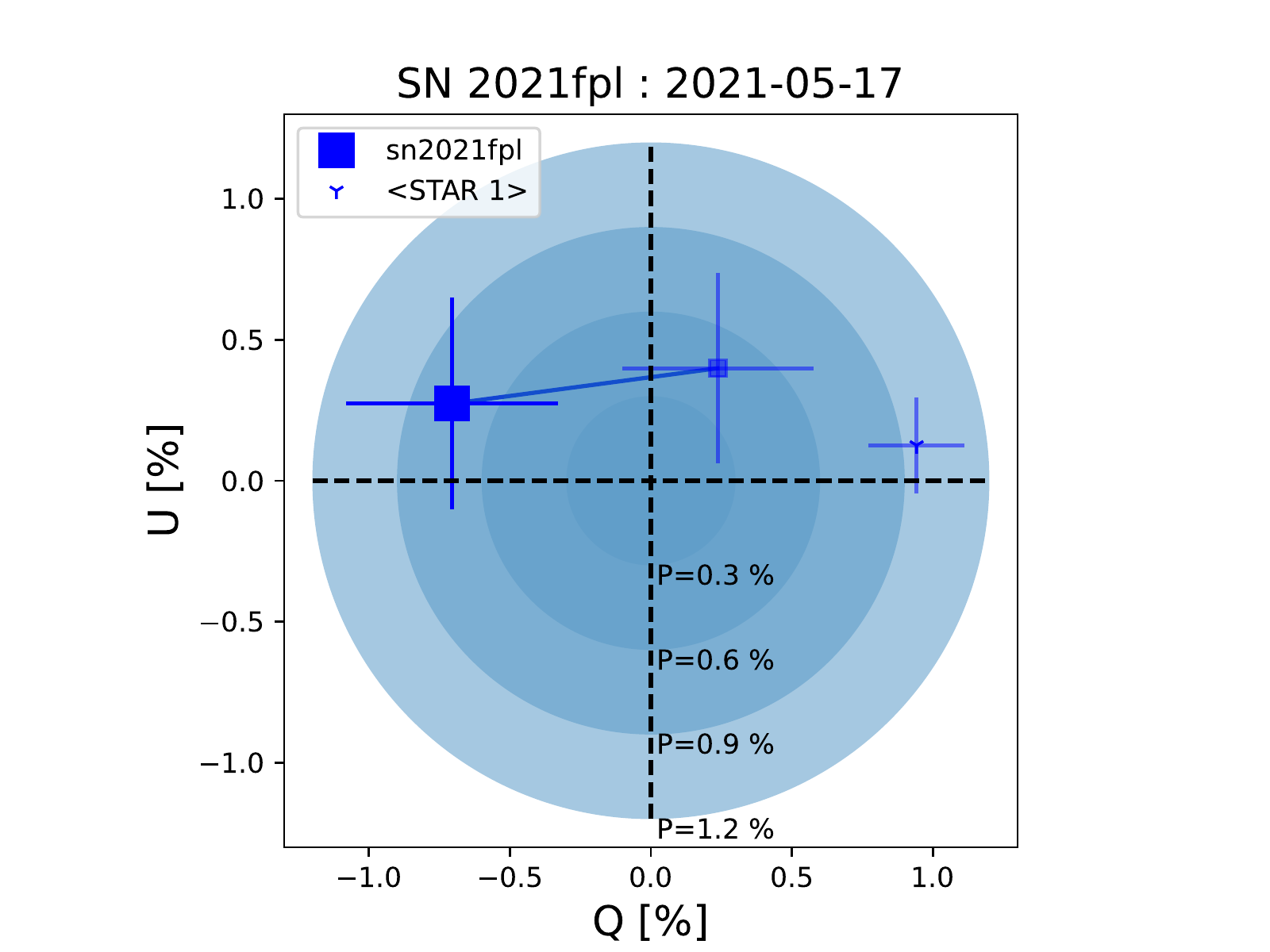}
\includegraphics[width=80mm,angle=0]{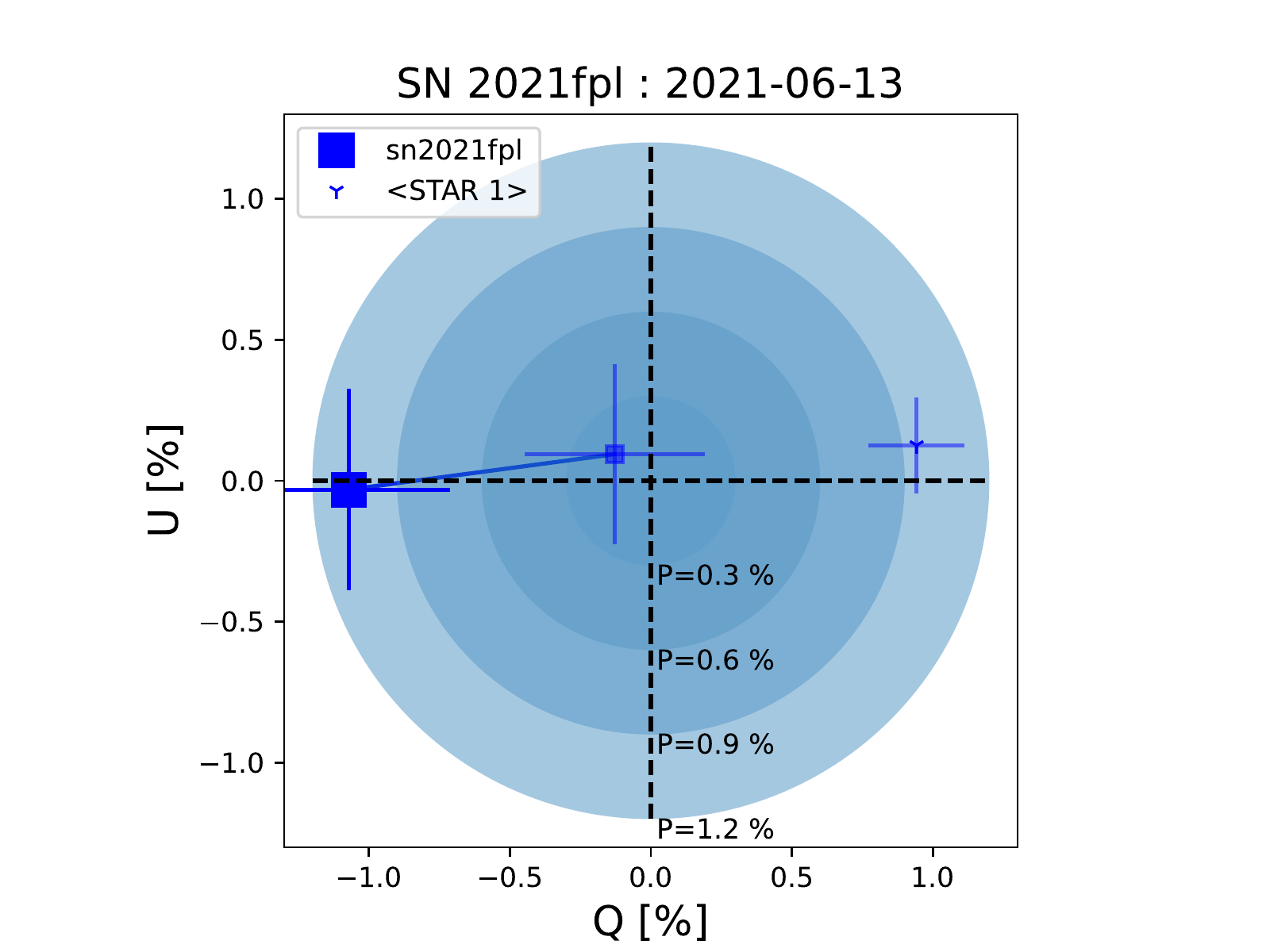}
\includegraphics[width=80mm,angle=0]{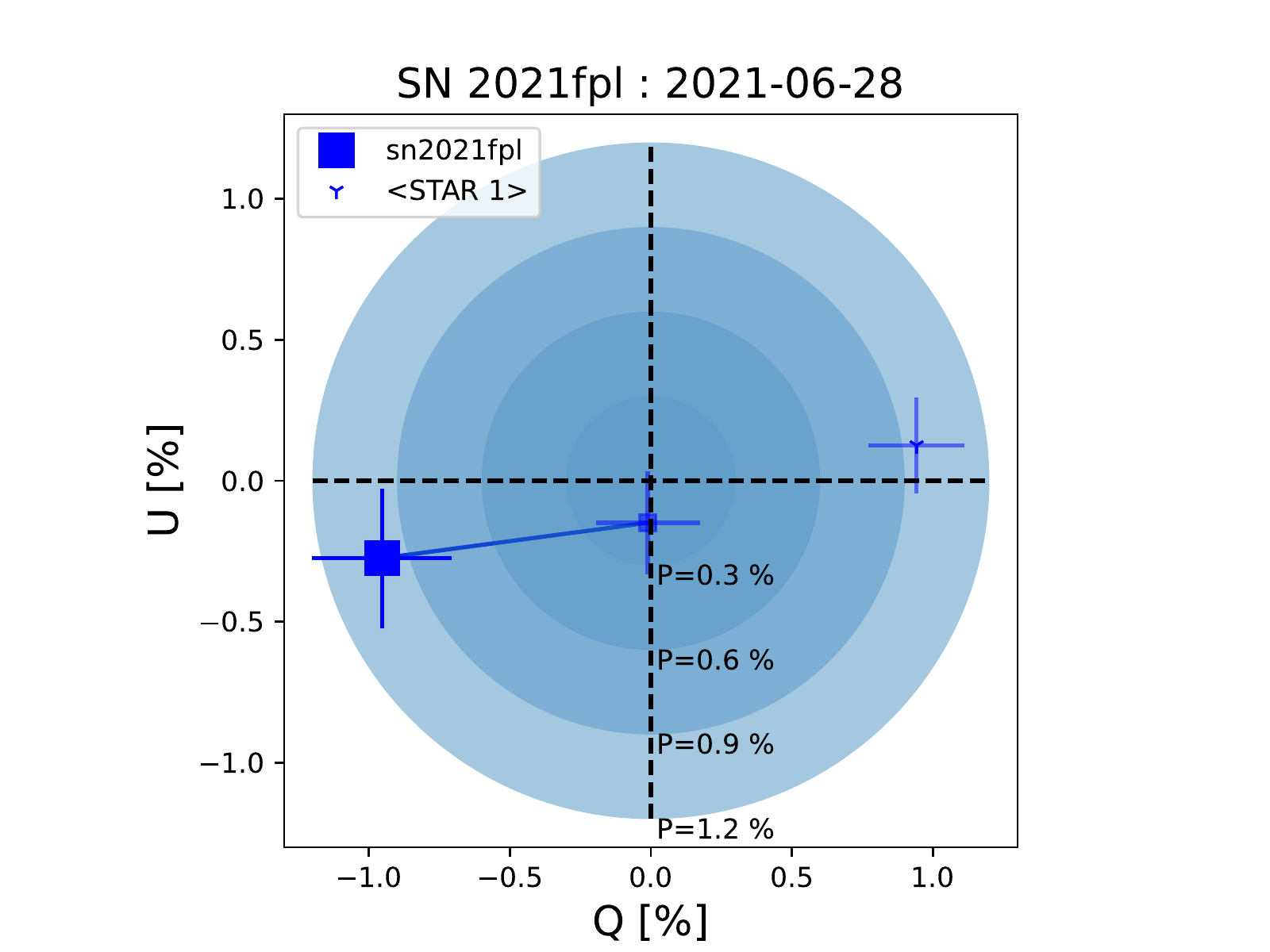}
\includegraphics[width=80mm,angle=0]{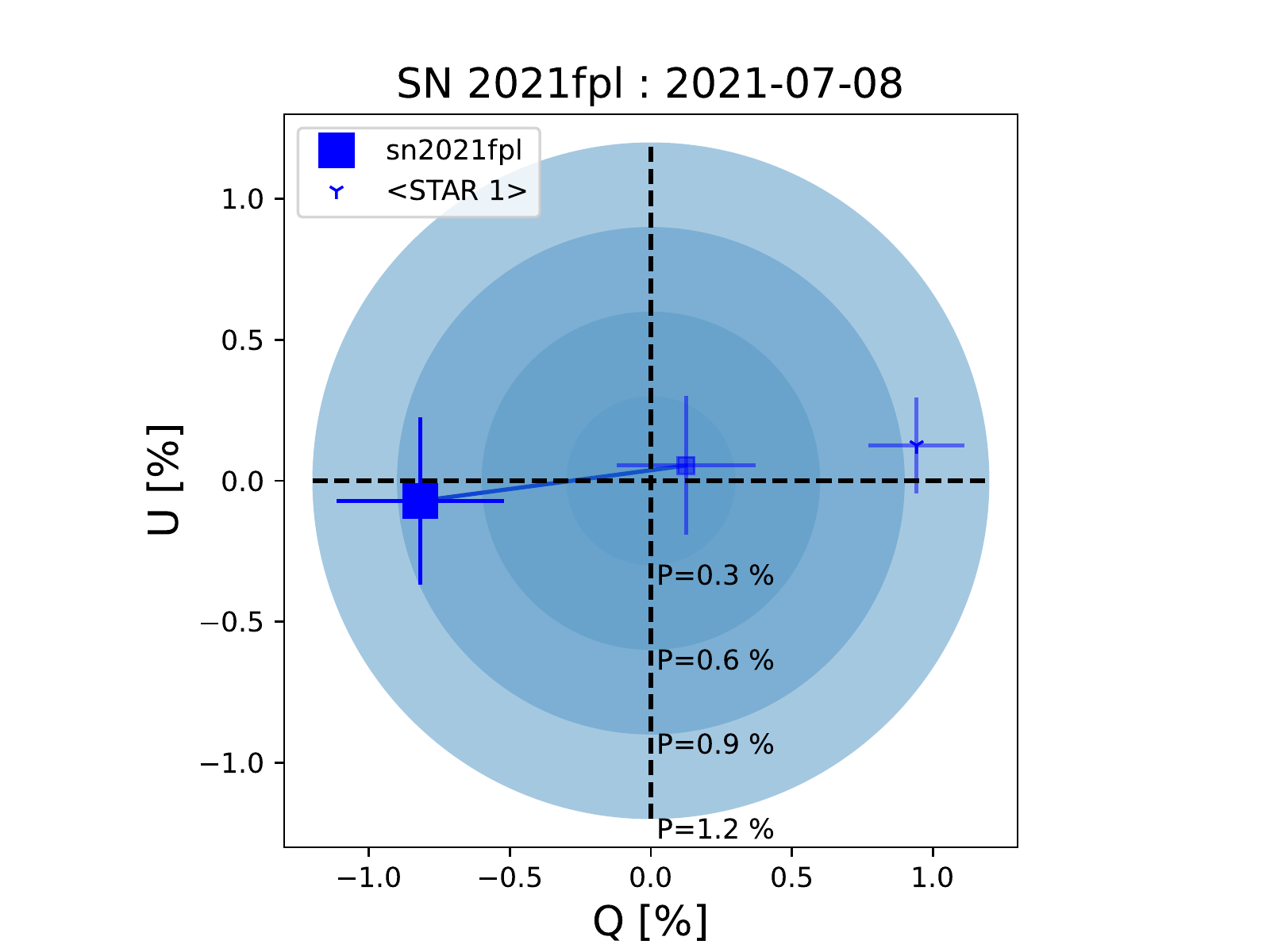}
\vspace*{0.7cm}
\caption{First row from top: SN 2021bnw (left) and its host (right), in the $Q-U$ plane, in the plane-of-sky reference frame after instrumental polarization ($IP$) and zero polarization angle ($ZPA$) corrections without Interstellar Polarization (ISP) correction. Concentric discs show polarization degrees ranges up to 0.3 $\%$, 0.6 $\%$ and 0.9 $\%$.
Second and third row from top: SN 2020fpl ISP corrected $Q-U$ estimates have been obtained after subtraction of the $IP$, $ZPA$ corrected STAR 1 $Q-U$ estimate averaged over the four epochs (see Section~\ref{pol_analysis} and Table~\ref{tab:pol_results} for details). The final values are shown with the double sized square symbols. The double sized square symbols are connected to the half translucent square symbols to show the effect of the ISP correction in the $Q-U$ plane. Concentric discs show polarization degrees ranges up to 0.3 $\%$, 0.6 $\%$, 0.9 $\%$ and 1.2 $\%$.}
\label{fig:QU_plots}
\end{center}
\end{figure*}

Intrinsic polarization in the $Q-U$ plane of SN 2021bnw, SN 2021bnw's host and SN 2021fpl are shown in the plots displayed in Figure~\ref{fig:QU_plots} (see Section~\ref{pol_analysis}).

\bsp	
\label{lastpage}
\end{document}